\documentclass[11pt]{article}
\usepackage[english]{babel}
\usepackage[cp850]{inputenc}
\usepackage[dvips]{graphicx}
\usepackage{amsmath,amsfonts,amsthm,amssymb}
\usepackage{mathrsfs}
\usepackage[usenames, dvipsnames]{color}
\usepackage{fancyhdr}
\usepackage{stmaryrd}
\usepackage[colorlinks=true,citecolor=red,linkcolor=blue,urlcolor=blue,pdfstartview=FitH]{hyperref}
\usepackage{dsfont}
\usepackage{xcolor}

\font\tenmath=msbm10 scaled 1200

\font\sevenmath=msbm7 scaled 1200
\font\fivemath=msbm5 scaled 1200 

\newfam\mathfam \textfont\mathfam=\tenmath
\scriptfont\mathfam=\sevenmath \scriptscriptfont\mathfam=\fivemath

\def\R{{\mathbb{R}}}
\def\N{{\mathbb{N}}}
\def\E{{\mathbb{E}}}
\def\D{{\mathbb{D}}}

\def\P{{\mathbb{P}}}

\def\F{{\cal F}}
\def\I{\Phi}

\newtheorem{theo}{Theorem}[section]
\newtheorem{lem}{Lemma}[section]
\newtheorem{prop}{Proposition}[section]
\newtheorem{cor}{Corollary}[section]
\newtheorem{defi}{Definition}[section]

\newfam\mathfam \textfont\mathfam=\tenmath
\scriptfont\mathfam=\sevenmath \scriptscriptfont\mathfam=\fivemath

\def \cqfd{\quad\Box}

\def \bs{\bigskip}

\def \Cov{{\rm Cov}}

\title{\bf Optimal posting price of limit orders: learning by trading}

\author{{\sc Sophie Laruelle} \thanks{Laboratoire de Probabilit\'es et 
Mod\`eles al\'eatoires, UMR~7599, UPMC, 4, pl. Jussieu, F-75252 Paris Cedex 5, France. E-mail: {\tt  sophie.laruelle@upmc.fr}}  \and {\sc Charles-Albert Lehalle} \thanks{Head of Quantitative Research, Cr\'edit Agricole Cheuvreux, CALYON group ; 9 quai Paul Doumer, 92920 Paris La D\'efense. E-mail: {\tt clehalle@cheuvreux.com}} \and {\sc Gilles Pag\`es} \thanks{Laboratoire de Probabilit\'es et 
Mod\`eles al\'eatoires, UMR~7599, UPMC, case 188, 4, pl. Jussieu, F-75252 Paris Cedex 5, France. E-mail: {\tt  gilles.pages@upmc.fr}}
}
\date{}

\begin{document}
\maketitle 

\begin{abstract}
Considering that a trader or a trading algorithm interacting with markets during continuous auctions can be modeled by an iterating procedure adjusting the price at which he posts orders at a given rhythm, this paper proposes a procedure minimizing his costs. We prove the $a.s.$ convergence of the algorithm under assumptions on the cost function and give some practical criteria on model parameters to ensure that the conditions to use the algorithm are fulfilled (using notably the co-monotony principle). We illustrate our results with numerical experiments on both simulated data and using a financial market dataset.
\end{abstract}


\bs \noindent {\em 2010 AMS classification:} 62L20,
secondary: 
62P05, 
60G55, 
65C05.

\section{Introduction}

In recent years, with the growth of electronic trading, most of the transactions in the markets occur in \emph{Limit Order Books}. During the matching of electronic orders, traders send orders of two kinds to the market: passive ($i.e.$ limit  or patient orders) which will not give birth to a trade but will stay in the order book (sell orders at a higher price than the \emph{higher bid price} or buy orders at a lower price than the \emph{lower ask price} are passive orders) and aggressive orders ($i.e.$ market or impatient orders) which will generate a trade (sell orders at a lower price than the higher passive buy price or buy orders at a higher price than the lowest passive price). When a trader has to buy or sell a large number of shares, he cannot just send his large order at once (because it would consume all of the available liquidity in the order book, impacting the price at his disadvantage); he has first to schedule his trading rate to strike balance between the market risk and the market impact cost of being too aggressive (too many orders exhaust the order book and makes the price move). Several theoretical frameworks have been proposed for optimal scheduling of large orders (see~\cite{AlmChr},~\cite{BouDanLeh},~\cite{PreShaShr},~\cite{AlfFruSch}). Once this optimal trading rate is known, the trader has to send smaller orders in the (electronic) limit order book by alternating limit ($i.e.$ patient or passive) orders and market ($i.e.$ urgent or aggressive) orders. The optimal mix of limit and market orders for a trader has not been investigated in the quantitative literature even if it has been studied from a global economic efficiency viewpoint (see for instance~\cite{FouKadKan}). It has not either been investigated from the viewpoint of one trader trying to optimize his own interactions with other market participants. One of the difficulties from the trader prospective is that transactions obtained by inserting a passive order in a limit order book is a functional of its distance to the mid-price, giving birth to a large number of possible tactics in terms of reassessment of the price of such orders.

In this paper, we study the optimal distance to submit passive orders in a limit order book, without needing a model of the limit order book dynamics (see $e.g.$~\cite{AbeJed} or~\cite{GueLehRaz} for such models of limit order books). 

Optimal submission strategies have been studied in the microstructure literature using utility framework and optimal control (see~\cite{AveSto},~\cite{GueFTLeh},~\cite{BayLud},~\cite{GuiMniPha} and~\cite{GuiPha}). The authors of such papers consider an agent who plays the role of a market maker, $i.e.$ he provides liquidity on the exchange by quoting bid and ask prices at which he is willing to buy and sell a specific quantity of assets. Strategies for bid and ask orders are derived by maximizing his utility function. 

Our approach is different: we consider an agent who wants to buy (or sell) during a short period $[0,T]$ a quantity $Q_T$ of traded assets and we look for the optimal distance where he has to post his order to minimize the execution cost. 

We are typically at a smaller time scale than in usual optimal liquidation frameworks. In fact order posting strategies derived from the viewpoint presented below can be ``plugged'' into any larger scale strategy. We are modeling the market impact of an aggressive order using a penalization function $\kappa\cdot\Phi(Q)$, where $Q$ is the size of the market order.

If a stochastic algorithm approach has been already proposed in \cite{LarLehPag} by the authors for optimal spatial split of orders across different Dark Pools, here the purpose is not to control fractions of the size of orders, but to adjust successive posting prices to converge to an optimal price. Qualitatively, this framework can be used as soon as a trader wants to trade a given quantity $Q_T$ over a given time interval $[0,T]$ with no firm constraint on its trading rate between $0$ and $T$. It is typically the case for small \emph{Implementation Shortfall} benchmarked orders. The trader can post his order very close to the ``\emph{fair price}''$(S_t)_{t\in[0,T]}$ (which can be seen as the fundamental price, the mid price of the available trading venues or any other reference price). In this case he will be exposed to the risk to trade too fast at a ``bad price'' and being adversely selected. Conversely he can post it far away from the fair price; in that case he will be exposed to never obtain a transaction for the whole quantity $Q_T$, but only for a part of it (say the positive part of $Q_T-N_T$, where $N_T$ is the quantity that the trading flow allowed him to trade). He will then have to consume aggressively liquidity with the remaining quantity, disturbing the market and paying not only the current market price $S_T$, but also a market impact (say $S_T\,\I( Q_T-N_T)$ where $\I$ is a market impact penalization function).

The approach presented here follows the mechanism of a ``learning trader''. He will try to guess the optimal posting distance to the fair price achieving the balance between being too demanding in price and too impatient, by successive trials, errors and corrections. 
The optimal recursive procedure derived from our framework gives the best price adjustment to apply to an order on given stopping time (reassessment dates) given the observed past on the market. We provide proofs of the convergence of the procedure and of its optimality.

To this end, we model the execution process of orders by a Poisson process $(N_t^{(\delta)})_{0\leq t\leq T}$ which intensity $\Lambda_T(\delta,S)$ depends on the fair price $\left(S_t\right)_{t\geq0}$ and the distance of order submission $\delta$. The execution cost results from the sum of the price of the executed quantity and a penalization function depending on the remaining quantity to be executed at the end of the period $[0,T]$. This penalty $\kappa\cdot\Phi(Q)$ models the over-cost induced by crossing the spread and the resulting market impact of this execution. The aim is to find the optimal distance $\delta^*\in[0,\delta_{\rm{max}}]$, where $\delta_{\rm{max}}$ is the depth of the limit order book, which minimizes the execution cost. In practice, the prices are constrained to be on a ``tick size grid''(see~\cite{RobRos}), instead of being on the real line. We will follow the approach of papers on market making \cite{HoSto2,AveSto,GueFTLeh} assuming the the tick size is small enough to not change the dynamics of the price and of the bid-ask spread~\cite{McCul}. Nevertheless, a ``rounding effect'' would not change the nature of our results, since any projection ($e.g.$ to nearest-neighbor) is a non-decreasing transform and the co-monotony principle still holds (by slightly adapting the proofs). This leads to an optimization problem under constraints which we solve by using a recursive stochastic procedure with projection (this particular class of algorithm is studied in~\cite{KusCla} and~\cite{KusYin}). We prove the $a.s.$ convergence of the constrained algorithm under additional assumptions on the execution cost function. From a practical point of view, it is not easy to check the conditions on the cost function. So we give criteria on the model parameters which ensure the viability of the algorithm which relies on the co-monotony principle assumed to be satisfied by the ``fair price'' process $(S_t)_{t\in[0,T]}$. This principle will be detailed in the Appendix Section \ref{DistQuatre}.
We conclude this paper by some numerical experiments with simulated and real data. We consider the Poisson intensity presented in~\cite{AveSto} and use a Brownian motion to model the fair price dynamics. We plot the cost function and its derivative and show the convergence of the algorithm to its target $\delta^*$. 

\medskip
The paper is organized as follows: in Section~\ref{DistDeux}, we first propose a model for the execution process of posted orders, then we define a penalized execution cost function (including the market impact at the terminal execution date). Then we devise the stochastic recursive procedure under constraint to solve the resulting optimization problem in terms of optimal posting distance on the limit order book. We state the main convergence result and provide operating criteria that ensure this convergence, based on a co-monotony principle for one dimensional diffusions. Section \ref{DistTrois} establishes the representations as expectations of the cost function and its derivatives which allow to define the mean function of the algorithm. 
Section \ref{DistCinq} presents the convergence criteria (which ensure that the optimization is well-posed) derived from the principle of co-monotony established in Section \ref{DistQuatre} of the appendix. Finally Section \ref{DistSix} illustrates with numerical experiments the convergence of the recursive procedure towards its target.

\medskip
\noindent{\bf Notations.} $\bullet$ $(x)_+=\max\left\{x,0\right\}$ denotes the positive part of $x$, $\left\lfloor x\right\rfloor=\max\left\{k\in\N\,:\,k\leq x\right\}$ $\llbracket 0,x\rrbracket:=\{y\in[0,1]^d\,:\,0\leq y\leq x\}=\prod_{i=1}^d\left[0,x_i\right]$, $\overline{\N}=\N\cup\{\infty\}$.

\smallskip
\noindent $\bullet$$\left\langle\cdot\left.\right|\cdot\right\rangle$ denotes the canonical inner product on $\R^d$.

\smallskip
\noindent $\bullet$ $\overset{(\R^d)}{\Longrightarrow}$ denotes the weak convergence on $\R^d$ and $\overset{{\cal L}}{\longrightarrow}$ denotes the convergence in distribution.

\smallskip
\noindent $\bullet$ ${\cal C}([0,T],A):=\left\{f:[0,T]\to A \mbox{ continuous}\right\}$ (equipped with the supnorm topology) and   $\D([0,T],A):=\left\{f:[0,T]\to A \mbox{ c\`adl\`ag}\right\}$ (equipped with the Skorokhod topology when necessary see~\cite{JacShi}) where c\`adl\`ag means right continuous with left limits and $A=\R^q$, $\R_+^q$, etc. They are equipped with the standard Borel $\sigma$-field $\sigma(\alpha\mapsto\alpha(t), t\in[0,T])$.

\smallskip
\noindent $\bullet$ $\P$-${\rm esssup}f=\inf\{a\in\R:\P\left(\{x:f(x)>a\}\right)=0\}$, $\left\|\alpha\right\|_{\infty}=\sup_{t\in[0,T]}\left|\alpha(t)\right|$, $\alpha\in\D([0,T],\R)$ and $f'_{\ell}$ denotes the left derivative of $f$.

\section{Design of the execution procedure and main results}\label{DistDeux}
\subsection{Modeling and design of the algorithm}

We focus our work on the problem of optimal trading with limit orders on one security without needing to model the limit order book dynamics. We only model the execution flow which reaches the price where the limit order is posted with a general price dynamics $(S_t)_{t\in[0,T]}$ since we intend to use real data. However there will be two frameworks for the price dynamics: either $(S_t)_{t\in[0,T]}$ is a process bounded by a constant $L$ (which is obviously an unusual assumption but not unrealistic on a short time scale see Section~\ref{SecAveraging}) or $(S_t)_{t\in[0,T]}$ is ruled by a Brownian diffusion model (see Section~\ref{Seciid}).

We consider on a short period $T$, say a dozen of seconds, a Poisson process modeling the execution of posted passive buy orders on the market
\begin{equation}\label{PoissonDist}
\big(N_t^{(\delta)}\big)_{0\leq t\leq T} \quad\mbox{with intensity}\quad \Lambda_T(\delta,S):=\int_0^T\lambda(S_t-(S_0-\delta))dt
\end{equation}
where $0\leq\delta\leq\delta_{\rm{max}}$ with $\delta_{\rm{max}}$ the depth of the order book, $\delta_{\rm{max}}\in(0,S_0)$, and $\left(S_t\right)_{t\geq0}$ is a stochastic process modeling the dynamics of the ``fair price'' of a security stock (from an economic point of view). In practice one may consider that $S_t$ represents the best opposite price at time $t$. 

One way to build $N^{(\delta)}$ is to set
$$N^{(\delta)}_t=\widetilde{N}_{\int_0^T\lambda(S_t-(S_0-\delta))dt}$$
where $\widetilde{N}$ is a Poisson process with intensity 1 independent of the price $(S_t)_{t\in[0,T]}$.

This representation underlines the fact that for one given trajectory of the price $S_t$, the intensity of the Point process $N$ is decreasing with $\delta$: in fact the above representation for $N^{(\delta)}$ is even pathwise consistent in the sense that if $0<\delta<\delta'$ then
$$\P\mbox{-}a.s. \quad \left(\forall t\in[0,T], \quad N^{(\delta)}_t\leq N^{(\delta')}_t\right).$$
The natural question how to account for the actually real possibility of simultaneously placing limit orders at different prices may be much harder to handle and is not studied in this paper. In particular, due to interacting impact features, it would need a more sophisticated approach then simply considering $(N^{\delta(k)})_{1\leq k\leq K}$, processes as above with $\delta(1)<\delta(2)<\ldots<\delta(K)$ (with the same $N$).

\begin{figure}[!ht]
\centering
\includegraphics[width=10cm]{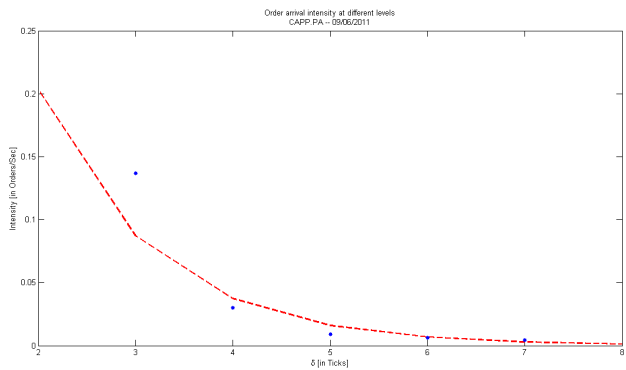}
\vskip-0.5cm
\caption{Empirical probabilities of execution (blue stars) and its fit with an exponential law (red dotted line) with respect to the distance to the ``fair price''.}
\label{FigProbaEmp}
\end{figure}

We assume that the function $\lambda$ is defined on $[-S_0,+\infty)$ as a finite non-increasing convex function. Its specification will rely on parametric or non parametric statistical estimation based on former obtained transactions (see~Figure~\ref{FigProbaEmp} below and Section~\ref{DistSix}). At time $t=0$, buy orders are posted in the limit order book at price $S_0-\delta$. Between $t$ and $t+\Delta t$, the probability for such an order to be executed is $\lambda(S_t-(S_0-\delta))\Delta t$ where $S_t-(S_0-\delta)$ is the distance to the current fair price of our posted order at time $t$. The further the order is at time $t$, the lower is the probability for this order to be executed since $\lambda$ is decreasing on $[-S_0,+\infty)$. Empirical tests strongly confirm this kind of relationship with a convex function $\lambda$ (even close to an exponential shape, see Figure \ref{FigProbaEmp}).
Over the period $[0, T]$, we aim at executing a portfolio of size $Q_T\in\N$ invested in the asset $S$. The execution cost for a distance $\delta$ is $\E\left[(S_0-\delta)\left(Q_{T}\wedge N_{T}^{(\delta)}\right)\right]$. We add to this execution cost a penalization depending on the remaining quantity to be executed, namely at the end of the period $T$, we want to have $Q_T$ assets in the portfolio, so we buy the remaining quantity $\Big(Q_{T}-N_{T}^{(\delta)}\Big)_+$ at price $S_T$. 

At this stage, we introduce a {\em market impact penalization function} $\I:\R\mapsto\R_+$, non-decreasing and convex, with $\I(0)=0$ to model the additional cost of the execution of the remaining quantity (including the market impact). Then the resulting cost of execution on a period $[0,T]$ reads
\begin{equation}\label{coutDist}
C(\delta):=\E\left[(S_0-\delta)\Big(Q_{T}\wedge N_{T}^{(\delta)}\Big)+\kappa \,S_{T}\,\I\Big(\big(Q_{T}-N_{T}^{(\delta)}\big)_+\Big)\right]
\end{equation}
where $\kappa>0$ is a free tuning parameter. When $\I(Q)=1$, we just consider that we buy the remaining quantity at the end price $S_T$. Introducing a market impact penalization function $\I(x)=(1+\eta(x))x$, where $\eta\geq0$, $\eta\not\equiv0$, models the market impact induced by the execution of $\big(Q_{T}-N_{T}^{(\delta)}\big)_+$ at time $T$ whereas we neglect the market impact of the execution process via limit orders over $[0,T)$. Our aim is then to minimize this cost by choosing the distance to post at, namely to solve the following optimization problem
\begin{equation}\label{MinCoutDist}
\min_{0\leq\delta\leq \delta_{\rm{max}}}C(\delta).
\end{equation}

Our strategy to solve numerically~(\ref{MinCoutDist}) using a large enough dataset is to take advantage of the representation of $C$ and its first two derivatives as expectations to devise a recursive stochastic algorithm, namely a stochastic gradient procedure, to find the minimum of the (penalized) cost function (see below). Furthermore we will show that under natural assumptions on the quantity $Q_T$ to be executed and on the parameter $\kappa$, the function $C$ is twice differentiable, {\em strictly convex on $[0,\delta_{\rm{max}}]$ with $C'(0)<0$}. Consequently, 
$$
{\rm argmin}_{\delta\in[0,\delta_{\rm{max}}]}C(\delta)=\{\delta^*\}, \quad\delta^*\in(0,\delta_{\rm{max}}]
$$
and 
$$
\delta^*=\delta_{\rm{max}}\quad\mbox{iff}\quad C \mbox{ is non-increasing on } [0,\delta_{\rm{max}}].
$$
Criteria involving $\kappa$ and based on both the risky asset $S$ and the trading process especially the execution intensity $\lambda$, are established further on in Proposition \ref{upperbound} and Proposition \ref{upperbound2}. We specify representations as expectations  of the   function $C$ and its  derivatives  $C'$ and $C''$. In particular we will exhibit a Borel functional
$$
H:[0,\delta_{\rm{max}}]\times\D\left([0,T],\R\right)\longrightarrow\R
$$
such that 
$$
\forall\delta\in[0,\delta_{\rm{max}}],\quad C'(\delta)=\E\Big[H\big(\delta,(S_t)_{t\in[0,T]}\big)\Big].
$$
The functional $H$ has an explicit form given in Proposition \ref{DerivExpect}, Equations~(\ref{Cprime1}) or~(\ref{Cprime2}), involving integrals over $[0,T]$ of the intensity $\lambda(S_t-S_0+\delta)$ of the Poisson process $(N_t^{(\delta)})_{t\in[0,T]}$. In particular, any quantity $H\left(\delta,(S_t)_{t\in[0,T]}\right)$ can be simulated, up to a natural time  discretization, either from a true dataset (of past executed orders) or from the stepwise constant discretization scheme of a formerly calibrated diffusion process modeling $(S_t)_{t\in[0,T]}$ (see below). This will lead us to replace for practical implementations the continuous time process $(S_t)_{t\in [0,T]}$ over $[0,T]$, either by a discrete time sample, $i.e.$ a finite dimensional $\R^{m+1}$-valued random vector $(S_{t_i})_{0\le i\le m}$ (where $t_0=0$ and $t_m=T$) or by a time discretization scheme with step $\frac Tm$ (typically the Euler scheme when $(S_t)_{t\in[0,T]}$ is a diffusion).

\bigskip
\noindent {\sc A theoretical stochastic learning procedure:} Based on this representation~(\ref{Cprime1}) of $C'$, we can formally devise a recursive stochastic gradient descent $a.s.$ converging toward $\delta^*$. However to make it consistent, we need to  introduce constrain so that it  lives in $[0,\delta_{\rm{max}}]$.  In the classical  literature on Stochastic Approximation Theory  (see~\cite{KusCla} and~\cite{KusYin}) this amounts to consider a variant {\em with projection  on  the ``order book depth interval"} $[0,\delta_{\max}]$, namely 
\begin{equation}\label{ASDist}
\delta_{n+1}={\rm Proj}_{[0,\delta_{{\rm max}}]}\Big(\delta_n-\gamma_{n+1}H\left(\delta_n,\big(S_{t}^{(n+1)}\big)_{t\in [0,T]}\Big)\right), \; n\geq0,\; \delta_0\in (0,\delta_{{\rm max}}),
\end{equation}
where 
\begin{itemize}
\item ${\rm Proj}_{[0,\delta_{{\rm max}}]}$ denotes the  projection on the (nonempty closed convex) $[0,\delta_{{\rm max}}]$, 
\item $(\gamma_n)_{n\geq1}$ is a positive step sequence satisfying (at least) the minimal  {\em decreasing step assumption} $\sum_{n\geq1}\gamma_n=+\infty$ and $\gamma_n \to 0$.
\item the sequence $\left\{(S^{(n)}_t)_{t\in [0,T]}, n\ge 0\right\}$, is the ``innovation" sequence of the procedure : ideally it is either a sequence of simulable independent copies of $(S_t)_{t\in [0,T]}$ or a sequence sharing some ergodic (or averaging)  properties with respect to the distribution of $(S_t)_{t\in [0,T]}$.
\end{itemize}
The case of independent copies can be understood as a framework where the dynamics of $S$ is   typically a Brownian diffusion solution to an stochastic differential equation,  which has been calibrated beforehand on a dataset in order to be simulated on a computer. The case of ergodic copies corresponds to a dataset which is directly plugged into the procedure $i.e.$ $S^{(n)}_t = S_{t-n\Delta t}$, $t\in [0, T]$, $n\ge 0$, where $\Delta t>0$ is a fixed shift parameter. To make this second approach consistent, we need to make the assumption that  at least within a laps of a few minutes, the dynamics of the asset $S$ (starting in the past) is {\em stationary} and shares $e.g.$ mixing properties.

\bigskip
\noindent {\sc The resulting  implementable  procedure:} In practice, the above procedure cannot be implemented since the full path $(S_t(\omega))_{t\in [0,T]}$ of a continuous process cannot be simulated nor a functional $H(\delta, (S_t(\omega))_{t\in [0,T]})$ of such a path can be computed. So we are led in practice  to replace the ``copies" $S^{(n)}$ by copies of a time discretization of step, say $\Delta t=\frac Tm$, ($m\in \N^*$). The time discretizations are formally defined in continuous time as follows
\[
\bar S_t = \bar S_{t_i}, \; t\in [t_i, t_{i+1}), \; i=0,\ldots,m-1\; \mbox{ with } t_i =\frac{iT}{m}, \; i=0,\ldots,m,
\]
where $(\bar S_{t_i})_{0\le i\le m}= (S_{t_i})_{0\le k\le m}$ if $(S_{t_i})_{0\le i\le m}$ can be simulated (see $e.g.$~\cite{BesRob} for 1D-Brownian diffusions processes). The sequence $(\bar S_{t_i})_{0\le i\le m}$ can also be a time discretization scheme (at times $t_i$) of $(S_t)_{t\in [0,T})$, typically an Euler scheme with step $\frac Tm$. 

Then, with an obvious abuse of notation for the function $H$, we can write the {\em implementable procedure} as follows:
\begin{equation}\label{ASDistbis}
\delta_{n+1}={\rm Proj}_{[0,\delta_{{\rm max}}]}\Big(\delta_n-\gamma_{n+1}H\left(\delta_n,\big(\bar{S}_{t_i}^{(n+1)}\big)_{0\leq i\leq m}\Big)\right), \; n\geq0,\; \delta_0\in[0,\delta_{{\rm max}}]
\end{equation}
where $\big(\bar{S}_{t_i}^{(n)}\big)_{0\leq i\leq m}$ are copies of $(\bar{S}_{t_i}\big)_{0\leq i\leq m}$ either independent or sharing ``ergodic" properties, namely some averaging properties in the sense of~\cite{LarPag}. In the first case, one will think about simulated data after a calibration process and in the second case to a direct implementation of a historical high frequency data base of best opposite prices of the asset $S$ (with $e.g.$ $\bar{S}^{(n)}_{t_i}=S_{t_i-n\frac Tm}$).


\subsection{Main convergence results}

The following theorems give $a.s.$ convergence results for the stochastic procedure (\ref{ASDist}): the first one for i.i.d. sequences and the second one for ``averaging'' sequences (see~\cite{LarPag}).

\subsubsection{I.i.d. simulated data from a formerly calibrated model}\label{Seciid}

In this section, we consider that the innovation process $\big\{\big(\bar{S}_{t_i}^{(n)}\big)_{0\leq i\leq m}, n\ge 0\big\}$ comes from a diffusion model beforehand {\em calibrated on real data} which can be simulated at time $t_i$, $0\leq i\leq m$, either exactly or via a stepwise constant time discretization scheme.

\begin{theo}\label{Convergence} $(a)$ {\sc Theoretical procedure.} Assume that $C$ is strictly convex  $[0, \delta_{\max}]$ with $C'(0)<0$. Let  $\big(S_{t}^{(n)}\big)_{t\in [0,T]}$, $n\ge 1$, be a sequence of i.i.d. copies of $(S_t)_{t\in[0,T]}$. Furthermore, assume that the decreasing step sequence satisfies the standard ``decreasing step assumption''
\begin{equation}\label{gamDistiid}
\sum_{n\geq1}\gamma_n=+\infty \quad \mbox{and} \quad \sum_{n\geq1}\gamma^2_n<+\infty. 
\end{equation}
%
Then the recursive procedure defined by (\ref{ASDist}) converges $a.s.$ towards its target $\delta^*={\rm argmin}_{\delta\in[0,\delta_{{\rm max}}]}C(\delta)$:
\[
\delta_n\mathop{\longrightarrow}^{a.s.}\delta^*.
\]
\noindent $(b)$ {\sc Implementable procedure.} Assume the cost function  $\bar C$  related to  the discretization scheme $(\bar S_t)_{t\in [0,T]}$ is strictly convex $[0, \delta_{\max}]$ with $\bar C'(0)<0$ and the  step sequence satisfies the ``decreasing step" assumption. Let $\big(\bar S_{t_i}^{(n)}\big)_{0\le i\le m}$, $n\ge 1$, be a sequence of i.i.d. copies of $\big(\bar S_{t_i}\big)_{0\le i\le m}$, then the recursive procedure defined by (\ref{ASDistbis}) converges $a.s.$ towards its target $\bar \delta^*={\rm argmin}_{\delta \in[0,\delta_{{\rm max}}]}\bar C(\delta)$.
\end{theo}

This theorem is a straightforward  application of the classical $a.s.$ convergence for constrained stochastic algorithms (see Appendix \ref{AppADist}). In particular, the fact that in the original theorem the innovation process takes values in a finite dimensional space $\R^q$ plays no role in the proof.


\subsubsection{Direct implementation on a historical high frequency dataset sharing averaging properties}\label{SecAveraging}

In this framework we will focus on the time discretized procedure $i.e.$ on $(\bar S_t)_{t\in [0,T]}$ rather than on $(S_t)_{t\in [0,T]}$ itself. Keep in mind that, when directly implementing  a high frequency dataset, then 
\[
\bar S_{t}=S_{t_i}\; t\in [t_i, t_{i+1}), i=0,\ldots, m \; \mbox{ and } \bar S_{T}=S_T.
\]
and that the sequence $(\bar S^{(n)}_{t_i})_{0\le i\le m}$, $n\ge 1$, is usually obtained by shifting the data as follows:
 if $\Delta t>0$ 
denotes a fixed time shift parameter such that $t_i-t_{i-1}=\Delta t=\frac Tm$, we set
 $$
 \forall\, t\in [0,T],\quad \bar S^{(n)}_{t_i}= \bar S_{t_i-n\Delta t}=\bar S_{t_{i-n}}.
$$

We will assume that the sequence $(\bar S^{(n)}_{t_i})_{0\le i\le m}$ shares an averaging property with respect to a distribution $\nu$ as developed in~\cite{LarPag}. The definition is recalled below.

\begin{defi} Let $m\in \N$ and $\nu$ be a probability measure on $([0,L]^{m+1},{\cal B}or([0,L]^{m+1}))$. A $[0,L]^{m+1}$-valued sequence $(\xi_n)_{n\geq1}$ is $\nu$-averaging if
$$
\frac{1}{n}\sum_{k=1}^n\delta_{\xi_k}\stackrel{\left({\R}^{m+1}\right)}{\Longrightarrow}\nu \quad \mbox{as} \ n\rightarrow\infty.
$$
\end{defi}
\noindent Then $(\xi_n)_{n\geq1}$ satisfies
$$
D^*_n(\xi):=\sup_{x\in[0,L]^{m+1}}\Big|\frac{1}{n}\sum_{k=1}^n\mathds{1}_{\llbracket 0,x\rrbracket}(\xi_k)-\nu(\llbracket 0,x\rrbracket)\Big|\longrightarrow0 \quad \mbox{as} \ n\rightarrow\infty,
$$
where $D^*_n(\xi)$ is called the discrepancy at the origin or star discrepancy.

\medskip
The resulting execution cost  function $\bar C$ is defined by~(\ref{coutDist}) where $S$ is replaced by $(\bar S_t)_{t\in [0,T]}$ whose distribution is entirely characterized by the distribution $\nu$. In some sense this function $\bar C$ is {\em the best possible approximation of the true execution function $C$} that we can get from the high frequency database.

In this setting, we apply the previous results to the price sequence $\big\{\big(\bar{S}_{t_i}^{(n)}\big)_{0\leq i\leq m}, n\ge 0\big\}$, $i.e.$ we set for every $n\geq1$, $\xi_n=\big(\bar{S}_{t_i}^{(n)}\big)_{0\leq i\leq m}$. In particular we will make the assumption that the dataset is bounded by a real number $L\in(0,+\infty)$ so that $\xi_n\in[0,L]^{m+1}$ for every $n\geq1$ . Moreover, we will need to prove the existence of a {\em pathwise} Lyapunov function, which means in this one dimensional setting that $H(\cdot,(\bar (s_{t_i})_{0\le i\le m})$ is non-decreasing for every $(s_{t_i})_{0\le i\le m}\in \R_+^{m+1}$, $n\geq1$.

\begin{theo}\label{Convergence2}{\sc Implementable procedure}. Let $\lambda(x)=Ae^{-kx}$, $A>0$, $k>0$. Assume $\big(\bar{S}^{(n)}\big)_{n\geq1}$ is an $[0,L]^{m+1}$-valued $\nu$-averaging sequence where $\nu$ is a probability measure on $(\R^{m+1}, {\cal B}or(\R^{m+1}))$. Assume that the execution cost function  $C$ is strictly convex over $[0, \delta_{\max}]$ with $\bar C'(0)<0$ and $\bar C'(\delta_{{\rm max}})>0$.  Finally assume that the step sequence  $(\gamma_n)_{n\geq1}$ is a positive non-increasing sequence satisfying
\begin{equation}\label{gammaASM}
	\sum_{n\geq1}\gamma_n=+\infty, \quad nD^*_n(\bar{S})\gamma_n\underset{n\rightarrow\infty}{\longrightarrow} 0, \quad \mbox{and} \quad \sum_{n\geq1}nD^*_n(\bar{S})\max\left(\gamma_n^2,\left|\Delta\gamma_{n+1}\right|\right)<+\infty.
\end{equation}
Furthermore (having in mind that $\bar S_0=S_0$), assume that  
\begin{equation}\label{CondHcrois}
Q_T\geq2T\lambda(-\bar S_0)\quad\mbox{ and }\quad\kappa\leq\displaystyle\frac{1+k(\bar S_0-\delta_{{\rm max}})}{k\left\|\bar S\right\|_{\infty}(\I(Q_T)-\I(Q_T-1))}
\end{equation}
Then the recursive procedure defined by (\ref{ASDistbis}) converges $a.s.$ towards its target $\bar \delta^*={\rm argmin}_{\delta\in[0,\delta_{{\rm max}}]}\bar C(\delta)$:
\[
\delta_n\mathop{\longrightarrow}^{a.s.}\bar\delta^*.
\]
\end{theo}

\medskip \noindent {\bf Proof.} We apply Theorem 2.1 Section 2 of~\cite{LarPag} in a QMC framework similar to Appendix~\ref{AppADist}. First we have that $H(\bar\delta^*,\cdot)\in L^1(\P)$. Note that $\bar \delta^*\in(0,\delta_{{\rm max}})$ since $\bar C'(0)<0$ and $\bar C'(\delta_{{\rm max}})>0$ so we can extend $\bar C'$ as a convex function on the whole real line. Moreover, by using the proof of Proposition~\ref{upperbound}$(b)$, we prove that if $Q_T\geq2T\lambda(-\bar S_0)$ and~(\ref{CondHcrois}) is satisfied, then $H$ is non-decreasing in $\delta$ so that $H$ satisfies the strict pathwise Lyapunov assumption with $L(\delta)=\frac 12\left|\delta-\bar\delta^*\right|^2$, namely
$$\forall \delta\in\R\backslash\{\delta^*\}, \ \forall y\in{\R}^{m+1}, \quad	\left\langle H(\delta,y)-H(\bar\delta^*,y) \left|\right. \delta-\bar\delta^*\right\rangle>0.$$
It remains to check the averaging rate assumption for $H(\bar\delta^*,\cdot)$: as $(s_i)_{1\leq i\leq m}\mapsto H(\bar\delta^*,(s_i)_{1\leq i\leq m})$ is a non-decreasing function (for $(s_i)_{1\leq i\leq m}\leq(s'_i)_{1\leq i\leq m}$, $i.e.$ $\forall 1\leq i\leq m$, $s_i\leq s'_i$, $i.e.$ $\llbracket 0,s\rrbracket\subset\llbracket 0,s'\rrbracket$), then $H(\bar\delta^*,\cdot)$ has finite variation and by using the Koksma-Hlawka Inequality, we get
$$
\Big|\frac{1}{n}\sum_{k=1}^n H(\bar\delta^*,\bar{S}^{(k)})-\int_{[0,L]^{m+1}}H(\bar\delta^*,s)\nu(ds)\Big|\leq (H(\bar\delta^*,L)-H(\bar\delta^*,0))D^*_n(\bar{S}),
$$
so that $H(\bar\delta^*,\cdot)$ is $\nu$-averaging at rate $\varepsilon_n=D_n^*(\bar{S})$. Finally, Theorem 2.1 of Section 2 from~\cite{LarPag} yields
$$
\hskip7.5cm\delta_n\mathop{\longrightarrow}^{a.s.}\bar\delta^*.\hskip7cm\cqfd
$$


\paragraph{Practical comments of the needed bounds.}
\begin{itemize}
	\item The constraint $Q_T\geq 2T\lambda(-S_0)$ is structural: it only involves parameters of the model and the asked quantity $Q_T$. It means that $Q_T$ does have some chances not to be   fully executed before the end of a slice of duration $T$ ($i.e.$ the intensity of trades obtained very far away from the current price is smaller than $Q_T/2$).
	\item The criterion involving the free parameter $\kappa$ is two-folded depending on the modeling of the ``market impact''. 
	\begin{itemize}
		\item {\em The market impact does not depend on the remaining quantity to be traded} ($i.e.$ when $ \I=\operatorname{id}$ or $\eta\equiv0$ which implies that $\I(Q_T)-\I(Q_T-1)=1$). This setting is appropriate for executing very small quantities or trading very liquid assets (like equity futures). Then the criterion on the free parameter $\kappa$ reads 
$$
\kappa\leq\frac{\bar S_0-\delta_{{\rm max}}}{\left\|\bar S\right\|_{\infty}}+\frac{1}{k \left\|\bar S\right\|_{\infty}}.
$$ 
It states that in this case, the constant premium to pay for the remaining quantity ($i.e.$ $\kappa$, in basis points) has to be lower than the price range inside which we wish to trade ($i.e.$ $(S_0-\delta_{{\rm max}})/\left\|S\right\|_{\infty}$) plus a {\em margin}, namely $1/(k \left\|S\right\|_{\infty})$. It can be seen as a \emph{symmetry argument}: a model where one cannot imagine buying at a lower price than a given threshold, one cannot either accept to pay (on the other side) more market impact than this very threshold.
		\item {\em The market impact of the remaining quantity is a function of the quantity}. The interpretation is very similar to the previous one. In this case a quantity homogeneous to the market impact ($i.e.$ $\kappa\cdot(\I(Q_T)-\I(Q_T-1))$ in basis points) should not exceed $(\bar  S_0-\delta_{{\rm max}})/\left\|\bar S\right\|_{\infty}+1/(k \left\|\bar S\right\|_{\infty})$. Here again it is a symmetry argument: the trader cannot consider to pay more market impact for almost one share ($i.e.$ $\I(Q_T)-\I(Q_T-1)$ plays the role of $\I(1)$ but ``taken around $Q_T$'') than his reasonable trading range ($(\bar S_0-\delta_{{\rm max}})/\left\|\bar S\right\|_{\infty}$ in basis points again), plus a margin. Looking more carefully at this margin: $1/(k \left\|\bar S\right\|_{\infty})$  (the same as in the constant market impact case), it can be read that it is in basis points, and means that if one considers large intensities of fill rates ($i.e.$ $k$ is small) then the trader can be lazy on the market impact constraint (because his margin, proportional to $1/k$, is large in such a case), mainly because he will have to pay market impact not that often. If on the contrary, $k$ is large ($i.e.$ he will have remaining quantities) then he needs to really fulfill his constraint on market impact.
	\end{itemize}
\end{itemize}
The needed bounds to obtain the convergence of $\delta_n$ toward $\delta^*$ are thus not only fulfilled naturally. They also emphasis the consistency of the model and the mechanisms of the proofs.

\subsection{Criteria for the convexity and monotony at the origin} 

In this section, we look for simple criteria involving the parameter $\kappa$, that imply the requested assumption on the execution cost function  $C$ (or $\bar C$). One important feature of this section is that we will never need to really specify the process $S$. For notational convenience we will drop the $\bar C$ notation.

Checking that the assumptions on the function $C$ ($i.e.$ $C$ convex with $C'(0)<0$) in Theorem \ref{Convergence} are satisfied on $[0,\delta_{\max}]$ is a nontrivial task: in fact, as emphasized further on in Figures~\ref{CoutsPen} and~\ref{CoutsPenRealData} in Section~\ref{DistSix}, the  function $C$ in~(\ref{coutDist}) is never convex on the whole non-negative real line, so we need reasonably simple criteria involving  the market impact function  $\I$, $Q_T$ and  the parameter $\kappa$ and others quantities related to  the asset dynamics  which ensure that the required  conditions are fulfilled by the function $C$. These criteria should take the form of upper bounds on  the free parameter $\kappa$.

Their original form, typically those derived by simply writing $C'(0)<0$ and $C''(0)\ge 0$,  are not really operating since they involve ratios of expectations of functionals combining both the dynamics of the asset $S$ and the execution parameters in a highly nonlinear way.  
A large part of this paper is devoted to establish simpler criteria (although slightly more conservative) when $(S_t)_{[0,T]}$ is a continuous process satisfying a functional co-monotony principle. 

We still need an additional assumption, this time on the function $\lambda$. Roughly speaking we need that the functional $\Lambda$ depends on the distance parameter $\delta$  {\em essentially exponentially} in the following sense (which depends on $S$): 
\vskip-0.9cm
\begin{eqnarray}\label{K1Dist}
0<\underline{k}_1:=\P\mbox{-}\mathop{\rm essinf}_{\delta\in[0,\delta_{{\rm max}}]}\left(-\displaystyle\frac{\frac{\partial}{\partial\delta}\Lambda_T(\delta,S)}{\Lambda_T(\delta,S)}\right)&\le&  \overline{k}_1:=\P\mbox{-}\mathop{\rm esssup}_{\delta\in[0,\delta_{{\rm max}}]}\left(-\displaystyle\frac{\frac{\partial}{\partial\delta}\Lambda_T(\delta,S)}{\Lambda_T(\delta,S)}\right)<+\infty, \\ \label{K2Dist}
0<\underline{k}_2:=\P\mbox{-}\mathop{\rm essinf}_{\delta\in[0,\delta_{{\rm max}}]}\left(-\displaystyle\frac{\frac{\partial^2}{\partial\delta^2}\Lambda_T(\delta,S)}{\frac{\partial}{\partial\delta}\Lambda_T(\delta,S)}\right)&\le& \overline{k}_2:=\P\mbox{-}\mathop{\rm esssup}_{\delta\in[0,\delta_{{\rm max}}]}\left(-\displaystyle\frac{\frac{\partial^2}{\partial\delta^2}\Lambda_T(\delta,S)}{\frac{\partial}{\partial\delta}\Lambda_T(\delta,S)}\right)<+\infty.
\end{eqnarray}

Note that the above assumption implies
\begin{equation}\label{K0Dist}
\overline{k}_0:=\P\mbox{-}{\rm esssup}\left(-\frac{\frac{\partial}{\partial\delta}\Lambda_T(0,S)}{\Lambda_T(0,S)}\right)\geq\underline{k}_0:=\P\mbox{-}{\rm essinf}\left(-\frac{\frac{\partial}{\partial\delta}\Lambda_T(0,S)}{\Lambda_T(0,S)}\right)\ge \underline{k}_1>0.
\end{equation}

Although this assumption is stated on the functional $\Lambda$ (and subsequently depends on $S$), this is mainly an assumption  on the intensity function $\lambda$. In particular, the above assumptions are satisfied by intensity functions $\lambda$ of the form
\[
\lambda_k(x)= e^{-kx}, \quad x\in \R, \quad k\in (0,+\infty).
\]
For these functions $\lambda_k$, one checks that $\underline{k}_0=\overline{k}_0=\underline{k}_1=\overline{k}_1=\underline{k}_2=\overline{k}_2=k$.\\

The key to establish the criteria is the {\em functional co-monotony principle} that we establish in the Appendix~\ref{DistQuatre} for a wide class of diffusions and their associated time discretization schemes. 

A Borel functional $F:\D([0,T],\R)\to\R$ is non-decreasing if
$$\forall\alpha,\beta\in\D([0,T],\R),\quad(\forall t\in[0,T],\, \alpha(t)\leq\beta(t))\Longrightarrow F(\alpha)\leq F(\beta).$$
It is monotonic if $F$ or $-F$ is non-decreasing. Two functionals $F$ and $G$ on $\D([0,T],\R)$ are co-monotonic if they are monotonic with the same monotony. 

A functional $F$ has polynomial growth if
$$\exists r>0 \quad\mbox{s.t.}\quad\forall\alpha\in\D([0,T],\R), \quad \left|F(\alpha)\right|\leq K\left(1+\left\|\alpha\right\|^r_{\infty}\right).$$

\begin{defi}
A stepwise constant c\`adl\`ag (resp. continuous) process $(S_t)_{t\in[0,T]}$ satisfies a functional co-monotony principle if for every pair $F$,$G:\D([0,T],\R)\to\R$ of co-monotonic Borel functionals (resp. continuous at every $\alpha\in{\cal C}([0,T],\R)$ for the sup-norm) with polynomial growth such that $F(S)$, $G(S)$ and $F(S)G(S)\in L^1$, one has
\begin{equation}\label{ComonotonyPrinciple}
\E\left[F\big(\left(S_t\right)_{t\in[0,T]}\big)G\big(\left(S_t\right)_{t\in[0,T]}\big)\right]\geq \E\left[F\big(\left(S_t\right)_{t\in[0,T]}\big)\right]\E\left[G\big(\left(S_t\right)_{t\in[0,T]}\big)\right].
\end{equation}
\end{defi}
We will use in the proofs below this principle for the price process $(S_t)_{t\in[0,T]}$ for monotonic functionals with opposite monotony. Thus Inequality in~(\ref{ComonotonyPrinciple}) is reversed (all we have to do is to replace $F$ by $-F$). This co-monotony principle is established and proved in Appendix~\ref{DistQuatre} for a wide class of Brownian diffusions, their discrete time samples and their Euler schemes. \\

The main dynamics in which we are interested are the following
\begin{enumerate}
	\item An ``admissible'' Brownian diffusion in the sense od Definition~\ref{DefAdmissible} satisfies a functional co-monotony principle. So does its Euler scheme with step $\frac Tm$, at least for $m$ large enough, say $m\geq m_{b,\sigma}$, where $m_{b,\sigma}$ only depends on (the Lipschitz coefficients of) the drift $b$ and the diffusion coefficient $\sigma$ of the SDE that $(S_t)_{t\in[0,T]}$ is solution to.
	\item Finally any discrete time sample $(S_{t_i})_{0\leq i\leq m}$ of a continuous process satisfying a functional co-monotony principle also satisfies a co-monotony principle for continuous functions $f,g:\R^{m+1}\to\R$ monotonic in each of their variables with the same monotony such that $f((S_{t_i})_{0\leq i\leq m})$, $g((S_{t_i})_{0\leq i\leq m})$, $fg((S_{t_i})_{0\leq i\leq m})\in L^1$.
	\item 
If $(S_{t_i})_{0\leq i\leq m}$ is a Markov chain whose transitions
$$P_ig(x)=\E\left[g(S_{t_{i+1}}\left.\right|S_{t_i}=x\right],\quad i=0,\ldots,m,$$
preserve monotony ($g$ non-decreasing implies $P_ig$ is non-decreasing), then $(S_{t_i})_{0\leq i\leq m}$ also satisfies a co-monotony principle in the same sense as for discrete time sample.
\end{enumerate}

The proof of 1. is postponed in Appendix~\ref{DistQuatre} (Theorem~\ref{thm:IneqCovDiff}). Admissible diffusions include standard Brownian motion, geometrical Brownian motion and most models of dynamics traded assets (see Section B.2.3).

Claim 2. follows by associating to a function $f$ the functional $F(\alpha)=f((\alpha(t_i))_{0\leq i\leq m})$.

For claim 3., we refer to~\cite{Pag3} Proposition 2.1.

\begin{theo}\label{Criteria} Assume that $(S_t)_{t\in[0,T]}$ satisfies a functional co-monotony principle. Assume that the function $\lambda$ is essentially exponential in the sense of~(\ref{K1Dist}) and ~(\ref{K2Dist}) above. Then the following monotony and convexity criteria hold true.

\medskip
\noindent$(a)$ {\sc Monotony at the origin:} The derivative $C'(0)<0$ as soon as
$$
Q_T\geq 2T\lambda(-S_0)\quad\mbox{ and }\quad \kappa\leq\displaystyle\frac{1+\underline{k}_0S_0}{\overline{k}_0\E\left[S_{T}\right]\left(\I(Q_T)-\I(Q_T-1)\right)}.
$$

\noindent $(b)$ {\sc Convexity.} Let $\rho_Q\in\left(0,1-\frac{\P\left(N^{\mu}=Q_{T}-1\right)}{\P\left(N^{\mu}\leq Q_{T}-1\right)}_{\left.\right|\mu=T\lambda(-S_0)}\right)$. If $\I\neq \operatorname{id}$, assume that $\I$ satisfies
$$
\forall x\in[1,Q_T-1], \quad \I(x)-\I(x-1)\leq\rho_Q(\I(x+1)-\I(x)).
$$
$$\mbox{If}\hskip3cm
Q_T\geq \Big(2\vee\big(1+\frac{\overline{k}_1^2}{\underline{k}_1\underline{k}_2}\big)\Big)T\lambda(-S_0)\quad \mbox{ and }\quad 
\kappa\leq\displaystyle\frac{2\underline{k}_1}{\overline{k}_1\overline{k}_2\E\left[S_{T}\right]\I'_{\ell}(Q_T)},\hskip4cm
$$
then $C''(\delta)\geq0$, $\delta\in[0,\delta_{{\rm max}}]$, so that $C$ is convex on $[0,\delta_{{\rm max}}]$. \\

\noindent $(c)$ The same inequalities hold for the Euler scheme of $(S_t)_{t\in[0,T]}$ with step $\frac Tm$, for $m\geq m_{b,\sigma}$, or for any $\R^{m+1}$-time discretization sequence $(S_{t_i})_{0\leq i\leq m}$ which satisfies a co-monotony principle. 
\end{theo}

\medskip \noindent {\bf Remark.} These conditions on the model parameters
 are conservative. Indeed, ``sharper" criteria can be given whose bounds involve ratios of expectation  which can be evaluated only by Monte Carlo simulations: 
$$C'(0)<0 \Longleftrightarrow 0<\kappa<b_2,$$
$$\mbox{where}\hskip1cm b_2=\frac{\E\left[-Q_{T}\P^{(0)}\left(N^{\mu}>Q_{T}\right)+\left(S_0\frac{\partial}{\partial\delta}\Lambda_T(0,S)-\Lambda_T(0,S)\right)\P^{(0)}\left(N^{\mu}\leq Q_{T}-1\right)\right]}{\E\left[S_{T}\frac{\partial}{\partial\delta}\Lambda_T(0,S)\varphi^{(0)}(\mu)\right]}\hskip2cm$$ 
$$\mbox{and}\hskip3cm\mbox{$C$ is convex on $[0, \delta_{{\rm max}}]$}\Longleftrightarrow0<\kappa<\min_{\delta\in{\cal D}_+}\frac{A(\delta)}{B(\delta)}\hskip4cm$$
\begin{eqnarray*}	  
\mbox{where}\hskip3cm A(\delta)&=&\E\left[\left(\left(S_0-\delta\right)\frac{\partial^2}{\partial\delta^2}\Lambda_T(\delta,S)-2\frac{\partial}{\partial\delta}\Lambda_T(\delta,S)\right)\P^{(\delta)}\left(N^{\mu}\leq Q_{T}-1\right)\right. \hskip2cm\\
	   & &-\left.\left(S_0-\delta\right)\left(\frac{\partial}{\partial\delta}\Lambda_T(\delta,S)\right)^2\P^{(\delta)}\left(N^{\mu}=Q_{T}-1\right)\right],
\end{eqnarray*}
$$B(\delta)=\E\left[S_{T}\left(\frac{\partial^2}{\partial\delta^2}\Lambda_T(\delta,S)\varphi^{(\delta)}(\mu)-\left(\frac{\partial}{\partial\delta}\Lambda_T(\delta,S)\right)^2\psi^{(\delta)}(\mu)\right)\right]\quad\mbox{and}\quad{\cal D}_+=\left\{\delta\in[0,\delta_{{\rm max}}]\left.\right|B(\delta)>0\right\}.$$

\section{Representations as expectations of \texorpdfstring{$C$}{C} and its derivatives}
\label{DistTrois}

First we briefly recall for convenience few basic facts on Poisson distributed variables that will be needed to compute the cost function $C$ and its derivatives $C'$ and $C''$ (proofs are left to the reader).

\begin{prop}\label{ptesDist}{\rm (Classical formulas).} Let $(N^{\mu})_{\mu>0}$ be a family of Poisson distributed random variables with parameter $\mu>0$.
\begin{enumerate}	
	\item[$(i)$] For every function $f:\N\rightarrow\R_+$ such that $\log f(n)=O(n)$, $$\frac{d}{d\mu}\E\left[f(N^{\mu})\right]=\E\left[f(N^{\mu}+1)-f(N^{\mu})\right]=\E\left[f(N^{\mu})\left(\frac{N^{\mu}}{\mu}-1\right)\right].$$ 
	In particular, for any $k\in\N$, $\frac{d}{d\mu}\P\left(N^{\mu}\leq k\right)=-\P\left(N^{\mu}=k\right)$.
\end{enumerate}	
For any $k\in\N^*$,
\begin{enumerate}	
	\item[$(ii)$] $\E\left[k\wedge N^{\mu}\right]=k\P\left(N^{\mu}> k\right)+\mu\P\left(N^{\mu}\leq k-1\right)$ and $\frac{d}{d\mu}\E\left[k\wedge N^{\mu}\right]=\P\left(N^{\mu}\leq k-1\right)$,
	\item[$(iii)$] $\E\left[\left(k-N^{\mu}\right)_+\right]=k\P\left(N^{\mu}\leq k\right)-\mu\P\left(N^{\mu}\leq k-1\right)$,
	\item[$(iv)$] $k\P\left(N^{\mu}=k\right)=\mu\P\left(N^{\mu}=k-1\right)$.	
\end{enumerate}
\end{prop}

To compute the cost function (or its gradient), it is convenient to proceed a pre-conditioning with respect to $\F_{T}^S:=\sigma\left(S_t,0\leq t\leq T\right)$. We come down to compute the above quantity when $N^{(\delta)}$ is replaced by a standard Poisson random variable with parameter $\mu$ denoted $N^{\mu}$. Therefore we have
\begin{eqnarray}
\label{EcoutDist}
	C(\delta)&=&\E\left[(S_0-\delta)\left(Q_{T}\wedge N_{T}^{(\delta)}\right)+\kappa S_{T}\I\left(\left(Q_{T}-N_{T}^{(\delta)}\right)_+\right)\right] \nonumber \\ 
	         &=&\E\left[(S_0-\delta)\E\left[\left(Q_{T}\wedge N_{T}^{(\delta)}\right)\left.\right|\F_{T}^S\right]+\kappa S_{T}\E\left[\I\left(\left(Q_{T}-N_{T}^{(\delta)}\right)_+\right)\left.\right|\F_{T}^S\right]\right] \nonumber \\ 
	         &=&\E\left[(S_0-\delta)\E\left[Q_{T}\wedge N^{\mu}\right]_{\left.\right|\mu=\Lambda_T(\delta,S)}+\kappa S_{T}\E\left[\I\left(\left(Q_{T}-N^{\mu}\right)_+\right)\right]_{\left.\right|\mu=\Lambda_T(\delta,S)}\right] \nonumber\\ 
	         &=&\E\left[\widetilde{C}\left(\delta,\Lambda_T(\delta,S),\left(S_t\right)_{0\leq t\leq T}\right)\right],
\end{eqnarray}
where for  every $x\in{\cal C}([0,T],\R_+)$ and every $\mu\in\R_+$,
$$
\widetilde{C}\left(\delta,\mu,x\right)=(x_0-\delta)\left(Q_{T}\P\left(N^{\mu}>Q_{T}\right)+\mu\P\left(N^{\mu}\leq Q_{T}-1\right)\right)+\kappa x_{T}\E\left[\I\left(\left(Q_{T}-N^{\mu}\right)_+\right)\right].$$
We introduce some notations for reading convenience: we set $$\P^{(\delta)}\left(N^{\mu}>Q_{T}\right)=\P\left(N^{\mu}>Q_{T}\right)_{\left.\right|\mu=\Lambda_T(\delta,S)} \quad \mbox{and} \quad \E^{(\delta)}\left[f(\mu)\right]=\E\left[f(\mu)\right]_{\left.\right|\mu=\Lambda_T(\delta,S)}.
$$
Now we are in position to compute the first and second derivatives of the cost function $C$.

\begin{prop}\label{DerivExpect} $(a)$ If $\I\neq\operatorname{id}$, then $C'(\delta)=\E\left[H(\delta,S)\right]$ with
\begin{eqnarray}\label{Cprime1}	
H(\delta,S)&=&-Q_{T}\P^{(\delta)}\left(N^{\mu}>Q_{T}\right)+\left(\frac{\partial}{\partial\delta}\Lambda_T(\delta,S)(S_0-\delta)-\Lambda_T(\delta,S)\right)\P^{(\delta)}\left(N^{\mu}\leq Q_{T}-1\right)\nonumber\\
					& &-\kappa S_{T}\frac{\partial}{\partial\delta}\Lambda_T(\delta,S)\varphi^{(\delta)}(\mu)           
\end{eqnarray}
where $\varphi^{(\delta)}(\mu)=\E^{(\delta)}\left[\left(\I\left(Q_{T}-N^{\mu}\right)-\I\left(Q_{T}-N^{\mu}-1\right)\right)\mathds{1}_{\{N^{\mu}\leq Q_T-1\}}\right]$ and
\begin{eqnarray}\label{Cseconde1}	
C''(\delta)&=&\E\left[\left((S_0-\delta)\frac{\partial^2}{\partial\delta^2}\Lambda_T(\delta,S)-2\frac{\partial}{\partial\delta}\Lambda_T(\delta,S)\right)\P^{(\delta)}\left(N^{\mu}\leq Q_{T}-1\right)-\kappa S_{T}\frac{\partial^2}{\partial\delta^2}\Lambda_T(\delta,S)\varphi^{(\delta)}(\mu)\right. \nonumber \\
	           & &\left.-(S_0-\delta)\left(\frac{\partial}{\partial\delta}\Lambda_T(\delta,S)\right)^2\P^{(\delta)}\left(N^{\mu}=Q_{T}-1\right)+\kappa S_{T}\left(\frac{\partial}{\partial\delta}\Lambda_T(\delta,S)\right)^2\psi^{(\delta)}(\mu)\right]	           
\end{eqnarray}
	where $\psi^{(\delta)}(\mu)=\E^{(\delta)}\left[\I\left(\left(Q_{T}-N^{\mu}-2\right)_+\right)-2\I\left(\left(Q_{T}-N^{\mu}-1\right)_+\right)+\I\left(\left(Q_{T}-N^{\mu}\right)_+\right)\right]$.

\medskip
\noindent $(b)$  If $\I=\operatorname{id}$, then $C'(\delta)=\E\left[H(\delta,S)\right]$ with
\begin{equation}\label{Cprime2}
	\hskip-0.3cm H(\delta,S)=-Q_{T}\P^{(\delta)}\left(N^{\mu}>Q_{T}\right)+\left((S_0-\delta-\kappa S_{T})\frac{\partial}{\partial\delta}\Lambda_T(\delta,S)-\Lambda_T(\delta,S)\right)\P^{(\delta)}\left(N^{\mu}\leq Q_{T}-1\right)
\end{equation}
\vskip-0.5cm
\begin{eqnarray}\label{Cseconde2}
	\mbox{and}\quad C''(\delta)&=&\E\left[\left((S_0-\delta-\kappa S_{T})\frac{\partial^2}{\partial\delta^2}\Lambda_T(\delta,S)-2\frac{\partial}{\partial\delta}\Lambda_T(\delta,S)\right)\P^{(\delta)}\left(N^{\mu}\leq Q_{T}-1\right)\right.\nonumber \\
	           & &-\left.(S_0-\delta-\kappa S_{T})\left(\frac{\partial}{\partial\delta}\Lambda_T(\delta,S)\right)^2\P^{(\delta)}\left(N^{\mu}=Q_{T}-1\right)\right]
\end{eqnarray}
\end{prop}

\noindent {\bf Proof.} Interchanging derivation and expectation in the representation (\ref{EcoutDist}) implies
$$C'(\delta)=\E\left[\frac{\partial}{\partial\delta}\widetilde{C}\left(\delta,\Lambda_T(\delta,S),\left(S_t\right)_{0\leq t\leq T}\right)+\frac{\partial}{\partial\mu}\widetilde{C}\left(\delta,\Lambda_T(\delta,S),\left(S_t\right)_{0\leq t\leq T}\right)\frac{\partial}{\partial\delta}\Lambda_T(\delta,S)\right].$$

\noindent $(a)$ We come down to compute the partial derivatives of $\widetilde{C}\left(\delta,\mu,x\right)$.
$$\frac{\partial \widetilde{C}}{\partial\delta}\left(\delta,\mu,x\right)=-\E\left[\left(Q_{T}\wedge N^{\mu}\right)\right]=-Q_{T}\P\left(N^{\mu}>Q_{T}\right)-\mu\P\left(N^{\mu}\leq Q_{T}-1\right),$$
\vskip-0.5cm
$$\frac{\partial \widetilde{C}}{\partial\mu}\left(\delta,\mu,x\right)=-(x_0-\delta)\frac{\partial}{\partial\mu}\E\left[\left(Q_{T}-N^{\mu}\right)_+\right]+\kappa x_{T}\frac{\partial}{\partial\mu}\E\left[\I\left(\left(Q_{T}-N^{\mu}\right)_+\right)\right].$$
\vskip-0.5cm
\begin{eqnarray*}
\mbox{We have}\hskip1cm\frac{\partial}{\partial\mu}\E\left[\left(Q_{T}-N^{\mu}\right)_+\right]&=&-Q_{T}\P\left(N^{\mu}=Q_{T}\right)-\P\left(N^{\mu}\leq Q_{T}-1\right) 
	+\mu\P\left(N^{\mu}=Q_{T}-1\right)\\
	&=&-\P\left(N^{\mu}\leq Q_{T}-1\right) \quad\mbox{thanks to $(iv)$ in Proposition \ref{ptesDist}}\hskip1.5cm
\end{eqnarray*}
\vskip-0.5cm
\begin{eqnarray*}
\mbox{and}\hskip1cm\frac{\partial}{\partial\mu}\E\left[\I\left(\left(Q_{T}-N^{\mu}\right)_+\right)\right]&=&\E\left[\I\left(\left(Q_{T}-N^{\mu}-1\right)_+\right)-\I\left(\left(Q_{T}-N^{\mu}\right)_+\right)\right] \\
&=&\E\left[\left(\I\left(Q_{T}-N^{\mu}-1\right)-\I\left(Q_{T}-N^{\mu}\right)\right)\mathds{1}_{\{N^{\mu}\leq Q_T-1\}}\right]:=-\varphi(\mu)\hskip1cm
\end{eqnarray*}
owing to $(v)$ in Proposition \ref{ptesDist}. Therefore
$$
\frac{\partial \widetilde{C}}{\partial\mu}\left(\delta,\mu,x\right)=\left(x_0-\delta\right)\P\left(N^{\mu}\leq Q_{T}-1\right)-\kappa x_T\varphi(\mu).
$$
Consequently
\begin{eqnarray}
\label{Cderiv1}
C'(\delta)&=&\E\left[-Q_{T}\P^{(\delta)}\left(N^{\mu}>Q_{T}\right)+\left(\frac{\partial}{\partial\delta}\Lambda_T(\delta,S)(S_0-\delta)-\Lambda_T(\delta,S)\right)\P^{(\delta)}\left(N^{\mu}\leq Q_{T}-1\right)\right. \nonumber \\
	           & &-\left.\kappa S_{T}\frac{\partial}{\partial\delta}\Lambda_T(\delta,S)\varphi^{(\delta)}(\mu)\right] \nonumber \\
	           &=&\E\left[\widehat{C}\left(\delta,\Lambda_T(\delta,S),\frac{\partial}{\partial\delta}\Lambda_T(\delta,S),(S_t)_{0\le t\le T}\right)\right],
\end{eqnarray}
where $\varphi^{(\delta)}(\mu):=\varphi(\mu)_{\left.\right|\mu=\Lambda_T(\delta,S)}$ and for every $x\in{\cal C}([0,T],\R_+)$ and every $\mu,\nu\in \R_+$,
\begin{equation*}
\widehat{C}\left(\delta,\mu,\nu,x\right)=-Q_{T}\P\left(N^{\mu}>Q_{T}\right)+\left(\nu(x_0-\delta)-\mu\right)\P\left(N^{\mu}\leq Q_{T}-1\right)-\kappa x_{T}\nu\varphi(\mu).
\end{equation*}         
Interchanging derivation and expectation in the representation (\ref{Cderiv1}) implies
\begin{eqnarray*}
C''(\delta)&=&\E\left[\frac{\partial}{\partial\delta}\widehat{C}\left(\delta,\Lambda_T(\delta,S),\frac{\partial}{\partial\delta}\Lambda_T(\delta,S),\left(S_t\right)_{0\leq t\leq T}\right) \right.\\
& &+\frac{\partial}{\partial\mu}\widehat{C}\left(\delta,\Lambda_T(\delta,S),\frac{\partial}{\partial\delta}\Lambda_T(\delta,S),\left(S_t\right)_{0\leq t\leq T}\right)\frac{\partial}{\partial\delta}\Lambda_T(\delta,S) \\
& &+\left.\frac{\partial}{\partial\nu}\widehat{C}\left(\delta,\Lambda_T(\delta,S),\frac{\partial}{\partial\delta}\Lambda_T(\delta,S),\left(S_t\right)_{0\leq t\leq T}\right)\frac{\partial^2}{\partial\delta^2}\Lambda_T(\delta,S)\right].
\end{eqnarray*}
We deal now with the partial derivatives of $\widehat{C}\left(\delta,\mu,\nu,x\right)$.
\begin{eqnarray*}
\frac{\partial \widehat{C}}{\partial\delta}\left(\delta,\mu,\nu,x\right)&=&-\nu\P\left(N^{\mu}\leq Q_{T}-1\right),\\
\frac{\partial \widehat{C}}{\partial\mu}\left(\delta,\mu,\nu,x\right)&=&-\P\left(N^{\mu}\leq Q_{T}-1\right)-(x_0-\delta)\nu\P\left(N^{\mu}=Q_{T}-1\right)+\kappa x_{T}\nu\psi(\mu), \\
\frac{\partial \widehat{C}}{\partial\nu}\left(\delta,\mu,\nu,x\right)&=&(x_0-\delta)\P\left(N^{\mu}\leq Q_{T}-1\right)-\kappa x_{T}\varphi(\mu).
\end{eqnarray*}
Consequently
\vskip-0.5cm
\begin{eqnarray*}	C''(\delta)&=&\E\left[\left((S_0-\delta)\frac{\partial^2}{\partial\delta^2}\Lambda_T(\delta,S)-2\frac{\partial}{\partial\delta}\Lambda_T(\delta,S)\right)\P^{(\delta)}\left(N^{\mu}\leq Q_{T}-1\right)-\kappa S_{T}\frac{\partial^2}{\partial\delta^2}\Lambda_T(\delta,S)\varphi^{(\delta)}(\mu) \right.\\
	& &\left.-(S_0-\delta)\left(\frac{\partial}{\partial\delta}\Lambda_T(\delta,S)\right)^2\P^{(\delta)}\left(N^{\mu}=Q_{T}-1\right)+\kappa S_{T}\left(\frac{\partial}{\partial\delta}\Lambda_T(\delta,S)\right)^2\psi^{(\delta)}(\mu)\right].
\end{eqnarray*}

\noindent $(b)$ If $\I=\operatorname{id}$ so that $\frac{\partial}{\partial\mu}\E\left[\I\left(\left(Q_T-N^{\mu}\right)_+\right)\right]=-\P\left(N^{\mu}\leq Q_T-1\right)$. Therefore
$$\frac{\partial \widetilde{C}}{\partial\delta}\left(\delta,\mu,x\right)=-Q_{T}\P\left(N^{\mu}>Q_{T}\right)-\mu\P\left(N^{\mu}\leq Q_{T}-1\right) \quad\mbox{and}\quad
\frac{\partial \widetilde{C}}{\partial\mu}\left(\delta,\mu,x\right)=(x_0-\delta-\kappa x_{T})\P\left(N^{\mu}\leq Q_{T}-1\right).$$
Consequently
\begin{eqnarray}\label{Cderiv2}
	C'(\delta)&=&\E\left[-Q_{T}\P^{(\delta)}\left(N^{\mu}>Q_{T}\right)+\left((S_0-\delta-\kappa S_{T})\frac{\partial}{\partial\delta}\Lambda_T(\delta,S)-\Lambda_T(\delta,S)\right)\P^{(\delta)}\left(N^{\mu}\leq Q_{T}-1\right)\right] \nonumber \\
	           &=&\E\left[\widehat{C}\left(\delta,\Lambda_T(\delta,S),\frac{\partial}{\partial\delta}\Lambda_T(\delta,S),(S_t)_{0\le t\le T}\right)\right],
\end{eqnarray}
where for every $x\in{\cal C}([0,T],\R_+)$ and every $\mu,\nu\in\R_+$,
$$\widehat{C}\left(\delta,\mu,\nu,x\right)=-Q_{T}\P\left(N^{\mu}>Q_{T}\right)+\left((x_0-\delta-\kappa x_{T})\nu-\mu\right)\P\left(N^{\mu}\leq Q_{T}-1\right).$$           
Interchanging derivation and expectation in the  the representation (\ref{Cderiv2}) implies
\begin{eqnarray*}
C''(\delta)&=&\E\left[\frac{\partial}{\partial\delta}\widehat{C}\left(\delta,\Lambda_T(\delta,S),\frac{\partial}{\partial\delta}\Lambda_T(\delta,S),\left(S_t\right)_{0\leq t\leq T}\right)\right. \\
& &+\frac{\partial}{\partial\mu}\widehat{C}\left(\delta,\Lambda_T(\delta,S),\frac{\partial}{\partial\delta}\Lambda_T(\delta,S),\left(S_t\right)_{0\leq t\leq T}\right)\frac{\partial}{\partial\delta}\Lambda_T(\delta,S) \\
& &+\left.\frac{\partial}{\partial\nu}\widehat{C}\left(\delta,\Lambda_T(\delta,S),\frac{\partial}{\partial\delta}\Lambda_T(\delta,S),\left(S_t\right)_{0\leq t\leq T}\right)\frac{\partial^2}{\partial\delta^2}\Lambda_T(\delta,S)\right].
\end{eqnarray*}
We come down to compute the partial derivatives of $\widehat{C}\left(\delta,\Lambda_T(\delta,S),\frac{\partial}{\partial\delta}\Lambda_T(\delta,S),\left(S_t\right)_{0\leq t\leq T}\right)$.
$$\frac{\partial \widehat{C}}{\partial\delta}\left(\delta,\mu,\nu,x\right)=-\nu\P\left(N^{\mu}\leq Q_{T}-1\right),\quad \frac{\partial \widehat{C}}{\partial\nu}\left(\delta,\mu,\nu,x\right)=(x_0-\delta-\kappa x_{T})\P\left(N^{\mu}\leq Q_{T}-1\right),$$
$$\frac{\partial \widehat{C}}{\partial\mu}\left(\delta,\mu,\nu,x\right)=-\P\left(N^{\mu}\leq Q_{T}-1\right)-(x_0-\delta-\kappa x_{T})\frac{\partial}{\partial\delta}\Lambda_T(\delta,S)\P\left(N^{\mu}=Q_{T}-1\right).$$
Consequently
\vskip-0.7cm
\begin{eqnarray*}
\hskip2cm	C''(\delta)&=&\E\left[\left((S_0-\delta-\kappa S_{T})\frac{\partial^2}{\partial\delta^2}\Lambda_T(\delta,S)-2\frac{\partial}{\partial\delta}\Lambda_T(\delta,S)\right)\P^{(\delta)}\left(N^{\mu}\leq Q_{T}-1\right)\right.\\
	           & &-\left.(S_0-\delta-\kappa S_{T})\left(\frac{\partial}{\partial\delta}\Lambda_T(\delta,S)\right)^2\P^{(\delta)}\left(N^{\mu}=Q_{T}-1\right)\right].\hskip2cm\cqfd
\end{eqnarray*}

\section{Convexity and monotony criteria for the cost function \texorpdfstring{$C$}{C}}
\label{DistCinq}

To ensure that the optimization problem is well-posed, namely that the cost function $C$ has a minimum on $[0, \delta_{{\rm max}}]$, we need some additional assumptions: the cost function $C$ must satisfy $C'(0)<0$. This leads to define bounds for the parameter $\kappa$ and this section is devoted to give sufficient condition on $\kappa$ to ensure that this two properties are satisfied. The computations of the bounds given below rely on the co-monotony principle introduced in the Appendix Section~\ref{DistQuatre}.

\subsection{Criteria for local and global monotony}

The proposition below gives bounds for the parameter $\kappa$ which ensure that the execution cost function $C$ has a minimum. The aim of this subsection is to obtain sufficient bounds, easy to compute, namely depending only of the model parameters. 

\begin{prop}\label{upperbound} Assume that $(S_t)_{t\in[0,T]}$ satisfies a co-monotony principle.

\noindent $(a)$ {\sc Monotony at the origin.} $C'(0)<0$ as soon as $Q_T\geq 2T\lambda(-S_0)$, $k_0={\rm infess}\left(-\frac{\frac{\partial}{\partial\delta}\Lambda_T(0,S)}{\Lambda_T(0,S)}\right)>0$ and 
\vskip-0.2cm
$$
\kappa\leq\displaystyle\frac{1+\underline{k}_0S_0}{\overline{k}_0\E\left[S_{T}\right]\left(\I(Q_T)-\I(Q_T-1)\right)}.
$$
In particular, when $\I\equiv \operatorname{id}$, the condition reduces to
$$
\kappa\leq\displaystyle\frac{1+\underline{k}_0S_0}{\overline{k}_0\E\left[S_{T}\right]}.
$$
\noindent $(b)$ {\sc Global monotony (exponential intensity).} Assume that $\displaystyle s^*:= \|\sup_{t\in[0,T]}S_t\|_{L^\infty}<+\infty$. 
If $\lambda(x)=Ae^{-kx}$, $A>0$, $k>0$, $Q_T\geq2T\lambda(-S_0)$ and
\begin{equation}\label{CritHcroissante}
\hskip-0.5cm
\kappa\leq\displaystyle\frac{1+k(S_0-\delta_{{\rm max}})}{k\,s^*} \; \mbox{ if } \; \I\neq\operatorname{id},\quad
\kappa\leq\displaystyle\frac{1+k(S_0-\delta_{{\rm max}})}{k\,s^*(\I(Q_T)-\I(Q_T-1))} \; \mbox{ if } \; \I=\operatorname{id},
\end{equation}
then $H(\cdot, (y_i)_{0\le i\le m})$ is non-decreasing on $[0, \delta_{\max}]$ for every $(y_i)_{0\le i\le m}\in [0,s^*]^{m+1}$.

\noindent $(c)$ The same inequality holds for the Euler scheme of $(S_t)_{t\in[0,T]}$ with step $\frac Tm$, for $m\geq m_{b,\sigma}$, or for any $\R^{m+1}$-time discretization sequence $(S_{t_i})_{0\leq i\leq m}$ which satisfies a co-monotony principle (see Theorem~\ref{Criteria}-$(c)$ and Corollary~\ref{cor:IneqCovDiff}).
\end{prop}

To prove this result, we need to establish the monotony of several functions of $\mu$ which appear in the expression of $C'$.

\begin{lem} $(i)$ The function $\mu\longmapsto\mu\P\left(N^{\mu}\leq Q\right)$ is non-decreasing on $\left[0,\left\lfloor \frac{Q+1}{2}\right\rfloor\right]$. 

\medskip
\noindent$(ii)$ The function $\mu\longmapsto\Theta(Q,\mu):=\E\left[\I(Q-N^{\mu})-\I(Q-N^{\mu}-1)\left.\right|N^{\mu}\leq Q-1\right]$ satisfies $\Theta(Q_T,\mu)\leq\Theta(Q_T,0)=\I(Q)-\I(Q-1)$ for all $\mu$.
\label{LemMonot}
\end{lem}

\noindent {\bf Proof of Lemma \ref{LemMonot}.} $(i)$ We have $\displaystyle\frac{d}{d\mu}\left(\mu\P\left(N^{\mu}\leq Q\right)\right)=\P\left(N^{\mu}\leq Q\right)-\mu\P\left(N^{\mu}=Q\right)$. Consequently 
\vskip-0.25cm
$$\frac{d}{d\mu}\left(\mu\P\left(N^{\mu}\leq Q\right)\right)\geq0 \quad \mbox{iff} \quad \sum_{k=0}^Q\frac{\mu^k}{k!}\geq\mu\displaystyle\frac{\mu^Q}{Q!}.$$ 
\vskip-0.25cm
But $k\longmapsto\displaystyle\frac{\mu^k}{k!}$ is non-decreasing on $\left\{0,1,\ldots,\lfloor \mu\rfloor\right\}$ and non-increasing on $\left\{\lfloor\mu\rfloor,\ldots\right\}$. 

\noindent Hence $\displaystyle\sum_{k=0}^Q\frac{\mu^k}{k!}\geq\sum_{k=\lfloor\mu\rfloor}^Q\frac{\mu^Q}{Q!}=\left(Q-\lfloor\mu\rfloor\right)\frac{\mu^Q}{Q!}$, so that $\displaystyle\sum_{k=0}^Q\frac{\mu^k}{k!}\geq\mu\frac{\mu^Q}{Q!}$ as soon as $Q\geq2\lfloor\mu\rfloor+1$. 

\medskip
\noindent $(ii)$ The function $\I$ is non-decreasing, non-negative and convex with $\I(0)=0$. If we look at the representation of $\mu\mapsto N^{\mu}$ by
\vskip-0.5cm
$$
N^{\mu}(\omega)=\max\Big\{n\in\N\left.\right|\prod_{i=1}^nU_i(\omega)>e^{-\mu}\Big\},
$$
where $U_i$ are i.i.d. uniformly distributed random variables on the probability space $\left(\Omega,{\cal A},\P\right)$, then $\mu\mapsto N^{\mu}$ is clearly non-decreasing, so $\mu\mapsto Q-N^{\mu}$ is non-increasing and $\mu\mapsto\varphi(\mu)=\I(Q-N^{\mu})-\I(Q-N^{\mu}-1)$ too (because of the convexity of $\I$).\hfill$\cqfd$

\bigskip \noindent {\bf Remark.} If $\mu\in(0,1)$, then $\mu\longmapsto\mu\,\P\left(N^{\mu}\leq Q\right)$ is always non-decreasing. If $\mu\in[1,2)$, then the function $\mu\longmapsto\mu\P\left(N^{\mu}=0\right)=\mu e^{-\mu}$ is not always non-decreasing clearly, but only on $[0,1]$.

\bigskip \noindent {\bf Proof of Proposition \ref{upperbound}.} $(a)$ In our problem the intensity parameter $\mu=\displaystyle\int_0^T\lambda(S_t-S_0+\delta)dt$ is continuous, non-increasing to zero when $\delta$ tends to $+\infty$ and bounded by assumption ($\lambda(-S_0)<+\infty$). Hence $\mu\in[0,\lambda(-S_0)T]$. 

\noindent $(i)$ From (\ref{Cderiv1}), we have for $\delta=0$,
\begin{eqnarray*}
 &&\widehat{C}\left(0,\Lambda_T(0,S),\frac{\partial}{\partial\delta}\Lambda_T(0,S),(S_t)_{0\le t\le T}\right)\\ 
&\leq&\left(S_0\frac{\partial}{\partial\delta}\Lambda_T(0,S)-\Lambda_T(0,S)\right)\P^{(0)}\left(N^{\mu}\leq Q_{T}-1\right)-\kappa S_T\frac{\partial}{\partial\delta}\Lambda_T(0,S)\varphi^{(0)}(\mu) \\
 &=&\left(\frac{\partial}{\partial\delta}\Lambda_T(0,S)\left(S_0-\kappa S_T\Theta(Q_T,\Lambda_T(0,S))\right)-\Lambda_T(0,S)\right)\P^{(0)}\left(N^{\mu}\leq Q_{T}-1\right),
\end{eqnarray*}
because $-Q_{T}\P\left(N^{\mu}>Q_{T}\right)_{\left.\right|\mu=\Lambda_T(0,S)}<0$ and small if $Q_{T}$ is large. Let $\underline{k}_0$ and $\overline{k}_0$ be defined by~(\ref{K0Dist}). We have that $\underline{k}_0>0$ $a.s.$ and $\overline{k}_0>0$ $a.s.$ by assumption, $i.e.$ $\frac{\partial}{\partial\delta}\Lambda_T(0,S)\leq -\underline{k}_0\Lambda_T(0,S)$ $a.s.$ and $\frac{\partial}{\partial\delta}\Lambda_T(0,S)\geq -\overline{k}_0\Lambda_T(0,S)$ $a.s.$. Then
\begin{eqnarray*}
\widehat{C}\left(0,\Lambda_T(0,S),\frac{\partial}{\partial\delta}\Lambda_T(0,S),(S_t)_{0\le t\le T}\right)\hskip10cm \\
\hskip5cm\leq-\left(1+\underline{k}_0S_0-\kappa S_{T}\overline{k}_0\Theta(Q_T,\Lambda_T(0,S))\right)\left(\mu\P\left(N^{\mu}\leq Q_{T}-1\right)\right)_{\left.\right|\mu=\Lambda_T(0,S)}.
\end{eqnarray*}

$$\mbox{Now, by Lemma \ref{LemMonot}}, \hskip2cm\Theta(Q_T,\mu)\leq\Theta(Q_T,0)=\I(Q)-\I(Q-1)=\varphi(0).\hskip5cm$$
\begin{eqnarray*}
\mbox{Therefore}& &\widehat{C}\Big(0,\Lambda_T(0,S),\frac{\partial}{\partial\delta}\Lambda_T(0,S),(S_t)_{0\le t\le T}\Big) \\
&\leq&-\left(1+\underline{k}_0S_0-\kappa S_{T}\overline{k}_0\left(\I(Q_T)-\I(Q_T-1)\right)\right)\left(\mu\P\left(N^{\mu}\leq Q_{T}-1\right)\right)_{\left.\right|\mu=\Lambda_T(0,S)}.\hskip2cm
\end{eqnarray*}
By Lemma \ref{LemMonot}, if $Q_T\geq 2T\lambda(-S_0)$, then $\mu\longmapsto\mu\P\left(N^{\mu}\leq Q-1\right)$ is non-decreasing. Moreover, the functionals
$$\begin{array}{rcl}
F:\D([0,T,\R)&\to&\R \\
  \alpha &\mapsto&\Lambda_T(0,\alpha)=\displaystyle\int_0^T\lambda(\alpha(t)-S_0)dt \\
\end{array}$$
\vskip-0.5cm
\noindent is non-increasing and 
$$\alpha(T)\longmapsto\left(-1-\underline{k}_0S_0+\kappa\alpha(T)\overline{k}_0\left(\I(Q_T)-\I(Q_T-1)\right)\right) \quad \mbox{is non-decreasing.}$$
Both are continuous (with respect to the sup-norm) at any $\alpha\in{\cal C}([0,T], \R)$. Conserquently, we obtain by the functional co-monotony principle 
\begin{eqnarray*}
&&\E\left[\left(-1-\underline{k}_0S_0+\kappa S_{T}\overline{k}_0\left(\I(Q_T)-\I(Q_T-1)\right)\right)\left(\mu\P\left(N^{\mu}\leq Q_{T}-1\right)\right)_{\left.\right|\mu=\Lambda_T(0,S)}\right] \\
&&\leq\E\left[-1-\underline{k}_0S_0+\kappa S_{T}\overline{k}_0\left(\I(Q_T)-\I(Q_T-1)\right)\right]\E\left[\left(\mu\P\left(N^{\mu}\leq Q_{T}-1\right)\right)_{\left.\right|\mu=\Lambda_T(0,S)}\right].
\end{eqnarray*}
In turn, this implies
$$
C'(0)\leq\E\left[-1-\underline{k}_0S_0+\kappa S_{T}\overline{k}_0\left(\I(Q_T)-\I(Q_T-1)\right)\right]\E\left[\left(\mu\P\left(N^{\mu}\leq Q_{T}-1\right)\right)_{\left.\right|\mu=\Lambda_T(0,S)}\right].
$$
As $\E\left[\left(\mu\P\left(N^{\mu}\leq Q_{T}-1\right)\right)_{\left.\right|\mu=\Lambda_T(0,S)}\right]\geq0$, then $C'(0)\leq0$ as soon as 
$$
\E\left[-1-\underline{k}_0S_0+\kappa S_{T}\overline{k}_0\left(\I(Q_T)-\I(Q_T-1)\right)\right]\leq0,$$
$$
i.e.\hskip5cm\kappa\leq\frac{1+\underline{k}_0S_0}{\overline{k}_0\E\left[S_{T}\right]\left(\I(Q_T)-\I(Q_T-1)\right)}.\hskip5cm
$$

\noindent$(b)$ From (\ref{Cderiv1}), the form of $\lambda$ and $(a)$, we get that for every $\delta\in[0,\delta_{{\rm max}}]$, $S=(S_i)_{1\leq i\leq m}\in\R^{m+1}$,
\vskip-0.2cm
$$H(\delta,S)=-Q_T\P^{(\delta)}\left(N^{\mu}>Q_{T}\right)+f(\delta,S)\left(\mu\P\left(N^{\mu}\leq Q_{T}-1\right)\right)_{\left.\right|\mu=\Lambda_T(\delta,S)},$$
where $f(\delta,S)=-1-k(S_0-\delta-\kappa S_T\Theta(Q_T-1,\Lambda_T(\delta,S)))$ if $\I\neq\operatorname{id}$ and $f(\delta,S)=-1-k(S_0-\delta-\kappa S_T)$ if $\I=\operatorname{id}$. Since $\delta\mapsto-Q_T\P^{(\delta)}\left(N^{\mu}>Q_{T}\right)$ is non-decreasing, $\delta\mapsto\left(\mu\P\left(N^{\mu}\leq Q_{T}-1\right)\right)_{\left.\right|\mu=\Lambda_T(\delta,S)}$ is non-increasing and non positive owing to Lemma \ref{LemMonot}~$(i)$ and $\delta\mapsto f(\delta,S)$ is non-decreasing owing to Lemma \ref{LemMonot}~$(ii)$, $\delta\mapsto H(\delta,S)$ is non-decreasing if $f(\delta,S)>0$, $\delta\in[0,\delta_{{\rm max}}]$, $S\in\R^{m+1}$, which leads to (\ref{CritHcroissante}). \hfill$\cqfd$

\subsection{Sufficient condition for the convexity condition}

We give below sufficient condition to insure the convexity of the execution cost function $C$, namely conservative upper bound for the free parameter $\kappa$. 

\begin{prop}\label{upperbound2} Assume that $(S_t)_{t\in[0,T]}$ satisfies a co-monotony principle.

\noindent $(i)$ If $\I\neq\operatorname{id}$, assume that there exists $\rho_Q\in\Big(0,1-\frac{\P\left(N^{\mu}=Q_{T}-1\right)}{\P\left(N^{\mu}\leq Q_{T}-1\right)}_{\left.\right|\mu=T\lambda(-S_0)}\Big)$ such that
$$
\forall x\in[1,Q_T-1], \quad \I(x)-\I(x-1)\leq\rho_Q(\I(x+1)-\I(x)).$$
$$\mbox{If} \hskip2cm Q_T\geq \Big(2\vee\big(1+\frac{\overline{k}_1^2}{\underline{k}_1\underline{k}_2}\big)\Big)T\lambda(-S_0) \qquad \mbox{and}\qquad
\kappa\leq\displaystyle\frac{2\underline{k}_1}{\overline{k}_1\overline{k}_2\E\left[S_{T}\right]\I'_{\ell}(Q_T)},\hskip3cm$$
where $\I'_{\ell}$ is the left derivative of the convex function $\I$, then $C''(\delta)\geq0$, $\delta\in[0,\delta_{{\rm max}}]$, so that $C$ is convex on $[0,\delta_{{\rm max}}]$. \\

\noindent $(ii)$ When $\I=\operatorname{id}$, the bound on $\kappa$ reads
$$\kappa\leq\displaystyle\frac{2\underline{k}_1}{\overline{k}_1\overline{k}_2\E\left[S_{T}\right]}.$$
\end{prop}

\paragraph{Remark.} Note that if we replace the diffusion process $(S_t)_{t\in[0,T]}$ by its stepwise constant Euler scheme or by a discrete time sample $(S_{t_i})_{0\leq i\leq m}$, we  have the same criteria owing to Corollary~\ref{cor:IneqCovDiff} (see also Theorem~\ref{Criteria}-$(c)$).\\

\noindent To prove the above Proposition, we need the following results
\begin{lem}\label{lemrho1} If $\mu\leq Q-1$, then $\mu\mapsto\frac{\P\left(N^{\mu}=Q-1\right)}{\P\left(N^{\mu}\leq Q-1\right)}$ is non-decreasing. 
\end{lem}

\smallskip

\noindent{\bf Proof of Lemma \ref{lemrho1}.}
\begin{eqnarray*}
\frac{d}{d\mu}\frac{\P\left(N^{\mu}=Q-1\right)}{\P\left(N^{\mu}\leq Q-1\right)}&=&\frac{\P\left(N^{\mu}=Q-2\right)-\P\left(N^{\mu}=Q-1\right)}{\P\left(N^{\mu}\leq Q-1\right)}+\frac{\P\left(N^{\mu}=Q-1\right)^2}{\P\left(N^{\mu}\leq Q-1\right)^2} \\
&=&\frac{\P\left(N^{\mu}=Q-1\right)\left(\P\left(N^{\mu}\leq Q-1\right)\left(\frac{Q-1}{\mu}-1\right)+\P\left(N^{\mu}=Q-1\right)\right)}{\P\left(N^{\mu}\leq Q-1\right)^2} \\
&\geq&0 \quad \mbox{if} \quad \mu\leq Q-1. \hskip8cm\cqfd
\end{eqnarray*}

\begin{lem}\label{lemrho2} Assume that $\I\neq\operatorname{id}$. If there exists $\rho_Q\in\Big(0,1-\frac{\P\left(N^{\mu}=Q_{T}-1\right)}{\P\left(N^{\mu}\leq Q_{T}-1\right)}_{\left.\right|\mu=T\lambda(-S_0)}\Big)$ such that
$$\forall x\in[1,Q_T-1], \quad \I(x)-\I(x-1)\leq\rho_Q(\I(x+1)-\I(x)),$$
then $\mu\mapsto\frac{\varphi(\mu)}{\P\left(N^{\mu}\leq Q_{T}-1\right)}$ is non-increasing where $\varphi(\mu)=\I(Q-N^{\mu})-\I(Q-N^{\mu}-1)$.
\end{lem}

\smallskip
\noindent{\bf Remark.} If $\I=\operatorname{id}$, then $\mu\mapsto\frac{\varphi(\mu)}{\P\left(N^{\mu}\leq Q_{T}-1\right)}\equiv1$, therefore we do not need the previous lemmas. 

\bigskip
\noindent{\bf Proof of Lemma \ref{lemrho2}.} We have
$$\frac{d}{d\mu}\frac{\varphi(\mu)}{\P\left(N^{\mu}\leq Q_{T}-1\right)}=\frac{\P\left(N^{\mu}=Q_{T}-1\right)\varphi(\mu)}{\P\left(N^{\mu}\leq Q_{T}-1\right)^2}-\frac{\psi(\mu)}{\P\left(N^{\mu}\leq Q_{T}-1\right)} \leq 0$$
where $\psi(\mu)=\I\left(\left(Q_{T}-N^{\mu}-2\right)_+\right)-2\I\left(\left(Q_{T}-N^{\mu}-1\right)_+\right)+\I\left(\left(Q_{T}-N^{\mu}\right)_+\right)$
$$\mbox{iff}\hskip3.5cm-\psi(\mu)\P\left(N^{\mu}\leq Q_{T}-1\right)+\P\left(N^{\mu}=Q_{T}-1\right)\varphi(\mu)\leq0\hskip4.5cm$$
$$\mbox{iff}\hskip0.5cm\P\left(N^{\mu}\leq Q_{T}-1\right)\E\left[\I\left(\left(Q_{T}-N^{\mu}-1\right)_+\right)-\I\left(\left(Q_{T}-N^{\mu}-2\right)_+\right)\right]
\leq\P\left(N^{\mu}\leq Q_T-2\right)\varphi(\mu).\hskip1cm$$
But
\begin{eqnarray*}
\I\left(\left(Q_{T}-N^{\mu}-1\right)_+\right)&-&\I\left(\left(Q_{T}-N^{\mu}-2\right)_+\right)\\
&\leq&\rho_Q\left(\I\left(\left(Q_{T}-N^{\mu}\right)_+\right)-\I\left(\left(Q_{T}-N^{\mu}-1\right)_+\right)\right)\mathds{1}_{\{N^{\mu}\leq Q_T-2\}} \\
&=&\rho_Q\left(\I\left(\left(Q_{T}-N^{\mu}\right)_+\right)-\I\left(\left(Q_{T}-N^{\mu}-1\right)_+\right)-\left(\I(1)-\I(0)\right)\mathds{1}_{\{N^{\mu}=Q_T-1\}}\right) \\
&\leq&\rho_Q\left(\I\left(\left(Q_{T}-N^{\mu}\right)_+\right)-\I\left(\left(Q_{T}-N^{\mu}-1\right)_+\right)\right)
\end{eqnarray*}
$$\mbox{since}\hskip4.5cm\left(\I(1)-\I(0)\right)\mathds{1}_{\{N^{\mu}=Q_T-1\}}\geq0 \quad a.s.\hskip5.5cm$$
Consequently
\begin{eqnarray*}
& &\P\left(N^{\mu}\leq Q_{T}-1\right)\E\left[\I\left(\left(Q_{T}-N^{\mu}-1\right)_+\right)-\I\left(\left(Q_{T}-N^{\mu}-2\right)_+\right)\right]-\P\left(N^{\mu}\leq Q_T-2\right)\varphi(\mu) \\
&\leq&\left(\rho_Q\P\left(N^{\mu}\leq Q_{T}-1\right)-\P\left(N^{\mu}\leq Q_{T}-2\right)\right)\varphi(\mu) \\
&=&\Big(\rho_Q-\Big(1-\frac{\P\left(N^{\mu}=Q_{T}-1\right)}{\P\left(N^{\mu}\leq Q_{T}-1\right)}\Big)\Big)\P\left(N^{\mu}\leq Q_{T}-1\right)\varphi(\mu)\leq0\quad\mbox{if}\quad\rho_Q\leq1-\frac{\P\left(N^{\mu}=Q_{T}-1\right)}{\P\left(N^{\mu}\leq Q_{T}-1\right)}. \cqfd
\end{eqnarray*}

\medskip

\noindent{\bf Proof of Proposition \ref{upperbound2}.} By using the notation (\ref{K1Dist})-(\ref{K2Dist}), we obtain the following bound for the second derivative of the cost function
\begin{eqnarray*}
C''(\delta)&\geq&\E\left[2\underline{k}_1\Lambda_T(\delta,S)\P^{(\delta)}\left(N^{\mu}\leq Q_{T}-1\right)\right. \\
           & &+(S_0-\delta)\underline{k}_1\underline{k}_2\Lambda_T(\delta,S)\Big(\P^{(\delta)}\left(N^{\mu}\leq Q_{T}-1\right)-\frac{\overline{k}_1^2}{\underline{k}_1\underline{k}_2}\Lambda_T(\delta,S)\P^{(\delta)}\left(N^{\mu}=Q_{T}-1\right)\Big)\\
           & &\left.-\kappa S_T\Lambda_T(\delta,S)\left(\overline{k}_1\overline{k}_2\phi^{(\delta)}(\mu)-\underline{k}_1^2\Lambda_T(\delta,S)\psi^{(\delta)}(\mu)\right)\right].
\end{eqnarray*}
By adapting the result of Lemma \ref{LemMonot}, we obtain, if $Q_T\geq \Big(1+\frac{\overline{k}_1^2}{\underline{k}_1\underline{k}_2}\Big)T\lambda(-S_0)$, that
$$\E\Big[\Big(\P\left(N^{\mu}\leq Q_{T}-1\right)-\frac{\overline{k}_1^2}{\underline{k}_1\underline{k}_2}\mu\P\left(N^{\mu}=Q_{T}-1\right)\Big)_{\left.\right|\mu=\Lambda_T(\delta,S)}\Big]\geq0$$
and by convexity of the penalty function $\I$, we have $\psi^{(\delta)}(\mu)\geq0$ $a.s.$ Then we obtain the following upper bound for $\kappa$,
\begin{equation}\label{BoundKappa}
\kappa\leq\frac{2\underline{k}_1\E\left[\left(\mu\P\left(N^{\mu}\leq Q_{T}-1\right)\right)_{\left.\right|\mu=\Lambda_T(\delta,S)}\right]}{\overline{k}_1\overline{k}_2\E\left[S_T\Lambda_T(\delta,S)\varphi^{(\delta)}(\mu)\right]}.
\end{equation}
By Lemma \ref{lemrho2}, $\mu\mapsto\frac{\varphi(\mu)}{\P\left(N^{\mu}\leq Q_{T}-1\right)}$ is non-increasing and by Lemma \ref{LemMonot} $\mu\mapsto\mu\P\left(N^{\mu}\leq Q_{T}-1\right)$ is non-decreasing for $Q_T\geq2\left\lfloor \mu\right\rfloor-1$. Furthermore $\alpha\mapsto\Lambda_T(\delta,\alpha)$ is non-increasing. By applying the functional co-monotony principle, we then have, for $Q_T\geq2T\lambda(-S_0)$, that
$$\E\left[S_T\Lambda_T(\delta,S)\varphi^{(\delta)}(\mu)\right]\leq\E\left[\left(\mu\P\left(N^{\mu}\leq Q_{T}-1\right)\right)_{\left.\right|\mu=\Lambda_T(\delta,S)}\right]\E\Big[\Big(\frac{\varphi(\mu)}{\P\left(N^{\mu}\leq Q_{T}-1\right)}\Big)_{\left.\right|\mu=\Lambda_T(\delta,S)}\Big].$$
Therefore (\ref{BoundKappa}) will be satisfied as soon as
$$\kappa\leq\frac{2\underline{k}_1}{\overline{k}_1\overline{k}_2\E\Big[S_T\left(\frac{\varphi(\mu)}{\P\left(N^{\mu}\leq Q_{T}-1\right)}\right)_{\left.\right|\mu=\Lambda_T(\delta,S)}\Big]}.$$
Since by Lemma \ref{lemrho2}, $\mu\mapsto\frac{\varphi(\mu)}{\P\left(N^{\mu}\leq Q_{T}-1\right)}$ is non-increasing, we get
$$\Big(\frac{\varphi(\mu)}{\P\left(N^{\mu}\leq Q_{T}-1\right)}\Big)_{\left.\right|\mu=\Lambda_T(\delta,S)}\leq\Big(\frac{\varphi(\mu)}{\P\left(N^{\mu}\leq Q_{T}-1\right)}\Big)_{\left.\right|\mu=0}=\I(Q_+)-\I((Q-1)_+)\leq\I'_{\ell}(Q),$$
which finally yields the announced (more stringent) criterion
$$\hskip6cm\kappa\leq\frac{2\underline{k}_1}{\overline{k}_1\overline{k}_2\E\left[S_T\right]\I'(Q)}.\hskip6.5cm\cqfd$$

\bigskip \noindent {\bf Remark.}  As $\delta\in[0,\delta_{{\rm max}}]$, then $(S_0-\delta)\in[S_0-\delta_{{\rm max}},S_0]$ and 
	\begin{eqnarray*}
	& &(S_0-\delta)\underline{k}_1\underline{k}_2\Lambda_T(\delta,S)\Big(\P^{(\delta)}\left(N^{\mu}\leq Q_{T}-1\right)-\frac{\overline{k}_1^2}{\underline{k}_1\underline{k}_2}\Lambda_T(\delta,S)\P^{(\delta)}\left(N^{\mu}=Q_{T}-1\right)\Big) \\
	&\geq&(S_0-\delta_{{\rm max}})\underline{k}_1\underline{k}_2\Lambda_T(\delta,S)\Big(\P^{(\delta)}\left(N^{\mu}\leq Q_{T}-1\right)-\frac{\overline{k}_1^2}{\underline{k}_1\underline{k}_2}\Lambda_T(\delta,S)\P^{(\delta)}\left(N^{\mu}=Q_{T}-1\right)\Big) \\
	&=&(S_0-\delta_{{\rm max}})\underline{k}_1\underline{k}_2\Big[\mu\Big(\P\left(N^{\mu}\leq Q_{T}-1\right)-\frac{\overline{k}_1^2}{\underline{k}_1\underline{k}_2}\mu\P\left(N^{\mu}=Q_{T}-1\right)\Big)\Big]_{\left|\right.\mu=\Lambda_T(\delta,S)}.
	\end{eqnarray*}
	Unfortunately we cannot use the functional co-monotony principle to improve the bound because, for $Q\geq \Big(2\vee\Big(1+\frac{\overline{k}_1^2}{\underline{k}_1\underline{k}_2}\Big)\Big)T\lambda(-S_0)$, the function $\mu\mapsto\mu\P\left(N^{\mu}\leq Q_{T}-1\right)$ is non-decreasing and $\mu\mapsto1-\frac{\mu\P\left(N^{\mu}=Q_{T}-1\right)}{\P\left(N^{\mu}\leq Q_{T}-1\right)}$ is non-increasing, and we need to obtain a lower bound for this expression but co-monotony naturally yields an upper bound.
	


\section{Numerical experiments}
\label{DistSix}

In this section, we present numerical results with simulated and real data. We first present the chosen model for the price dynamic and the penalization function. Within the numerical examples, we are modeling the optimal behavior of a ``learning trader'' reassessing the price of his passive order every 5 units of time (can be seconds or minutes) in the order books to adapt to the characteristics of the market (fair price moves $S_t$ and order flow dynamics $N_t$). During each  period of 5 seconds, the trader posts her order of size $Q_5$ in the book at a distance $\delta$ of the best opposite price ($\delta$ lower than the best ask for a buy order), and waits 5 seconds. If the order is not completely filled after these 5 seconds (say at time $T$), the trader cancels the remaining quantity $(Q_5-N_5)_+$ and buys it using a market order at price $S_T$ plus a market impact; she will buy at $\kappa S_T \, (1+\eta(( Q_5-N_5)_+))$.
Then she can reproduce the experiment choosing another value for the distance to the best opposite $\delta$.

The reassessment procedure used here is the one of formula (\ref{ASDist}) using the expectation representation of $C'$ given by Property~\ref{DerivExpect} to provide the proper form for the function $H$.

Then we plot the cost function and its derivative for  a trivial penalization function $\Phi= Id$ ($\eta \equiv 0$)  and for a nontrivial one. We conclude by giving the results of the recursive procedure in both cases,  either on  simulated data or on a  real data obtained by ``replaying" the market. 

\subsection{Simulated data}\label{sec:simd}

$$\mbox{We assume that}\hskip1cm
dS_t=\sigma dW_t, \quad S_0=s_0 \quad \mbox{and} \quad \Lambda_T(\delta,S)=A\displaystyle\int_0^{T}e^{-k(S_t-S_0+\delta)}dt\hskip3cm$$
where $(W_t)_{t\geq0}$ is a standard Brownian motion and $\sigma,A,k>0$ (this means that $\lambda(x)=Ae^{-kx}$). We denote here by $(\bar{S}_t)_{t\geq0}$ the Euler scheme with step $\frac Tm$ of $(S_t)_{t\in[0,T]}$ defined by
$$
\bar{S}_{k+1}:=\bar{S}_{k}+\sigma\sqrt{\frac{T}{m}}Z_{k+1}, \quad \bar{S}_0=s_0, \quad Z_{k+1}\sim{\cal N}(0,1),\quad k\geq0,
$$
and we approximate $\Lambda_T(\delta,S)$ by $\bar{\Lambda}_T(\delta,S)=A\frac{T}{m}\sum_{k=0}^m e^{-k(\bar{S}_k-S_0+\delta)}$. The market impact penalization function is $\I(x)=(1+\eta(x))x \quad \mbox{with} \quad \eta(x)=A'e^{k'x}.$ Now we present the cost function and its derivative for the following parameters: 
\begin{itemize}
	\item[$\bullet$] parameters of the asset dynamics: $s_0=100$ and $\sigma=0.01$,
	\item[$\bullet$] parameters of the intensity of the execution process: $A=5$ and $k=1$, 
	\item[$\bullet$] parameters of the execution: $T=5$ and $Q=10$,
	\item[$\bullet$] parameters of the penalization function: $\kappa=1$, $A'=0.1$ and $k'=0.05$.
\end{itemize}
We use $m=20$ for the Euler scheme and $M=10000$  Monte Carlo simulations. \\


\noindent\underline{{\bf{\em Settin}}}{\bf{\em g}}\underline{ {\bf{\em  1}}} ($\eta\not\equiv0$)
\begin{figure}[!ht]
\vskip-0.3cm
\centering
\includegraphics[width=14cm]{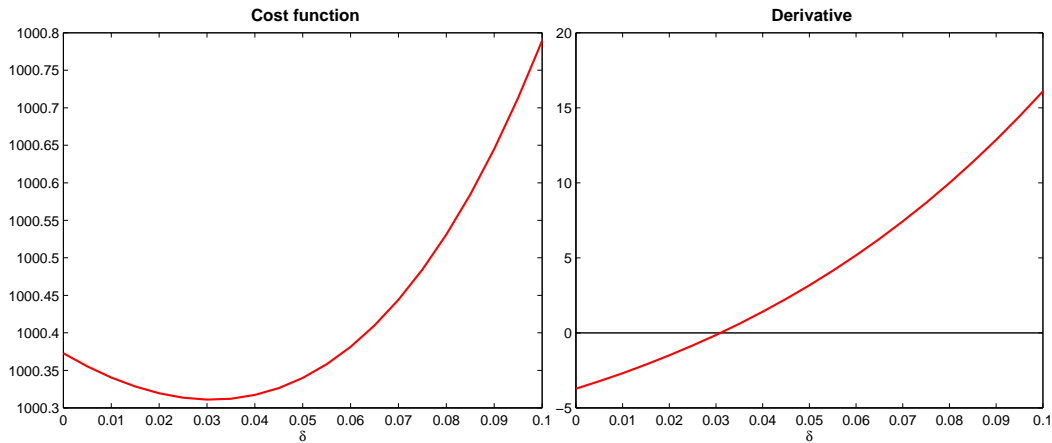}
\vskip-0.5cm
\caption{$\eta\not\equiv0$: $T=5$, $A=5$, $k=1$, $s_0=100$, $\sigma=0.01$, $Q=10$, $\kappa=6$, $A'=1$, $k'=0.01$, $m=20$ and $M=10000$.}
\label{CoutsPen}
\end{figure}

\clearpage

\noindent\underline{{\bf{\em Settin}}}{\bf{\em g}}\underline{ {\bf{\em  2}}} ($\eta\equiv0$)
\begin{figure}[!ht]
\vskip-0.3cm
\centering
\includegraphics[width=14cm]{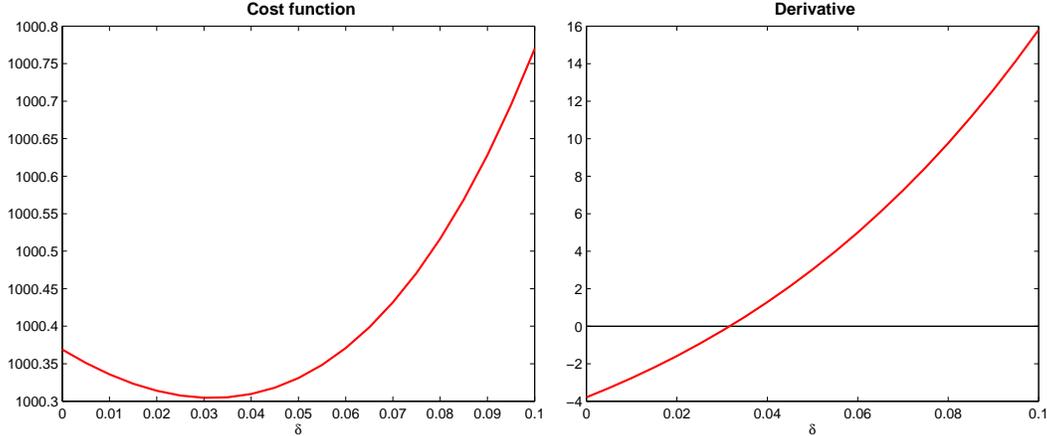}
\vskip-0.5cm
\caption{$\eta\equiv0$: $T=5$, $A=5$, $k=1$, $s_0=100$, $\sigma=0.01$, $Q=10$, $\kappa=12$, $m=20$ and $M=10000$.}
\label{Couts}
\end{figure}

\paragraph{Remark.} When one looks at the cost functions in Figures~\ref{CoutsPen} and~\ref{Couts} (left), one may think that it would be simpler to compute the cost functions to derive the minimum and the associated optimal distance. But the computing time of the costs is about 100 seconds which is too large compared to the length of the posting period $T=5$ seconds, whereas that of the stochastic recursive procedure is about 1 second. \\

Now we present the results of the stochastic recursive procedure for the two settings with
$$n=100 \quad \mbox{and} \quad \gamma_n=\frac{1}{100n}.$$


\noindent\underline{{\bf{\em Settin}}}{\bf{\em g}}\underline{ {\bf{\em  1}}} ($\eta\not\equiv0$)
\begin{figure}[!ht]
\vskip-0.3cm
\centering
\begin{tabular}{cc}
\includegraphics[width=8cm]{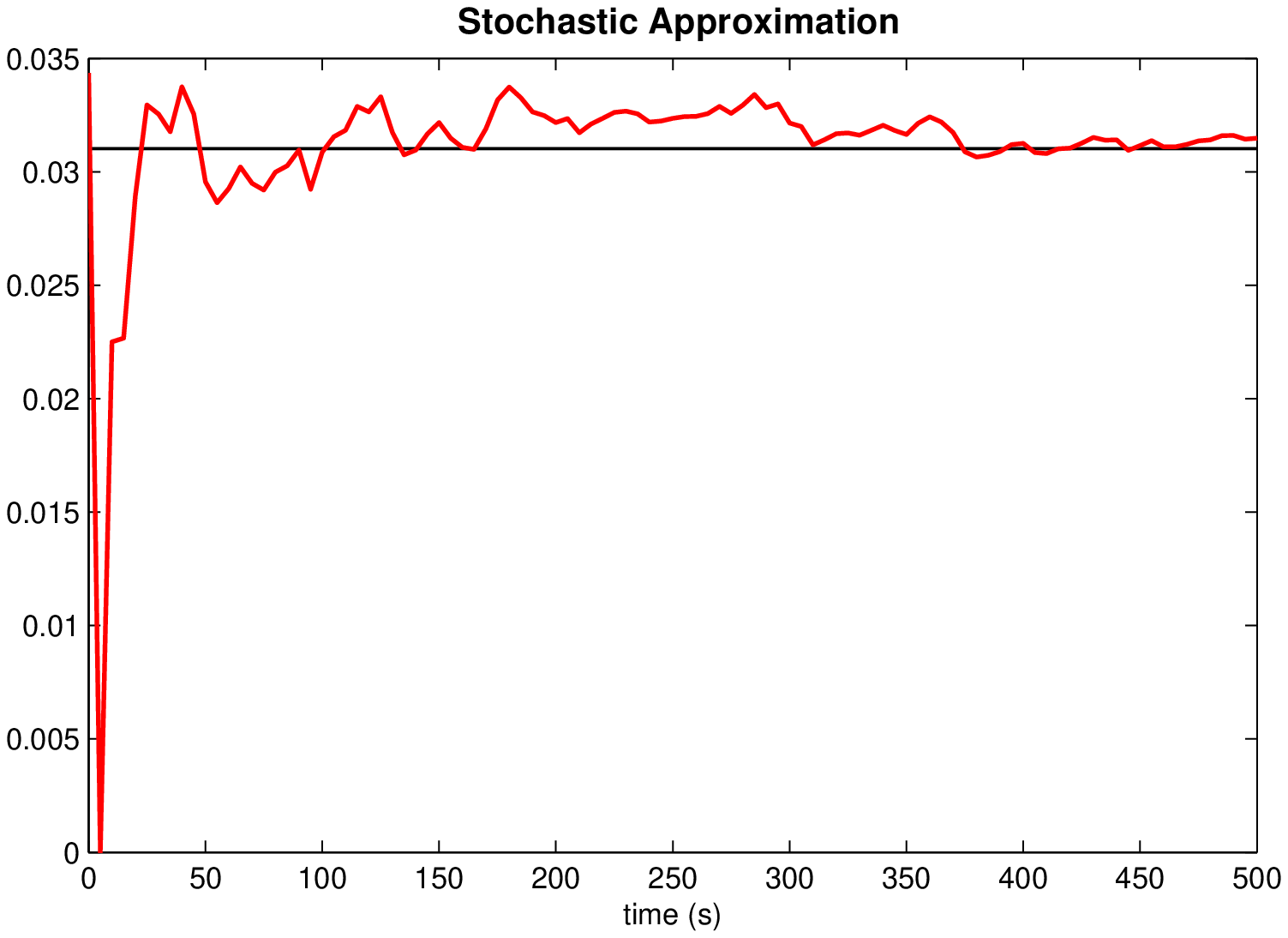} &
\includegraphics[width=8cm]{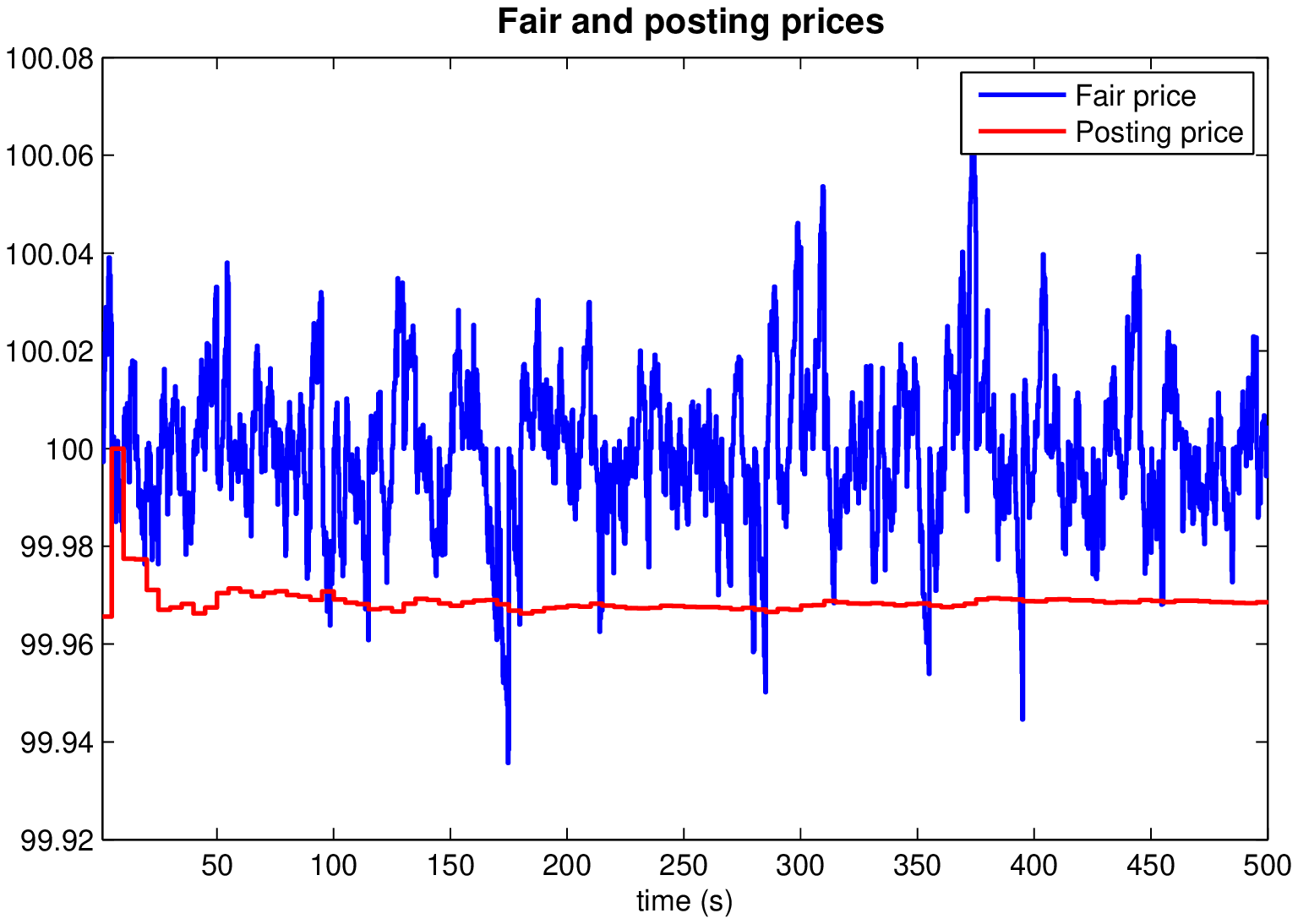}
\end{tabular}
\vskip-0.5cm
\caption{$\eta\not\equiv0$: $T=5$, $A=5$, $k=1$, $s_0=100$, $\sigma=0.01$, $Q=10$, $\kappa=6$, $A'=1$, $k'=0.01$, $m=20$ and $n=100$ }
\label{AlgoStoPen}
\end{figure}

\clearpage

\noindent\underline{{\bf{\em Settin}}}{\bf{\em g}}\underline{ {\bf{\em  2}}} ($\eta\equiv0$)
\begin{figure}[!ht]
\vskip-0.3cm
\centering
\begin{tabular}{cc}
\includegraphics[width=8cm]{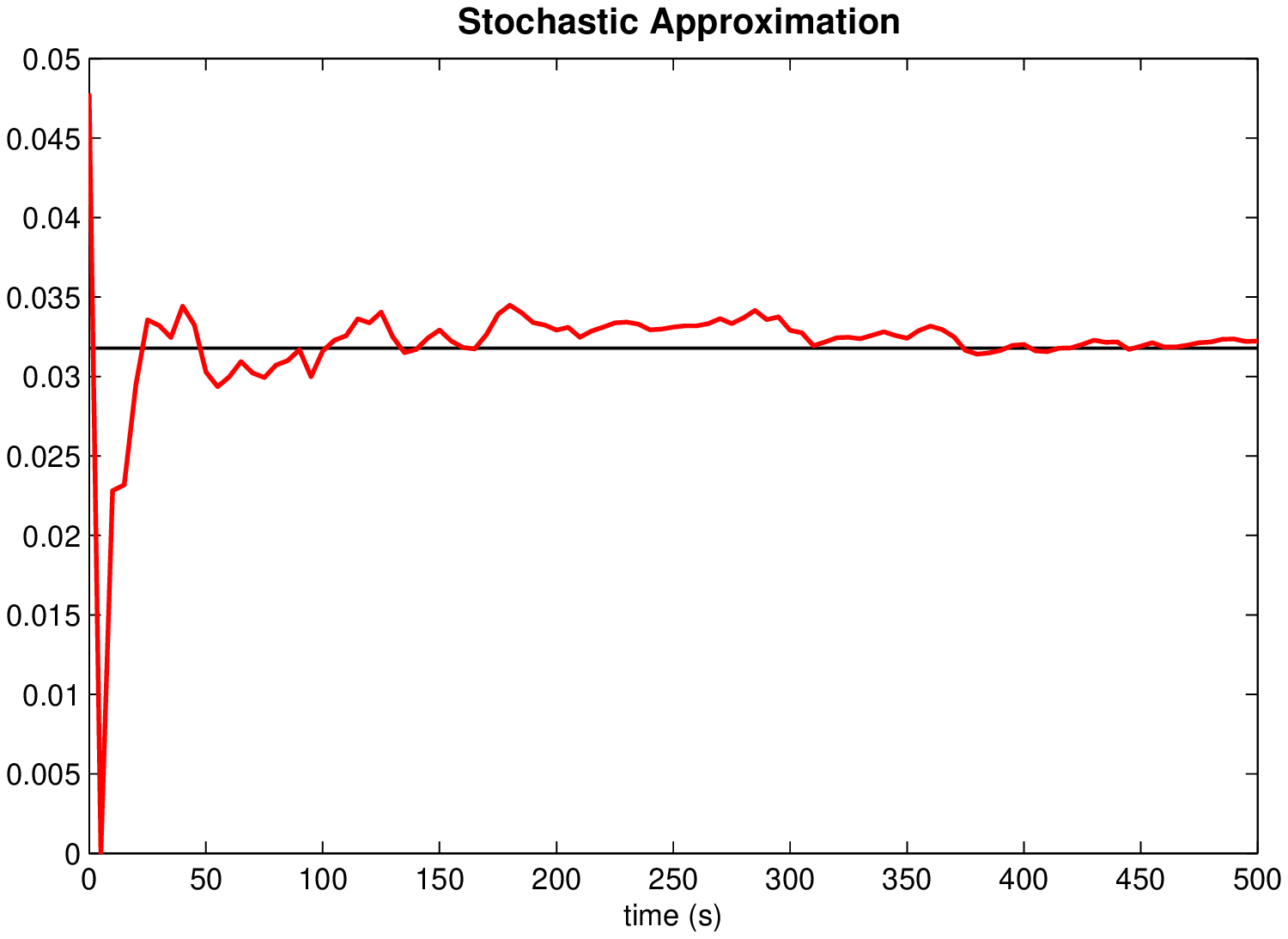} & 
\includegraphics[width=8cm]{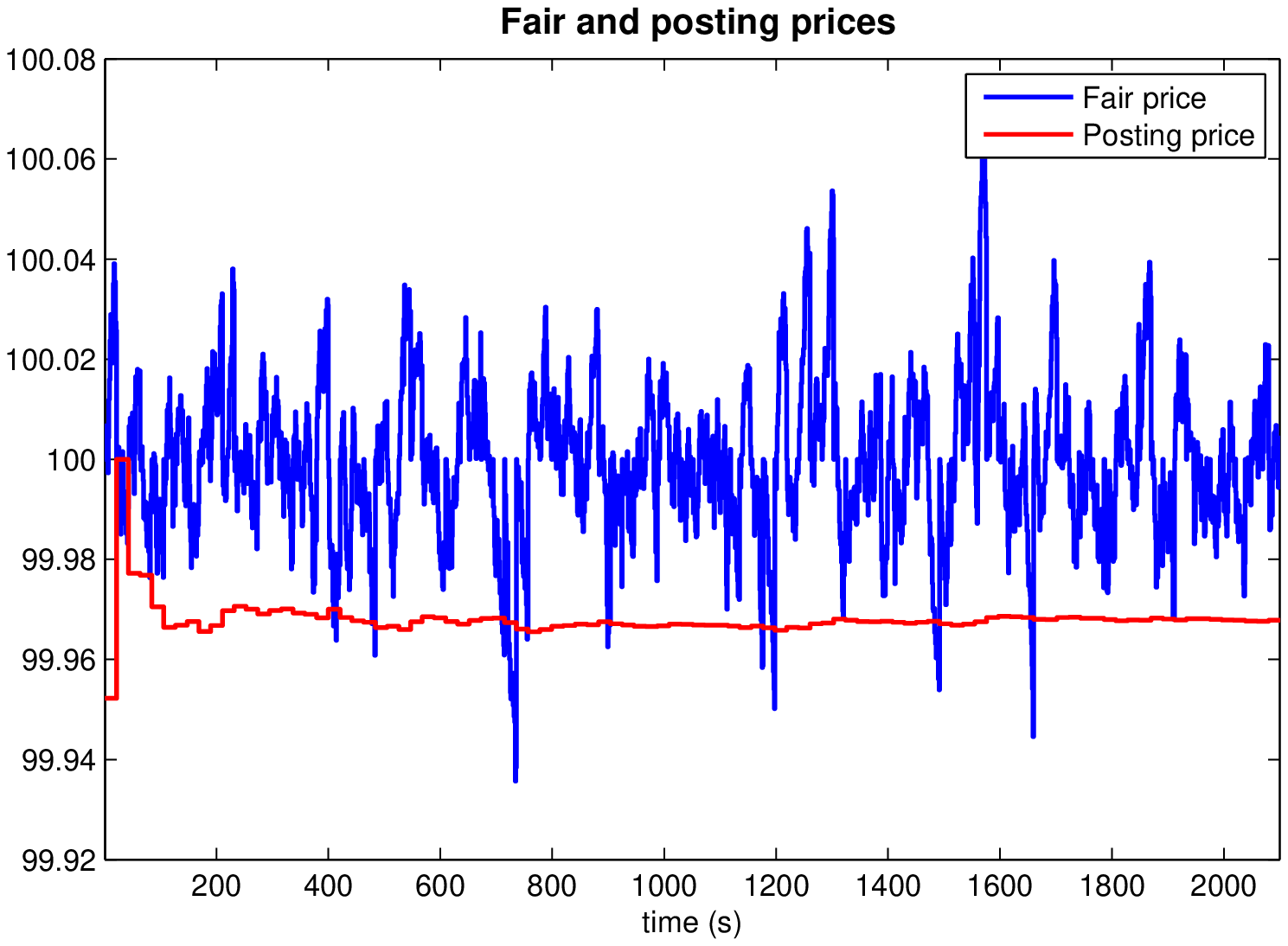}
\end{tabular}
\vskip-0.5cm
\caption{$\eta\equiv0$: $T=5$, $A=5$, $k=1$, $s_0=100$, $\sigma=0.01$, $Q=10$, $\kappa=12$, $m=20$ and $n=100$.}
\label{AlgoStoId}
\end{figure}

\subsection{Market data}\label{sec:marketd}

The self-adaptiveness nature of this recurrence procedure allows to implement it on real data, even if they are not exactly fulfilling the model assumptions. In the numerical example of this section, the trader reassess his order using the previously exposed recurrence procedure on real data on which the parameters of the models ($k$, $A$, $\kappa$, $k'$, $A'$) have been beforehand fitted.
 
\begin{figure}[!ht]
\vskip-0.3cm
\centering
\includegraphics[width=7cm]{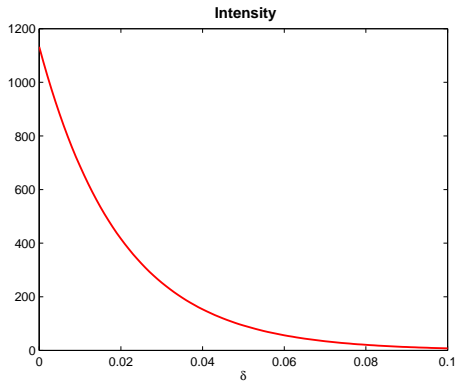}
\vskip-0.5cm
\caption{Fit of the exponential model on real data (Accor SA (ACCP.PA) 11/11/2010): $A=1/50$, $k=50$ and $N_{{\rm cycles}}=220$.}
\label{LambdaRealData}
\end{figure}

As market data, we use the bid prices of Accor SA (ACCP.PA) of 11/11/2010 for the fair price process $(S_t)_{t\in[0,T]}$. We divide the day into periods of 15 trades which will denote the steps of the stochastic procedure. Let $N_{{\rm cycles}}$ be the number of these periods. For every $n\in N_{{\rm cycles}}$, we have a sequence of bid prices $(S^{(n)}_{t_i})_{1\leq i\leq15}$ and we approximate the jump intensity of the Poisson process $\Lambda_{T^n}(\delta,S)$, where $T^n=\sum_{i=1}^{15}t_i$, by
\vskip-0.5cm 
$$\forall n\in\{1,\ldots,N_{{\rm cycles}}\}, \quad \Lambda_{T^n}(\delta,S)=A\sum_{i=2}^{15}e^{-k(S^{(n)}_{t_i}-S_{t_1}+\delta)}(t_i-t_{i-1}).$$
The empirical mean of the intensity function
\vskip-0.5cm 
$$\bar{\Lambda}(\delta,S)=\frac{1}{N_{{\rm cycles}}}\sum_{n=1}^{N_{{\rm cycles}}}\Lambda_{T^n}(\delta,S)$$
is plotted on Figure \ref{LambdaRealData}.

\clearpage

The penalization function has the following form
$$\I(x)=(1+\eta(x))x \quad \mbox{with} \quad \eta(x)=A'e^{k'x}.$$
Now we present the cost function and its derivative for the following parameter values: $A=1/50$, $k=50$, $Q=100$, $A'=0.001$ and $k'=0.0005$. 

\bigskip
\noindent\underline{{\bf{\em Settin}}}{\bf{\em g}}\underline{{\bf{\em  1}}} ($\eta\not\equiv0$)
\begin{figure}[!ht]
\vskip-0.3cm
\centering
\includegraphics[width=15cm]{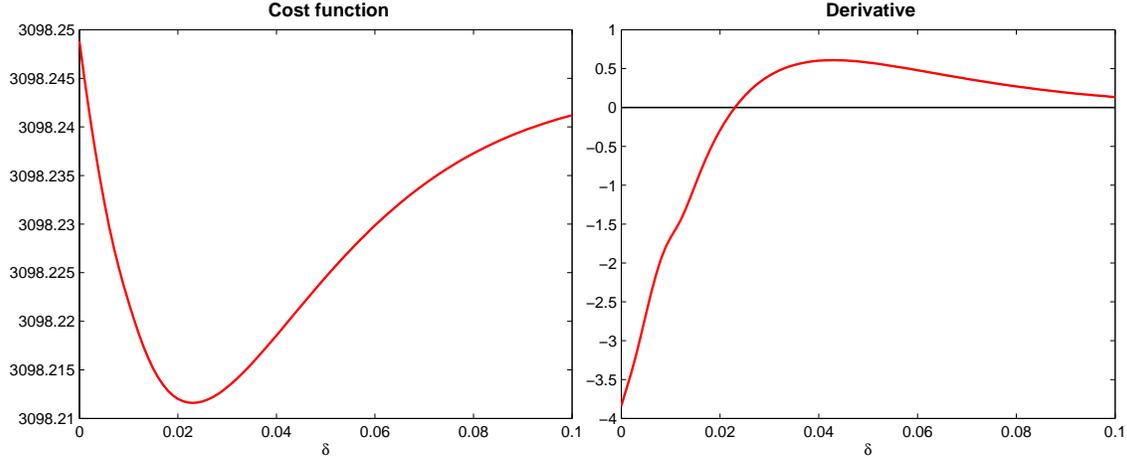}
\vskip-0.5cm
\caption{$\eta\not\equiv0$: $A=1/50$, $k=50$, $Q=100$, $\kappa=1$, $A'=0.001$, $k'=0.0005$ and $N_{{\rm cycles}}=220$.}
\label{CoutsPenRealData}
\end{figure}


\noindent\underline{{\bf{\em Settin}}}{\bf{\em g}}\underline{{\bf{\em  2 }}}($\eta\equiv0$)
\begin{figure}[!ht]
\vskip-0.3cm
\centering
\includegraphics[width=15cm]{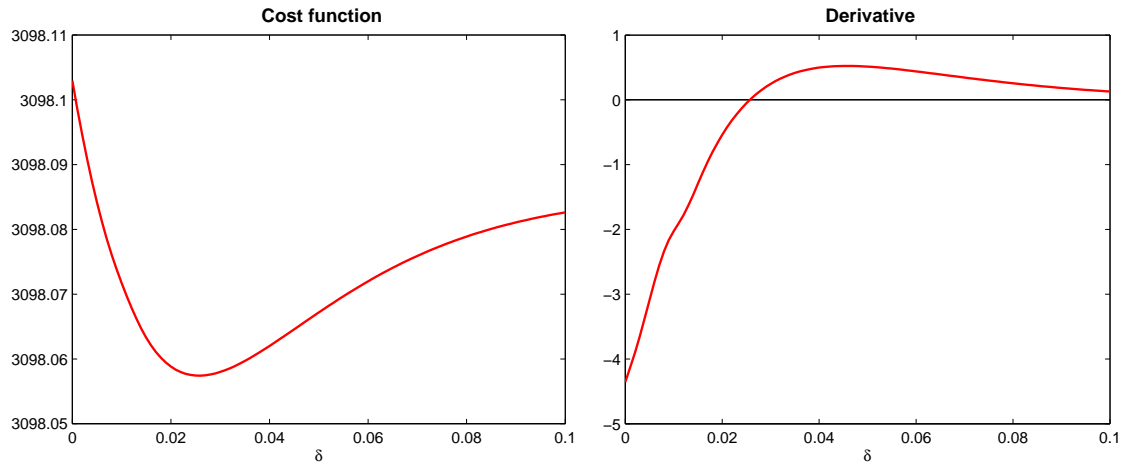}
\vskip-0.52cm
\caption{$\eta\equiv0$: $A=1/50$, $k=50$, $Q=100$, $\kappa=1.001$ and $N_{{\rm cycles}}=220$.}
\label{CoutsRealData}
\end{figure}

Now we present the results of the stochastic recursive procedure for two cases. To smoothen the behavior of the stochastic algorithm, we use Ruppert and Poliak's averaging principle  (see~\cite{Duf}). In short, this principle is two-folded: 

\smallskip
-- {\em Phase~1}: Implement the original zero search procedure with $\gamma_n =\frac{\gamma_1}{n^{\rho}}$, $\frac 12 <\rho<1$, $\gamma_1>0$,

\smallskip
-- {\em Phase~2}: Compute the arithmetic mean at each step $n$ of all the past values of the procedure, namely
\[
\bar \delta_n = \frac{1}{n+1} \sum_{k=0}^n\delta_k,\; n\ge 1.
\]
It has been shown by several authors that this procedure under appropriate assumptions is ruled by a $CLT$ having a minimal asymptotic variance (among recursive procedures).

\bigskip
\noindent\underline{{\bf{\em Settin}}}{\bf{\em g}}\underline{ {\bf{\em  1}}} ($\eta\not\equiv0$)
\begin{figure}[!ht]
\vskip-0.3cm
\centering
\begin{tabular}{cc}
\includegraphics[width=8cm]{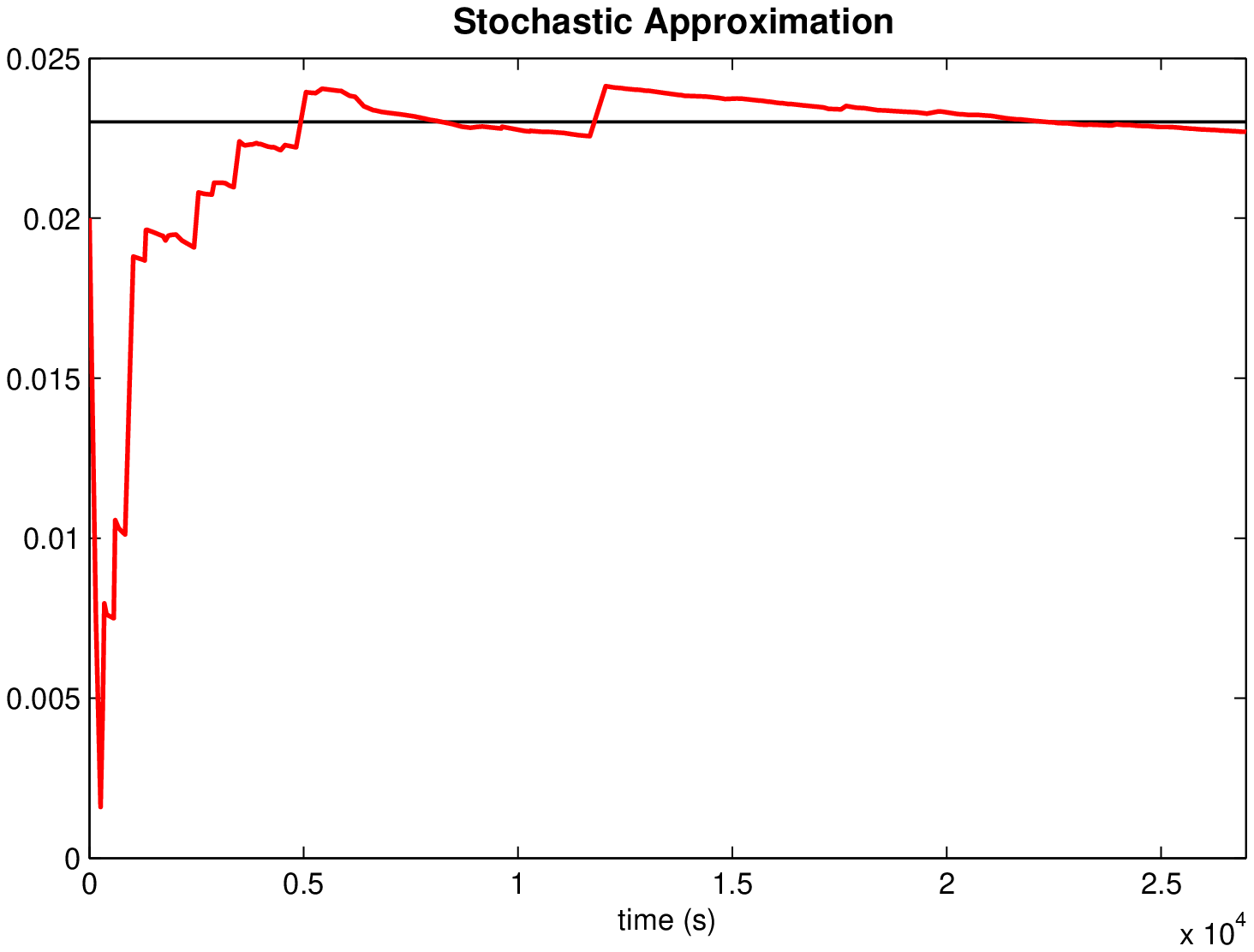}
&
\includegraphics[width=8cm]{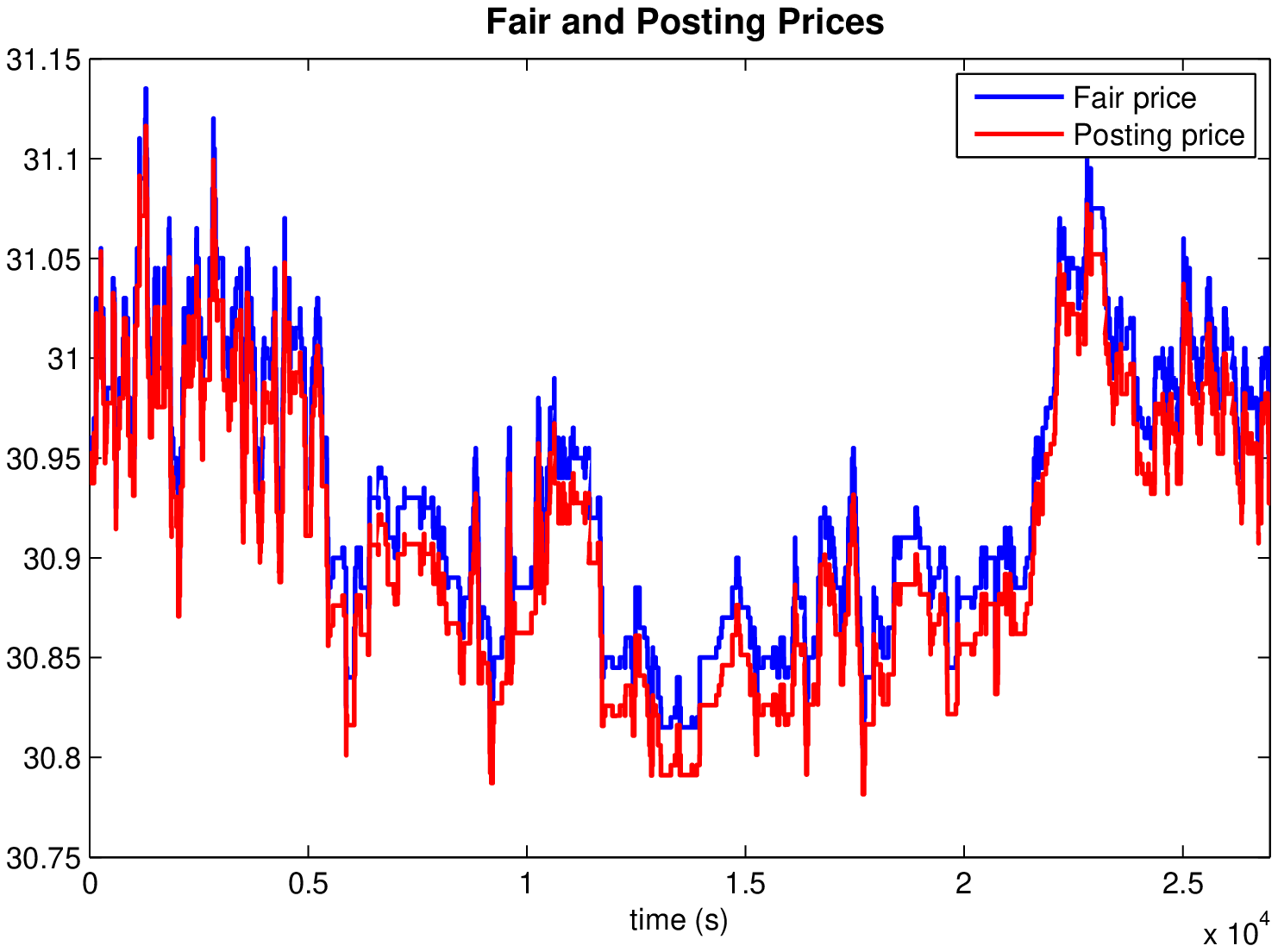}
\end{tabular}
\vskip-0.52cm
\caption{$\eta\not\equiv0$: $A=1/50$, $k=50$, $Q=100$, $\kappa=1$, $A'=0.001$, $k'=0.0005$ and $N_{{\rm cycles}}=220$. Crude algorithm with $\gamma_n=\frac{1}{550n}$.}
\label{AlgoStoPenRealData}
\end{figure}


\begin{figure}[!ht]
\vskip-0.3cm
\centering
\begin{tabular}{cc}
\includegraphics[width=8cm]{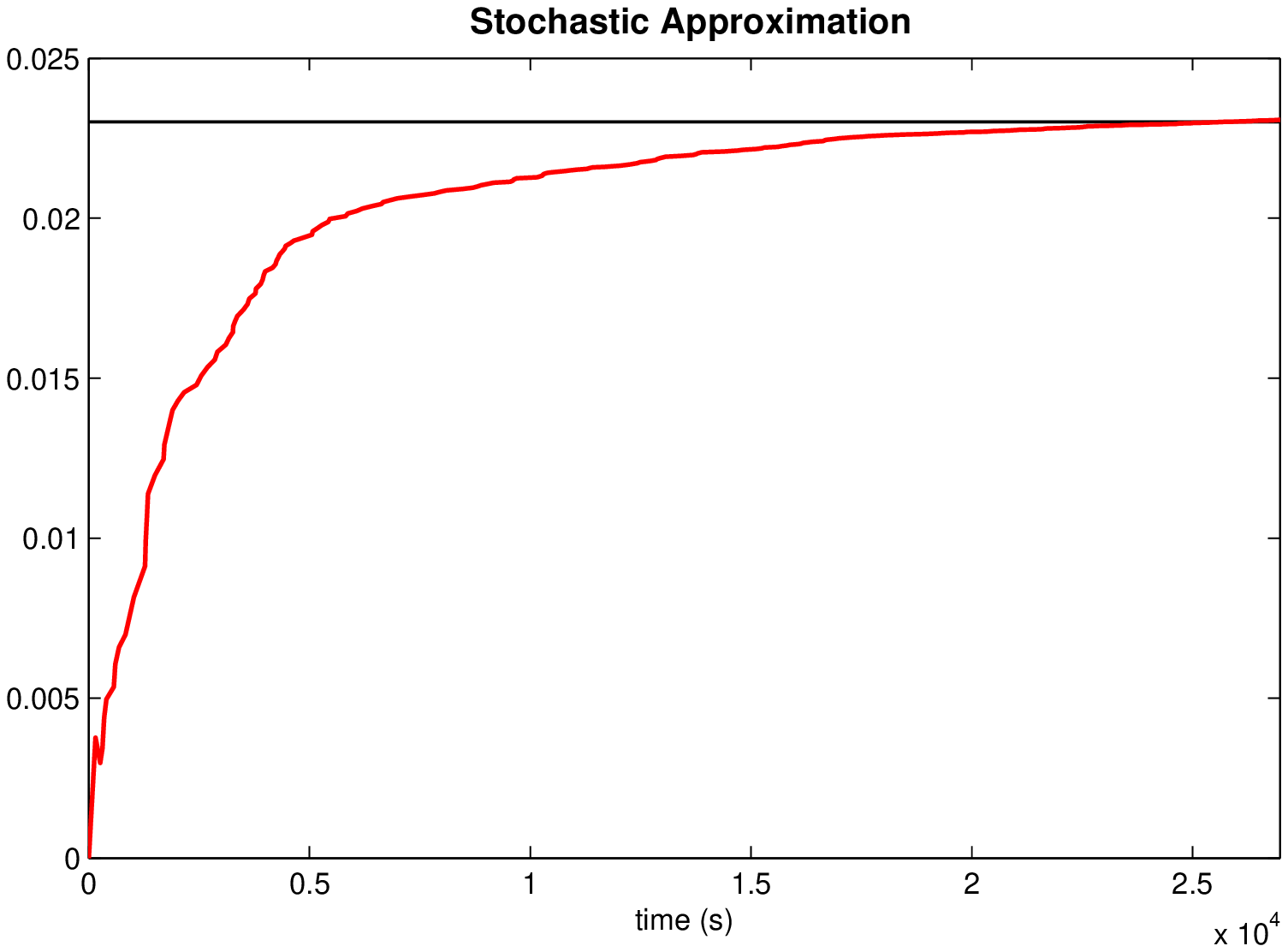}
&
\includegraphics[width=8cm]{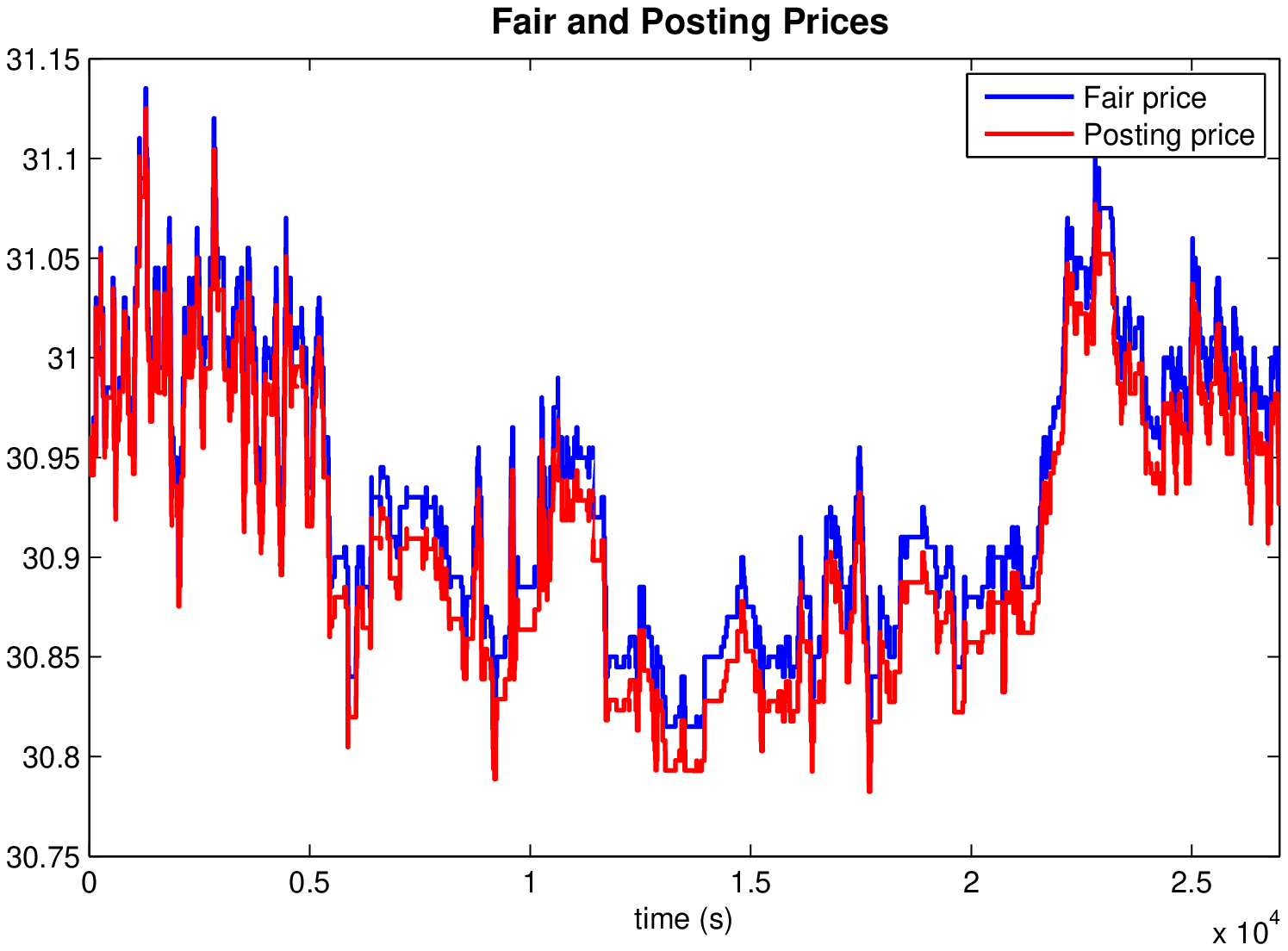}
\end{tabular}
\vskip-0.52cm
\caption{$\eta\not\equiv0$: $A=1/50$, $k=50$, $Q=100$, $\kappa=1$, $A'=0.001$, $k'=0.0005$ and $N_{{\rm cycles}}=220$. Ruppert and Poliak's averaging algorithm  with $\gamma_n=\frac{1}{550n^{0.95}}$.}
\label{AlgoStoPenRealDataRP}
\end{figure}

\clearpage

\noindent\underline{{\bf{\em Settin}}}{\bf{\em g}}\underline{ {\bf{\em  2}}} ($\eta\equiv0$)
\begin{figure}[!ht]
\vskip-0.3cm
\centering
\begin{tabular}{cc}
\includegraphics[width=8cm]{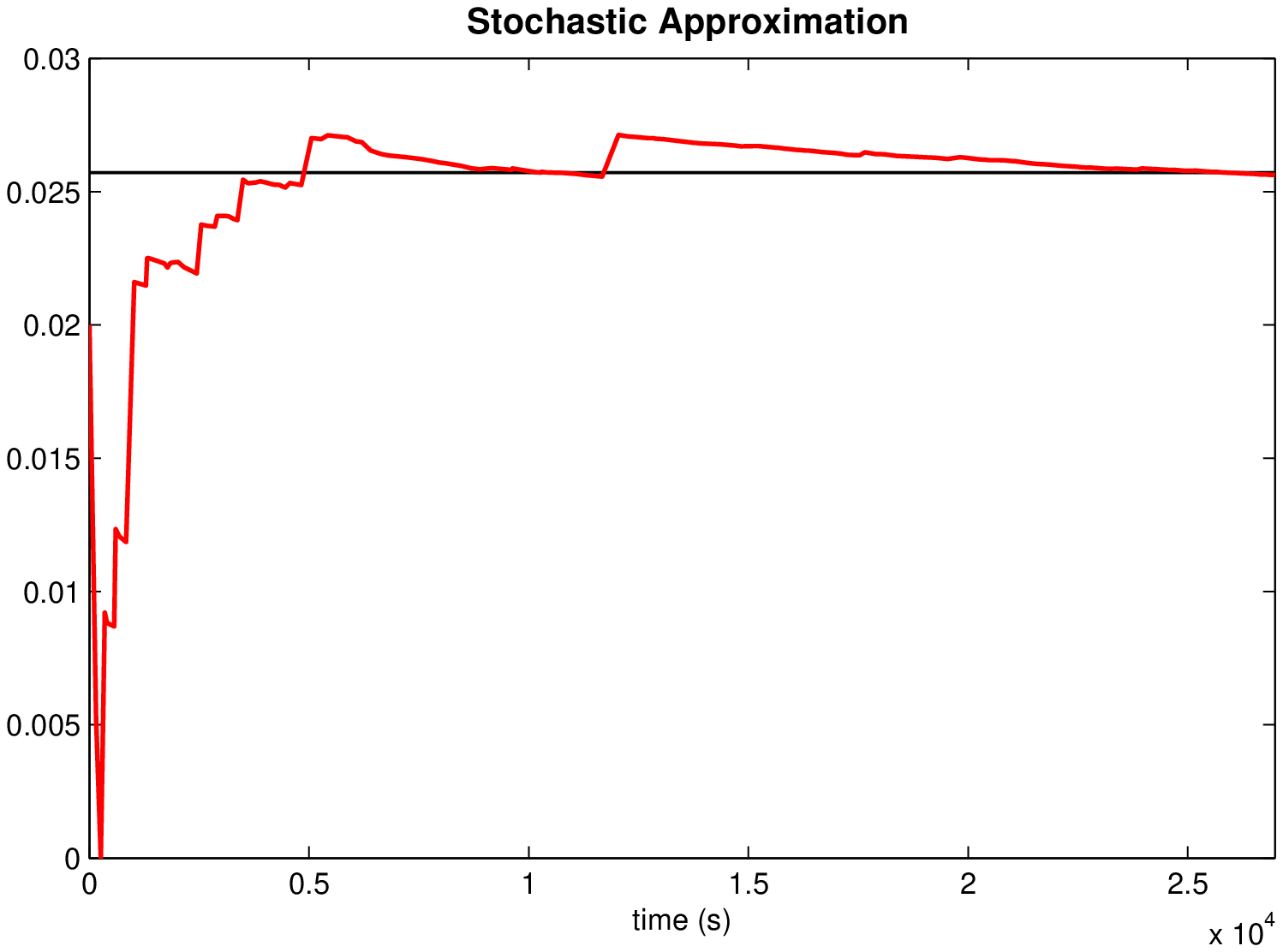} &
\includegraphics[width=8cm]{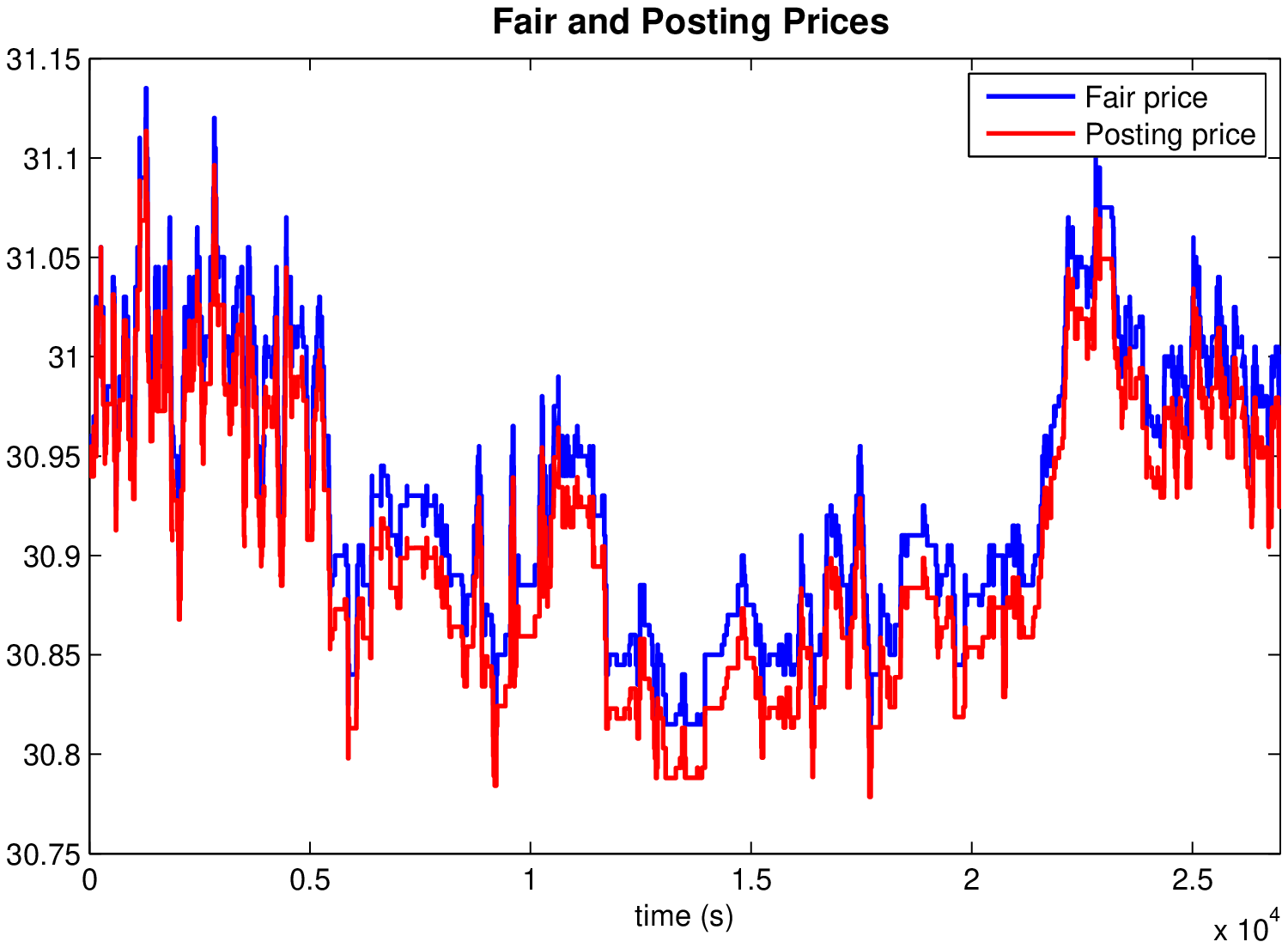}
\end{tabular}
\vskip-0.5cm
\caption{$\eta\equiv0$: $A=1/50$, $k=50$, $Q=100$, $\kappa=1.001$, $\gamma_n=\frac{1}{450n}$ and $N_{{\rm cycles}}=220$.}
\label{AlgoStoIdRealData}
\end{figure}

\begin{figure}[!ht]
\vskip-0.4cm
\centering
\begin{tabular}{cc}
\includegraphics[width=8cm]{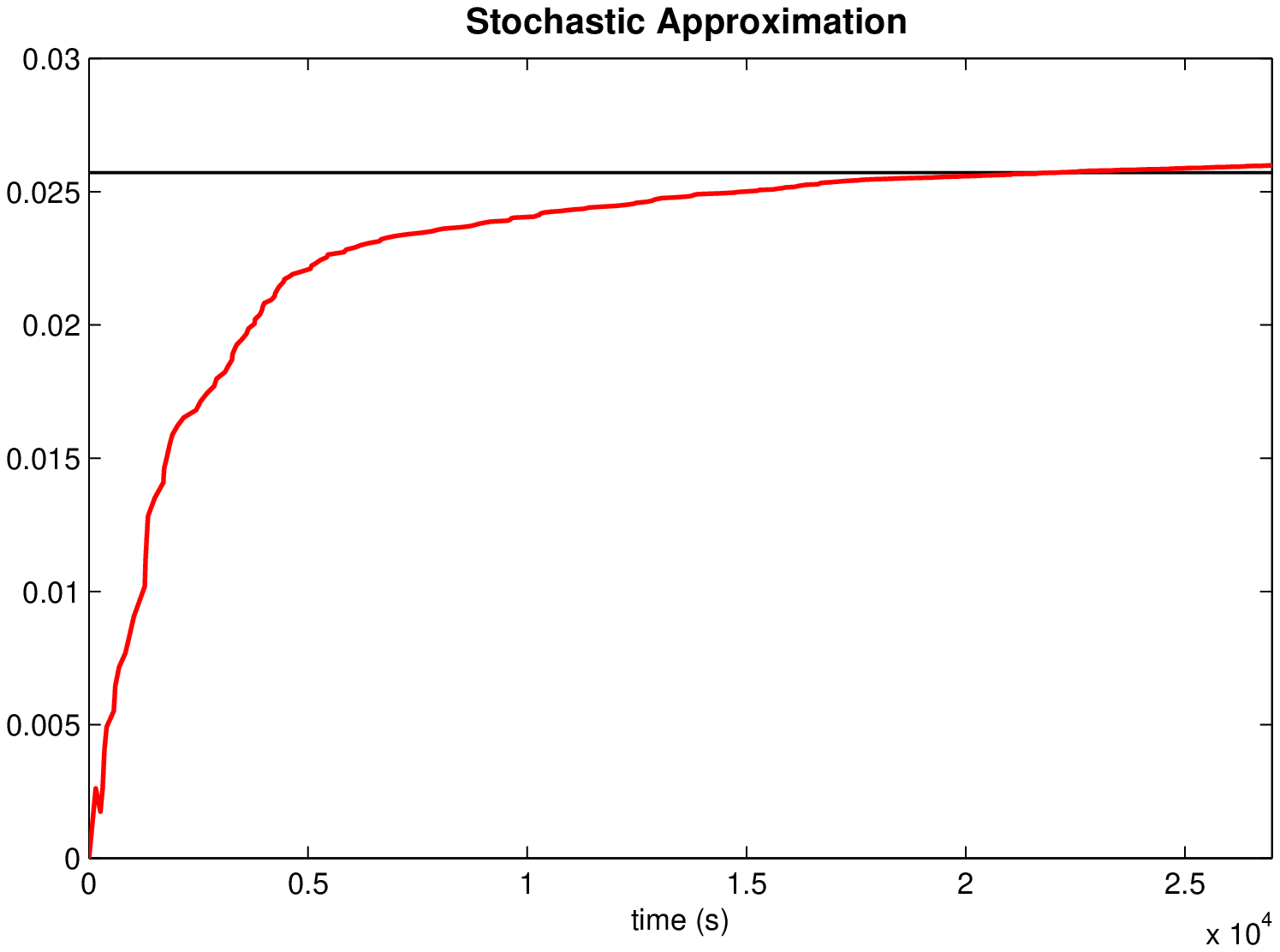}
&
\includegraphics[width=8cm]{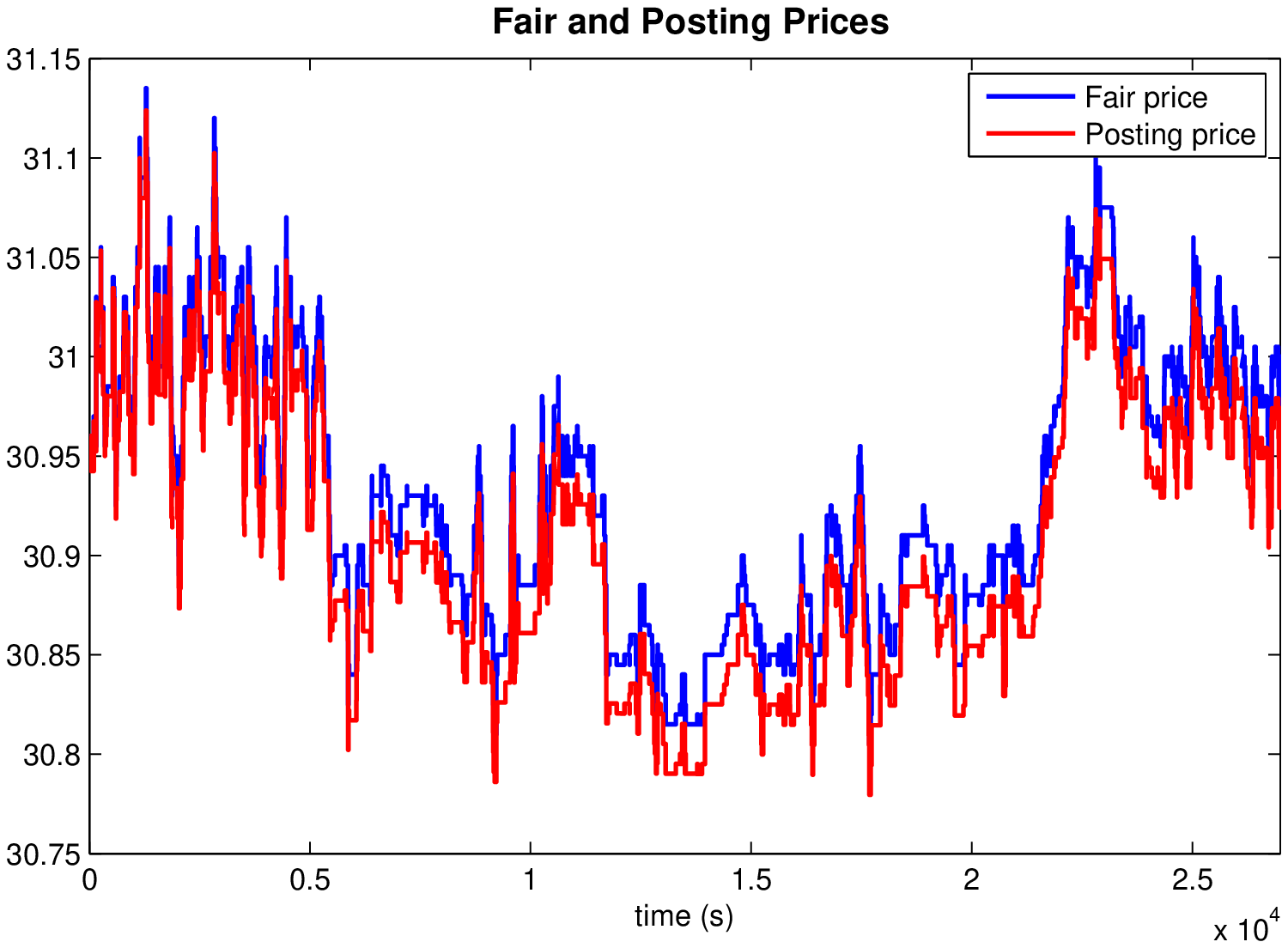}
\end{tabular}
\vskip-0.52cm
\caption{$\eta\not\equiv0$: $A=1/50$, $k=50$, $Q=100$, $\kappa=1.001$ and $N_{{\rm cycles}}=220$. Ruppert and Poliak's averaging algorithm  with $\gamma_n=\frac{1}{450n^{0.95}}$.}
\label{AlgoStoIdRealDataRP}
\end{figure}

We see on Figures \ref{AlgoStoPenRealDataRP} (for $\eta\not\equiv0$) and \ref{AlgoStoIdRealDataRP} (for $\eta\equiv0$) that the recursive procedures converge toward their respective targets, namely the minimum of the execution cost functions presented in Figures \ref{CoutsPenRealData} left (for $\eta\not\equiv0$) and \ref{CoutsRealData} left (for $\eta\equiv0$).

\subsection{Comparisons and comment on the efficiency of  the reassessment rules}

Practically, the stochastic algorithm defined by the dynamics of equation (\ref{ASDist}) has to be read as a reassessment rule to apply to the distance to the reference price (which can be taken as the best opposite) at a given frequency (that can be expressed in calendar time or business time, $i.e.$ number of trades or traded quantities): each cycle (of given 5 seconds), the trading algorithm updates $\delta_n$ and modifies the price of its limit order to be the best opposite price minus $\delta_n$ (for a buy order).

The change from $\delta_{n-1}$ to $\delta_n$ is computed using the expression (\ref{Cprime1}) for $H$, and a typical choice for the step is $\gamma_n=\gamma_0/n^\rho$ ($0<\rho<1$).

Comparing the results obtained on simulated data (subsection \ref{sec:simd}) to the ones obtained on real data (subsection \ref{sec:marketd}), the first point to notice is that the convergence is ten times slower on real data than on simulated ones (less than 10 minutes on real data $vs$ less than one minute on simulated ones).

The second element is that once the algorithm succeeded in being close to the optimal value $\delta^*$, it oscillates around it without any need to wait again from one to ten minutes to converge. It means that using a heuristic rule to choose $\delta_0$ rather than a random value can practically avoid ``paying'' this first step.

Another important element to highlight is that the posting price dynamics on simulated data are really different than the ones on real data. On simulated data, the price evolution diagrams are unrealistic for any trader. It mainly comes from the fact that the price does not behave like a Brownian motion at a very short time scale. On real data, the posting price adapts to it, being also far more realistic by following the last prices more closely than in a simulated environment.

It shows the adaptiveness property of a stochastic algorithm, that may need more time to catch a reasonable value for $\delta_n$, but then keep very close to it, despite the price moves that are less smooth than a classical diffusion.
\\

Going back to the expression of $H$, it must be qualitatively said that:
\begin{itemize}
	\item the first term $-Q_{T}\P^{(\delta)}\left(N^{\mu}>Q_{T}\right)$ will push the price away when the order has been completely filled before the end of the reload period $T$;
	\item the second term $\left(\frac{\partial}{\partial\delta}\Lambda_T(\delta,S)(S_0-\delta)-\Lambda_T(\delta,S)\right)\P^{(\delta)}\left(N^{\mu}\leq Q_{T}-1\right)$ will attract the price to the best opposite as far as the price move between $S_{t}$ and $S_{t+T}$ does not change the intensity of the ``order flow'' $\Lambda(\delta_n,S_{t+T})$ it will be exposed to;
	\item the last component of the reassessment policy $-\kappa S_{T}\frac{\partial}{\partial\delta}\Lambda_T(\delta,S)\varphi^{(\delta)}(\mu)$ attracts the price to the best opposite when the ``price impact cost'' ($i.e.$ coming into the equation via $\kappa\varphi^{(\delta)}$, which can bee seen as the first derivative of the market impact $\kappa\Phi$) to pay when the expected fill rate has not been obtain, is too high.
\end{itemize}
\medskip

This qualitative interpretation of the update rule obtained by rigorous derivation of criteria (\ref{coutDist}) shows how to estimate and optimally mix these three natural effects  that a trader would like to see into any price reassessment policy. It shows the value of a stochastic algorithm approach can bring to optimal trading at every time scale: (1) clearly links an objective criteria to the reassessment policies, (2) exposing the assumptions needed to guarantee its convergence.
It is clearly the risk-control role that one can expect from a formalized approach of trading.

\section{Conclusion}

This paper presents a rigorous proof of convergence of a reassessment scheme for a trading tactic aiming at capturing liquidity ``\emph{around the book}''. It implements a \emph{learning by trading} approach, validated inside our class of model (which  is quite general):
\begin{itemize}
	\item the distance to a reference point (in practice it can be the best opposite, the mid point, or any efficient price estimate) is fixed at $\delta_n$ during few seconds or market trades,
	\item the tactic observes the market feedback resulting from the combination of the natural diffusion of the reference price and a point process filling my order with an intensity  depending on my instantaneous distance to the reference price (which varies),
	\item our formal results give the optimal way to adjust the distance $\delta_{n+1}$ to the reference price, given then marginal variations of the different market components which can be anticipated to an increase or decrease of posting distance.
\end{itemize}

The robustness of the approach is not only guaranteed theoretically (provided that the market impact is in a realistic range, as   commented on section \ref{Seciid}), but also confirmed and emphasized  by tests carried out    on real data (partially reproduced in  Section \ref{sec:marketd}). This benchmark  shows that even if the real data behave differently from Monte-Carlo generated scenarios (see Section~\ref{sec:simd}), the convergence still occurs.

This paper strongly suggests that such iterative trading procedures, very often used  by practitioners because of the way it can be efficiently fitted {\em on line}  to real time data and  providing  optimal reassessment rules. This has to be compared to a stochastic control approach which needs to be calibrated on  using data on longer time frames inducing  somehow an ``averaging'' effect of the instantaneous liquidity effects that can occur on the real markets.

With the recent modifications of market microstructure following fragmentation of markets and emergence of high frequency trading, it is clear that algorithmic traders will need to devise more reactive and short term tactics. Covering this aspect of trading, this paper opens the door to other researches (like multi trading pools and multi asset reassessment of limit prices), and to applications for practitioners. Traders can use such scheme as sub-tactics of a brokerage algorithm, of an high frequency market making mechanism, or of any intraday arbitrage automated process.

\clearpage
\appendix

\begin{center}
\Large{{\bf Appendix}}
\end{center}

\section{Convergence theorem for constrained algorithms}
\label{AppADist}

The aim is to determine an element of the set $\{\theta\in\Theta\,:\,h(\theta)=\E\left[H(\theta,Y)\right]=0\}$ (zeros of $h$ in $\Theta$) where $\Theta\subset\R^d$ is a closed convex set, $h:\R^d\rightarrow\R^d$ and $H:\R^d\times\R^q\rightarrow\R^d$. For $\theta_0\in\Theta$, we consider the $\R^d$-valued sequence $(\theta_n)_{n\geq0}$ defined by
\begin{equation}\label{ASProj}
\theta_{n+1}={\rm Proj}_{\Theta}\left(\theta_n-\gamma_{n+1}H(\theta_n,Y_{n+1})\right),
\end{equation}
where $(Y_n)_{n\geq1}$ is an i.i.d. sequence with the same law as $Y$, $(\gamma_n)_{n\geq1}$ is a positive sequence of real numbers and ${\rm Proj}_{\Theta}$ denotes the Euclidean projection on $\Theta$. The recursive procedure (\ref{ASProj}) can be rewritten as follows
\begin{equation}\label{ASProj2}
\theta_{n+1}=\theta_n-\gamma_{n+1}h(\theta_n)-\gamma_{n+1}\Delta M_{n+1}+\gamma_{n+1}p_{n+1},
\end{equation}
where $\Delta M_{n+1}=H(\theta_n,Y_{n+1})-h(\theta_n)$ is a martingale increment and $$p_{n+1}=\frac{1}{\gamma_{n+1}}{\rm Proj}_{\Theta}\left(\theta_n-\gamma_{n+1}H(\theta_n,Y_{n+1})\right)-\frac{\theta_n}{\gamma_{n+1}}+H(\theta_n,Y_{n+1}).$$

\begin{theo}\label{ASContraint}(see~\cite{KusCla} and~\cite{KusYin}) Let $(\theta_n)_{n\geq0}$ be the sequence defined by (\ref{ASProj2}). Assume that there exists a unique $\theta^*\in\Theta$ such that $h(\theta^*)=0$ and that the mean function satisfies on $\Theta$ the following mean-reverting property, namely
\begin{equation}\label{mr}
\forall\theta\neq\theta^*\in\Theta,\quad\left\langle h(\theta)\left.\right|\theta-\theta^*\right\rangle>0.
\end{equation}
Assume that the gain parameter sequence $(\gamma_n)_{n\geq1}$ satisfies
\begin{equation}\label{gamma}
\sum_{n\geq1}\gamma_n=+\infty \quad\mbox{and}\quad\sum_{n\geq1}\gamma^2_n<+\infty.
\end{equation}
If the function $H$ satisfies
\begin{equation}\label{HypH}
\exists\, K>0 \; \mbox{ such that }\;\forall\theta\in\Theta, \quad \E\left[\left|H(\theta,Y)\right|^2\right]\leq K(1+\left|\theta\right|^2),\end{equation}
then
$$
\theta_n\overset{a.s.}{\underset{n\rightarrow+\infty}{\longrightarrow}}\theta^*.
$$
\end{theo}

\smallskip \noindent {\bf Remark.} If $\Theta$ is bounded (\ref{HypH}) reads $
\sup_{\theta\in\Theta}\E\left[\left|H(\theta,Y)\right|^2\right]<+\infty$, 
which  is always satisfied if $\Theta$ is compact and $\theta\mapsto\E\left[\left|H(\theta,Y)\right|^2\right]$ is  continuous.

\clearpage
\section{Functional co-monotony principle for a class of one-dimensional diffusions}
\label{DistQuatre}

In this section, we present the principle of co-monotony, first for random vectors taking values in a nonempty interval $I$, then for one-dimensional diffusions lying in  $I$.

\subsection{Case of random variables and random vectors}

First we recall a classical result for random variables.
 
\begin{prop}\label{Comonotony1d} Let $f,g:I\subset\R\rightarrow\R$ be two monotonic functions with same monotony. Let $X:(\Omega,{\cal A},\P)\rightarrow I$ be a real valued random variable such that $f(X),g(X)\in L^2(\P)$. Then
$$
\Cov(f(X),g(X))\geq 0.
$$
\end{prop}

\noindent{\bf Proof.} Let $X,\,Y$ be two independent random variables defined on the same probability space with the same distribution $\P_X$. Then
$$
(f(X)-f(Y))(g(X)-g(Y))\geq0
$$
hence its expectation is non-negative too. Consequently
$$
\E\left[f(X)g(X)\right]-\E\left[f(X)g(Y)\right]-\E\left[f(Y)g(X)\right]+\E\left[f(Y)g(Y)\right]\geq0
$$
so, using that $Y\overset{(d)}{=}X$ and $Y$, $X$ are independent, yields
$$
2\E\left[f(X)g(X)\right]\geq\E\left[f(X)\right]\E\left[g(Y)\right]+\E\left[f(Y)\right]\E\left[g(X)\right]=2\E\left[f(X)\right]\E\left[g(X)\right]
$$
that is $\Cov(f(X),g(X))\geq 0$.\hfill$\cqfd$

\begin{prop}\label{ComonotonyMultid} Let $F,G:\R^d\rightarrow\R$ be two monotonic functions with same monotony in each of their variables, $i.e.$ for every $i\in\{1,\ldots,d\}$, $x_i\longmapsto F(x_1,\ldots,x_i,\ldots,x_d)$ and $x_i\longmapsto G(x_1,\ldots,x_i,\ldots,x_d)$ are monotonic with the same monotony which may depend on $i$ (but does not depend on $(x_1,\ldots,x_{i-1},$ $x_{i+1},\ldots,x_d)\in\R^{d-1}$). Let $X_1,\ldots,X_d$ be independent real valued random variables defined on a probability space $(\Omega,{\cal A},\P)$ such that $F(X_1,\ldots,X_d),G(X_1,\ldots,X_d)\in L^2(\P)$. Then
$$\Cov\left(F(X_1,\ldots,X_d),G(X_1,\ldots,X_d)\right)\geq0.$$
\end{prop}

\noindent{\bf Proof.} The proof of the above proposition is made by induction on $d$. The case $d=1$ is given by Proposition \ref{Comonotony1d}. We give here the proof for $d=2$ for notational convenience, but the general case of dimension $d$ follows likewise. By the monotonic assumption on $F$ and $G$, we have for every $x_2\in\R$, if $X'_1\overset{d}{=}X_1$ with $X'_1$, $X_1$ independent, that
$$\left(F(X_1,x_2)-F(X'_1,x_2)\right)\left(G(X_1,x_2)-G(X'_1,x_2)\right)\geq0.$$
This implies that (see Proposition \ref{Comonotony1d})
$$\Cov\left(F(X_1,x_2)G(X_1,x_2)\right)\geq0.$$
If $X_1$ and $X_2$ are independent, using Fubini's Theorem and what precedes, we have
\begin{eqnarray*}
\E\left[F(X_1,X_2)G(X_1,X_2)\right]&=&\int_{\R}\P_{X_2}(dx_2)\E\left[F(X_1,x_2)G(X_1,x_2)\right] \\
   &\geq&\int_{\R}\P_{X_2}(dx_2)\E\left[F(X_1,x_2)\right]\E\left[G(X_1,x_2)\right].
\end{eqnarray*}   
By setting $\varphi(x_2)=\E\left[F(X_1,x_2)\right]$ and $\psi(x_2)=\E\left[G(X_1,x_2)\right]$ and using the monotonic assumptions on $F$ and $G$, we have that $\varphi$ and $\psi$ are monotonic with the same monotony so that
$$\int_{\R}\P_{X_2}(dx_2)\E\left[F(X_1,x_2)\right]\E\left[G(X_1,x_2)\right]=\E\left[\varphi(X_2)\psi(X_2)\right]\geq\E\left[\varphi(X_2)\right]\E\left[\psi(X_2)\right].$$
Combining these above two inequalities finally yields $\Cov\left(F(X_1,X_2)G(X_1,X_2)\right)\geq0$. \hfill$\cqfd$

\subsection{Case of (one-dimensional) diffusions}

This framework corresponds to the infinite dimensional case and we can not apply straightforwardly the result of Proposition \ref{Comonotony1d}: indeed, if we define the following natural order relation on $\D([0,T],\R)$
$$
\forall\alpha_1,\alpha_2\in\D([0,T],\R), \quad \alpha_1\leq\alpha_2\Longleftrightarrow\left(\forall t\in[0,T], \ \alpha_1(t)\leq\alpha_2(t)\right),
$$
this order is partial which makes the formal proof of Proposition~\ref{Comonotony1d} collapse. To establish a co-monotony principle for diffusions, we proceed in two steps: first, we use the Lamperti transform to ``force'' the diffusion coefficient to be equal to 1 and we establish the co-monotony principle for this class of diffusions. Then by the inverse Lamperti transform, we go back to the original process. 

In this section, we first present our framework in more details. Then we recall some weak convergence results for diffusion with diffusion coefficient equal to 1. Afterwards we present the Lamperti transform and we conclude by the general co-monotony principle. 

\medskip
Let $I$ be a nonempty open interval of $\R$. One considers a real-valued Brownian diffusion process
\begin{equation}\label{eds}
dX_t=b(t,X_t)dt+\sigma(t,X_t)dW_t, \quad X_0=x_0\in I,\quad t\in[0, T],
\end{equation}
where $b,\,\sigma : [0,T]\times I\rightarrow\R$ are Borel functions with at most linear growth such that the above Equation (\ref{eds}) admits at least one (weak) solution over $[0,T]$ and $W$ is a Brownian motion defined on a probability space $(\Omega, {\cal A}, \P)$. We assume that  the diffusion $X$ $a.s.$ does not explode and lives in the interval $I$. This implies assumptions on the function $b$ and $\sigma$ especially in the neighborhood (in $I$) of the endpoints of $I$ that we will not detail here. At a finite endpoint of $I$, these assumptions are strongly connected with the Feller classification for which we refer to~\cite{KarTay} (with $\sigma(t,\cdot)>0$ for every $t\in[0,T]$). We will simply make the classical linear growth assumption on $b$ and $\sigma$ (which prevents explosion at a finite time) that will be used for different purpose in what follows.

To ``remove'' the diffusion coefficient of the diffusion $X$, we will introduce the so-called {\em Lamperti transform} which requires additional assumptions on the drift $b$ and the diffusion coefficient~$\sigma$, namely
\begin{equation}\label{bsigma}
({\cal A}_{b,\sigma})\equiv\left\{\begin{array}{rl}
(i) & \sigma\in{\cal C}^1([0,T]\times I,\R), \\
\\
(ii)& \forall (t,x)\in[0,T]\times I, \quad \left|b(t,x)\right|\leq C(1+\left|x\right|) \quad \mbox{and} \quad 0<\sigma(t,x)\leq C(1+\left|x\right|),\\
\\
(iii)& \forall\, x \in I, \quad \int_{(-\infty,x]\cap I}\frac{d\xi}{\sigma(t,\xi)}=\int_{[x,+\infty,)\cap I}\frac{d\xi}{\sigma(t,\xi)}=+\infty
\end{array}\right.
\end{equation}

\bigskip \noindent {\bf Remark.}  Condition~$(iii)$ clearly does not depend on $x\in I$. Furthermore, if $I=\R$, $(iii)$ follows from $(ii)$ since $\frac{1}{\sigma(t,\xi)} \ge \frac 1C \frac{1}{1+|\xi|}$. \\

Before passing to a short background on  the Lamperti transform which will lead to the new diffusion deduced from (\ref{eds}) whose diffusion coefficient is equal to $1$, we need to recall (and adapt) some background on solution and discretization of such $SDE$.

\subsubsection{Background on diffusions with \texorpdfstring{$\sigma\equiv1$}{sigma=1} (weak solution, discretization).}

The following proposition gives a condition on the drift for the existence and the uniqueness of a weak solution of a SDE when $\sigma\equiv1$ (see~\cite{KarShr} Proposition 3.6, Chap. 5, p. 303 and Corollary 3.11, Chap. 5, p. 305).
\begin{prop}\label{weaksolution} Consider the stochastic differential equation
\begin{equation}\label{edsweak}
dY_t=\beta(t,Y_t)dt + dW_t, \quad t\in[0, T],
\end{equation}
where $T$ is a fixed positive number, $W$ is a one-dimensional Brownian motion and $\beta:[0,T]\times\R\to\R$ is a Borel-measurable function satisfying
$$
\left|\beta(t,y)\right|\leq K(1+\left|y\right|), \quad t\in[0, T],\quad y\in\R,\quad K>0.
$$
For any probability measure $\nu$ on $(\R,{\cal B}(\R))$, equation (\ref{edsweak}) has a weak solution with initial distribution $\nu$.

If, furthermore, the drift term $\beta$ satisfies one of the following conditions:
\begin{enumerate}
	\item[$(i)$] $\beta$ is bounded on $[0,T]\times\R$,
	\item[$(ii)$] $\beta$ is continuous, locally Lipschitz in $y\in \R$ uniformly in $t\in[0,T]$, 
\end{enumerate}
then this weak solution is unique (in fact $(ii)$ is a strong uniqueness assumption).
\end{prop}

Now we introduce the stepwise constant (Brownian) Euler scheme $\bar{Y}^m=\left(\bar{Y}_{\frac{kT}{m}}\right)_{0\leq k\leq m}$ with step $\frac{T}{m}$ of the process $Y=(Y_t)_{t\in[0,T]}$ defined by (\ref{edsweak}). It is defined by
\begin{equation}\label{Euler1}
\bar{Y}_{t^m_{k+1}}=\bar{Y}_{t^m_k}+\beta(t^m_k,\bar{Y}_{t^m_k})\frac{T}{m}+\sqrt{\frac{T}{m}}U_{k+1}, \quad \bar{Y}_0=Y_0=y_0, \quad k=0,\ldots,m-1,
\end{equation}
where $t^m_k=\frac{kT}{m}$, $k=0,\ldots,m$, and $(U_k)_{0\leq k\leq m}$ denotes a sequence of i.i.d. ${\cal N}(0,1)$-distributed random variables given by
$$
U_k=\sqrt{\frac{m}{T}}\left(W_{t^m_k}-W_{t^m_{k-1}}\right), \quad \quad k=1,\ldots,m.
$$
The following theorem gives a weak convergence result for the stepwise constant Euler scheme (\ref{Euler1}). Its proof is a straightforward consequence of the functional limit theorems for semi-martingales (to be precise Theorem 3.39, Chap. IX, p. 551 in~\cite{JacShi}).
\begin{theo}\label{weak} Let $\beta:[0,T]\times\R\rightarrow\R$ be a continuous function satisfying
$$
\exists\, K>0,\quad \left|\beta(t,y)\right|\leq K(1+\left|y\right|), \quad t\in[0, T],\quad y\in\R.
$$
Assume that the weak solution of equation (\ref{edsweak}) is unique. Then, the stepwise constant Euler scheme of (\ref{edsweak}) with step $\frac{T}{m}$ satisfies
$$
\bar{Y}^m\stackrel{{\cal L}}{\longrightarrow}Y \quad\mbox{for the Skorokhod topology as }m\to \infty.
$$
In particular, for every functional $F:\D([0,T],\R)\rightarrow\R$ $\P_Y$-$a.s.$ continuous at $\alpha\in{\cal C}([0,T],\R)$, with polynomial growth, we have
$$
\E F(\bar{Y}^m)\underset{m\rightarrow\infty}{\longrightarrow}\E F(Y)
$$
(by uniform integrability since $\sup_{t\in[0,T]}\left|\bar{Y}^m_t\right|\in\bigcap_{p>0}L^p$).
\end{theo}

\subsubsection{Background on the Lamperti transform}

We will introduce a new diffusion $Y_t:=L(t,X_t)$ which will satisfy a new SDE whose diffusion coefficient will be constant equal to $1$. This function $L$ defined on $[0,T]\times I$ is known in the literature as the Lamperti transform. It is defined for every $(t,x)\in[0,T]\times I$ by
\begin{equation}\label{lamperti}
L(t,x):=\int_{x_1}^x\frac{d\xi}{\sigma(t,\xi)}
\end{equation}
where $x_1$ is an arbitrary fixed value lying in $I$. The Lamperti transform clearly depends on the choice of $x_1$ in $I$  but  not its properties of interest. First, under $({\cal A}_{b,\sigma})$-$(i)$-$(ii)$, $L\in {\cal C}^{1,2}([0,T]\times I)$ with
$$
\frac{\partial L}{\partial t}(t,x)=-\int_{x_1}^x\frac{1}{\sigma^2(t,\xi)}\frac{\partial \sigma}{\partial t}(t,\xi)d\xi, \quad
\frac{\partial L}{\partial x}(t,x)=\frac{1}{\sigma(t,x)}>0 \quad \mbox{and} \quad \frac{\partial^2 L}{\partial x^2}(t,x)=-\frac{1}{\sigma^2(t,x)}\frac{\partial \sigma}{\partial x}(t,x).
$$

Let  $t\in[0,T]$, $L(t,\cdot)$ is an increasing ${\cal C}^{2}$-diffeomorphism from $I$ onto $\R = L(t,I)$ (the last claim follows from $({\cal A}_{b,\sigma})$-$(iii)$). Its inverse will be  denoted $L^{-1}(t,\cdot)$. 

Notice that, $(t,y)\mapsto L^{-1}(t,y)$ is continuous on $[0,T]\times I$ since both sets
$$
\left\{(t,y)\in[0,T]\times I\,:\, L^{-1}(t,y)\leq c\right\} =\{(t,y)\in [0,T]\times\R\,:\, L(t,c)\ge y\}
$$
and 
$$
 \left\{(t,y)\in[0,T]\times I\,:\, L^{-1}(t,y)\geq c\right\}=\{(t,y)\in [0,T]\times\R\,:\, L(t,c)\le y\}
$$
are both closed for every $c\in\R$. Therefore, if $({\cal A}_{b,\sigma})$ holds, the function $\beta:[0,T]\times I\mapsto\R$ defined by
\begin{equation}\label{beta}
\beta(t,y):=\left(\frac{b}{\sigma}-\int_{x_1}^{\cdot}\frac{1}{\sigma^2(t,\xi)}\frac{\partial \sigma}{\partial t}(t,\xi)d\xi-\frac{1}{2}\frac{\partial \sigma}{\partial x}\right)(t,L^{-1}(t,y))
\end{equation}
is a Borel function, continuous as soon as $b$ is.
$$
\mbox{Now, we set }\hskip4cm\forall\, t\in [0,T],\quad  Y_t:=L(t,X_t).\hskip6.5cm
$$
It\^o formula straightforwardly yields
\begin{equation}\label{edsY}
dY_t=\beta(t,Y_t)dt+dW_t, \quad Y_0=L(0,x_0)=:y_0\in \R.
\end{equation}

\smallskip \noindent {\bf Remarks.} $\bullet$ In the homogeneous case, which is the most important case for our applications, 
\begin{equation}\label{edsH}
dX_t=b(X_t)dt+\sigma(X_t)dW_t, \quad X_0=x_0\in\R,\quad t\in[0, T],
\end{equation}
we have
$$
L(t,x)=L(x):=\int_{x_1}^x\frac{d\xi}{\sigma(\xi)}.
$$
Then by setting $Y_t:=L(X_t)$, we obtain
$$dY_t=\beta(Y_t)dt+dW_t, \quad Y_0=L(x_0)=:y_0\quad\mbox{with}\quad\beta:=\Big(\frac{b}{\sigma}-\frac{\sigma'}{2}\Big)\circ L^{-1}.$$
Note that $\beta$ is bounded as soon as $\frac{b}{\sigma}-\frac{\sigma'}{2}$ is.

\medskip
\noindent $\bullet$ If  the partial derivative $b'_x$ exists on $[0,T]\times I$, one easily checks, using $(L^{-1})'_y(t,y)= \sigma(t,L^{-1}(t,y))$, that for every $(t,y)\in [0,T]\times I$, 
\begin{equation}\label{eq:betaprime}
\beta'_y(t,y)= \Big(b'_x-\frac{b\sigma'_x+\sigma'_t}{\sigma}-\frac{\sigma\sigma''_{x^2}}{2}\Big)(t,L^{-1}(t,y)).
\end{equation}
As a consequence, as soon as the function
\begin{equation}\label{eq:betaprime2}
b'_x-\frac{b\sigma'_x+\sigma'_t}{\sigma}-\frac{\sigma\sigma''_{x^2}}{2}
\left\{\begin{array}{l}
\mbox{ is bounded on $[0,T]\times I$, then $\beta$ satisfies the linear growth} \\
\mbox{ Lipschitz assumption}\\
\geq0,\mbox{ then $\beta$ is non-decreasing.}
\end{array}\right.
\end{equation}

\begin{defi}  The {\em functional Lamperti transform}, denoted $\Lambda$, is a functional from ${\cal C}([0,T],I)$ to ${\cal C}([0,T],\R)$ defined by
\[
\forall\, \alpha\in {\cal C}([0,T],I),\quad \Lambda(\alpha) = L(\cdot,\alpha(\cdot)).
\]
\end{defi}

\begin{prop}\label{prop: LamperCont} If the diffusion coefficient $\sigma$ satisfies $({\cal A}_{b,\sigma})$, the functional Lamperti transform is an homeomorphism from ${\cal C}([0,T],I)$ onto ${\cal C}([0,T],\R)$.
\end{prop}

\medskip \noindent {\bf Proof.} Let $\alpha\in {\cal C}([0,T],I)$. Since $\sigma$ is bounded away from $0$ on the compact set $[0,T]\times\alpha([0,T])$, standard arguments based on Lebesgue domination theorem, imply that $\Lambda(\alpha) \in {\cal C}([0,T],\R)$. 

\noindent Conversely, as $L(t,\cdot): I\to \R$ is an homeomorphism for every $t\in [0,T]$, $\Lambda$ admits an inverse defined by
\[
\forall\, \xi\in {\cal C}([0,T],\R),\quad \Lambda^{-1}(\xi) := \big (t\mapsto L^{-1}(t,\xi(t))\big) \in {\cal C}([0,T],I).
\]
Let $U_K$ denote the topology of the convergence on compact sets of $I$ on ${\cal C}([0,T],I)$. \\

\noindent $\rhd$ {\em $U_K$-Continuity of $\Lambda$ on $[0,T]\times I$:} If $\alpha_n\stackrel{U_K}{\longrightarrow}\alpha_{\infty}$, the set $K= [0,T]\times\bigcup_{n\in \overline{\N}}\alpha_n([0,T])$ is a compact set included in $I$. Hence $\sigma$ is bounded away from $0$ on $K$ so that
\[
\forall\, t\in [0,T],\quad |L(t,\alpha_n(t))-L(t,\alpha_{\infty}(t))|\le \frac{1}{\inf_K \sigma} |\alpha_n(t)-\alpha_{\infty}(t)|
\] 
\[
i.e.\hskip4cm\|\Lambda(\alpha_n)-\Lambda(\alpha_{\infty})\|_{\infty}\le \frac{1}{\inf_K \sigma} \|\alpha_n-\alpha_{\infty}\|_{\infty}.\hskip5cm
\]

\noindent $\rhd$ {\em $U_K$-Continuity of $\Lambda^{-1}$ on $[0,T]\times I$:}  by using $({\cal A}_{b,\sigma})$-$(ii)$, we have for a fixed $t\in[0,T]$,
$$
\forall x,x'\in I, \quad\left|L(t,x)-L(t,x')\right|\geq\frac{1}{C}\int_{x\wedge x'}^{x\vee x'}\frac{d\xi}{1+|\xi|}=\frac{1}{C}\left|\Phi(x)-\Phi(x')\right|,
$$
where $\Phi(z)=\mbox{sign}(z)\log(1+|z|)$. Thus, 
$$
\forall y,y'\in \R, \quad\left|\Phi(L^{-1}(t,y))-\Phi(L^{-1}(t,y'))\right|\leq C\left|y-y'\right|.
$$
Let $(\xi_n)_{n\geq1}$ be a sequence of functions of $\D([0,T],\R)$ such that $\xi_n\overset{U}{\underset{n\to+\infty}{\longrightarrow}}\xi\in{\cal C}([0,T],\R)$. Then, for every $t\in[0,T]$ and $n\geq1$,
$$
\left|\Phi(L^{-1}(t,\xi_n(t)))-\Phi(L^{-1}(t,0))\right|\leq C\left|\xi_n(t)\right|\leq C\left(\left\|\xi_n(t)-\xi\right\|+\|\xi\|\right)+\left|\Phi(x_0)\right|\leq C',
$$
since $L^{-1}(t,0)=x_0$. Consequently, for every $t\in[0,T]$ and every $n\geq1$, $L^{-1}(t,\xi_n(t))\in K':=\Phi^{-1}([-C',C'])$. The set $K'$ is compact (because the function $\Phi$ is continuous and proper ($\lim_{|z|\to\infty}\left|\Phi(z)\right|=+\infty$)). As $\inf_{K'} \Phi'>0$, we deduce that there exists $\eta_0>0$ such that
$$
\forall x,y\in I, \quad \left|\Phi(x)-\Phi(y)\right|>\eta_0|x-y|,
$$
$i.e.$
$$
\forall t\in[0,T], \ \forall u,v\in L(t,I), \quad\left|L^{-1}(t,u)-L^{-1}(t,v)\right|\leq C''\left|u-v\right|, \quad C''>0.
$$
Hence, one concludes that
\[
\hskip 5cm\|\Lambda^{-1}(\xi_n)-\Lambda^{-1}(\xi_{\infty})\|_{\infty}\le C'' \|\xi_n-\xi_{\infty})\|_{\infty}. \hskip 4cm \cqfd
\]

\subsubsection{Functional co-monotony principle for diffusion}

\begin{defi}\label{DefAdmissible}
The diffusion process (\ref{eds}) is {\rm admissible} if $({\cal A}_{b,\sigma})$ holds and
\begin{itemize}
	\item[(i)] for every starting value $x_0\in I$, (\ref{eds}) has a unique weak solution which lives in $I$ up to $t=+\infty$ (see~Proposition~\ref{weaksolution} for a criteria),
	\item[(ii)] the function $\beta$ defined by
$$
\beta(t,y):=\left(\frac{b}{\sigma}-\int_{x_1}^{\cdot}\frac{1}{\sigma^2(t,\xi)}\frac{\partial \sigma}{\partial t}(t,\xi)d\xi-\frac{1}{2}\frac{\partial \sigma}{\partial x}\right)(t,L^{-1}(t,y)),
$$
\noindent is continuous on $[0,T]\times\R$, non-decreasing in $y$ for every $t\in[0,T]$ or Lipschitz in $y$ uniformly in $t\in[0,T]$, and satisfies
$$
\exists\, K>0\;\mbox{ such that }\;\left|\beta(t,y)\right|\leq K(1+\left|y\right|), \;t\in[0, T],\; y\in\R.
$$
\end{itemize}
\end{defi}

\begin{defi}\label{DefFunctional} Let $F:\D([0,T],\R)\rightarrow\R$ be a functional. 
\begin{enumerate}
	\item[$(i)$] The functional $F$ is non-decreasing (resp. non-increasing) on $\D([0,T],\R)$ if
	$$
	\forall\alpha_1,\alpha_2\in\D([0,T],\R), \quad \left(\forall t\in[0,T], \ \alpha_1(t)\leq\alpha_2(t)\right)\Rightarrow F(\alpha_1)\leq F(\alpha_2) \ \mbox{(resp. $F(\alpha_1)\geq F(\alpha_2)$)}.
	$$
	\item[$(ii)$] The functional $F$ is continuous  at $\alpha\in{\cal C}([0,T],\R)$ if
	$$
	\forall\alpha_m\in\D([0,T],\R), \quad \alpha_m\overset{U}{\longrightarrow}\alpha\in{\cal C}([0,T],\R),\quad F(\alpha_m)\rightarrow F(\alpha).
	$$
	where $U$ denotes the uniform convergence of functions on $[0,T]$. The functional $F$ is $C$-continuous if it is continuous at every $\alpha\in{\cal C}([0,T],\R)$.
	\item[$(iii)$] The functional $F$ has polynomial growth if there exists a positive real number  $r>0$ such that 
	\begin{equation}\label{polygrowth}
	\forall\alpha\in\D([0,T],\R), \quad \left|F(\alpha)\right|\leq K\left(1+\left\|\alpha\right\|^r_{\infty}\right).
	\end{equation}
\end{enumerate}
\end{defi}

\noindent {\bf Remark.} Any $C$-continuous functional in the above sense is in particular $\P_Z$-$a.s.$ continuous for every process $Z$ with continuous paths.

\begin{defi}\label{DefMonotony} A process $(X_t)_{t\in[0,T]}$ with {\em continuous} (resp. {\em c\`adl\`ag stepwise constant}) paths defined on $(\Omega,{\cal A},\P)$ satisfies a functional co-monotony principle if for every $C$-continuous functionals (resp. measurable functionals on $\D([0,T],\R)$) $F,G$ monotonic with the same monotony satisfying (\ref{polygrowth}) such that $F(X)$, $G(X)$ and $F(X)G(X)\in L^1$, we have
$$
\Cov\left(F\big(\left(X_t\right)_{t\in[0,T]}\big),G\big(\left(X_t\right)_{t\in[0,T]}\big)\right)\geq 0.
$$
\end{defi}

The main result of this section is the following
\begin{theo}\label{thm:IneqCovDiff}
Assume that the real-valued diffusion process (\ref{eds}) is admissible (see Definition~\ref{DefAdmissible}).
Then it satisfies a co-monotony principle.
\end{theo}

\begin{cor}\label{cor:IneqCovDiff} 
Assume that the real-valued diffusion process (\ref{eds}) is admissible (see Defintion~\ref{DefAdmissible}).\\
\noindent$(a)$ Let $\left(\bar{X}_{t^m_k}\right)_{0\leq k\leq m}$ be its stepwise constant Euler scheme with step $\frac{T}{m}$ ($t^m_k=\frac{kT}{m}$, $0\leq k\leq m$). Then $\left(\bar{X}_{t^m_k}\right)_{0\leq k\leq m}$ satisfies a co-monotony principle.\\
\noindent$(b)$ Let $\left(\tilde{X}_{t_k}\right)_{0\leq k\leq m}$ be a sample of discrete time observations of $(X_t)_{t\in[0,T]}$ for a subdivision $(t_k)_{0\leq k\leq m}$ of $[0,T]$($0=t_0<\cdots<t_m=T$). Then $\left(\tilde{X}_{t_k}\right)_{0\leq k\leq m}$ satisfies a co-monotony principle.
\end{cor}

\noindent {\bf Remark.} The proof of Corollary~\ref{cor:IneqCovDiff} is contained in the proof of Theorem~\ref{thm:IneqCovDiff}. The only difference is that we do not need to transfer the co-monotony principle from the Euler scheme to the diffusion process.

\bigskip

Before passing to the proof of Theorem~\ref{thm:IneqCovDiff}, we need two lemmas: one  is a key step to transfer co-monotony from the Euler scheme to the diffusion process, the other aims at transferring uniqueness property for weak solutions. 
\begin{lem}\label{lem:B1} For every $\alpha\in\D([0,T],\R)$, set 
\begin{equation}\label{alphan}
\alpha^{(m)}=\sum_{k=0}^{m-1}\alpha(t_k^m)\mathds{1}_{[t_k^m,t_{k+1}^m)}+\alpha(T)\mathds{1}_{\{T\}}, \quad m\geq1,
\end{equation}
with $t^m_k:=\frac{kT}{m}$, $k=0,\ldots,m$. Then $\alpha^{(m)}\overset{U}{\longrightarrow}\alpha$ as $m\to\infty$. 

If $F:\D([0,T],\R)\rightarrow\R$ is $C$-continuous and non-decreasing (resp. non-increasing), then the unique function $F_m:\R^{m+1}\rightarrow\R$ satisfying $F(\alpha^{(m)})=F_m(\alpha(t^m_k),\,k=0,\ldots,m)$ is continuous and non-decreasing (resp. non-increasing) in each of its variables. Furthermore, if $F$ satisfies a polynomial growth assumption of the form
\[
\forall\,\alpha\in \D([0,T],\R),\quad |F(\alpha)|\le C(1+\|\alpha\|^r_{\infty})
\]
then, for every $m\ge 1$,
\[
|F_m(x_0,\ldots,x_m)|\le C(1+\max_{0\le k\le m}|x_k|^r)
\]
with the same real constant $C>0$.
\end{lem}

\begin{lem}\label{LamInj} Let $(S,d)$, $(T,\delta)$ be two Polish spaces and let $\Phi:S\mapsto T$ be a continuous injective function. Let $\mu$ and $\mu'$ be two probability measures on $(S,{\cal B}or(S))$. If $\mu\circ\Phi^{-1}=\mu'\circ\Phi^{-1}$, then $\mu=\mu'$.
\end{lem}

\medskip \noindent {\bf Proof of Lemma \ref{LamInj}.} For every Borel set $A$ of $S$, $\mu(A)=\sup\left\{\mu(K), \ K\subset A, \ K \mbox{ compact}\right\}$. Let $A\in{\cal B}or(S)$ such that $\mu(A)\neq\mu'(A)$. Then there exists a compact set $K$ of $A$ such that $\mu(K)\neq\mu'(K)$. But $\Phi(K)$ is a compact set of $S$ because $\Phi$ is continuous, so $\Phi^{-1}\left(\Phi(K)\right)$ is a Borel set of $S$ which contains $K$. As $\Phi$ is injective, $\Phi^{-1}\left(\Phi(K)\right)=K$. Therefore $\mu\left(\Phi^{-1}\left(\Phi(K)\right)\right)\neq\mu'\left(\Phi^{-1}\left(\Phi(K)\right)\right)$. We deduce that $\mu\circ\Phi^{-1}\neq\mu'\circ\Phi^{-1}$. \hfill$\cqfd$

\medskip \noindent {\bf Proof of Theorem~\ref{thm:IneqCovDiff}.} First we consider the Lamperti transform $(Y_t)_{t\geq0}$ (see (\ref{lamperti})) of the diffusion $X$  solution to~(\ref{edsweak}) with $X_0=x_0\in I$. Using the homeomorphism property of $\Lambda$ and calling upon the above Lemma~\ref{LamInj} with $\Lambda^{-1}$ and $\Lambda$, we see that existence and uniqueness assumptions on Equation~(\ref{edsweak}) can be transferred to~(\ref{edsY}) since $\Lambda$ is a one-to-one mapping between the solutions of these two SDE's. 

To fulfill condition $(ii)$ in Definition~\ref{DefAdmissible}, we need to introduce the smallest integer, denoted $m_{b,\sigma}$, such that $y\mapsto y+\frac{T}{m_{b,\sigma}}\beta(t,y)$ is non-decreasing in $y$ for every $t\in[0,T]$. Its existence follows from ${\cal A}_{b,\sigma}$-$(ii)$.Note that if $\beta$ is non-decreasing in $y$ for every $t\in[0,T]$, then $m_{b,\sigma}=1$. Then we introduce the stepwise constant (Brownian) Euler scheme $\bar{Y}^m=\big(\bar{Y}_{\frac{kT}{m}}\big)_{0\leq k\leq m}$ with step $\frac{T}{m}$ (defined by (\ref{Euler1})) of  $Y=(Y_t)_{t\in[0,T]}$ with $m\geq m_{b,\sigma}$. It is clear by induction on $k$ that there exists for every $k\in\{1,\ldots,m\}$ a function $\Theta_k:\R^{k+1}\rightarrow\R$ such that
$$
\bar{Y}_{t^m_k}=\Theta_k(y_0,\Delta W_{t^m_1},\ldots,\Delta W_{t^m_k})
$$
where for $(y_0,z_1,\ldots,z_k)\in\R^{k+1}$, 
\begin{eqnarray*}
	\Theta_k(y_0,z_1,\ldots,z_k)&=&\Theta_{k-1}(y_0,z_1,\ldots,z_{k-1})+\beta(t^m_{k-1},\Theta_{k-1}(y_0,z_1,\ldots,z_{k-1}))\frac{T}{m}+z_k \\
	                            &=&\left({\rm id}+\beta(t^m_{k-1},\cdot)\frac{T}{m}\right)\circ\Theta_{k-1}(y_0,z_1,\ldots,z_{k-1})+z_k.
\end{eqnarray*}	                            
Thus for every $i\in\{1,\ldots,k\}$,  $z_i\mapsto\Theta_k(y_0,z_1,\ldots,z_i,\ldots,z_k)$ is non-decreasing because $y\mapsto\left(y+\beta(t^m_{k-1},y)\frac{T}{m}\right)$ is non-decreasing for $m$ large enough, say $m\geq m_{b,\sigma}$. We deduce that if $F_m:\R^{m+1}\rightarrow\R$ is non-decreasing in each variables, then, for every $i\in\{1,\ldots,k\}$, 
$$z_i\mapsto F_m\left(y_0,\Theta_1(y_0,z_1),\ldots,\Theta_m(y_0,z_1,\ldots,z_m)\right) \mbox{ is non-decreasing}.$$
By the same reasoning, we deduce that for $G_m:\R^{m+1}\rightarrow\R$, non-increasing in each variables, we have for every $i\in\{1,\ldots,k\}$, 
$$z_i\mapsto G_m\left(y_0,\Theta_1(y_0,z_1),\ldots,\Theta_m(y_0,z_1,\ldots,z_m)\right) \mbox{ is non-increasing}.$$
Let $F_m$ and $G_m$ be the functions defined on $\R^{m+1}$ associated to $F$ and $G$ respectively by Lemma~\ref{lem:B1}. As $\beta$ has linear growth, $Y$ and its Euler scheme have polynomial moments at any order $p>0$. Then we can apply Proposition \ref{ComonotonyMultid} to deduce that
\begin{eqnarray*}
	\E\left[FG\left(\bar{Y}^m\right)\right]&=&\E\left[F_m\left(\left(\bar{Y}_{\frac{kT}{m}}\right)_{0\leq k\leq m}\right)G_m\left(\left(\bar{Y}_{\frac{kT}{m}}\right)_{0\leq k\leq m}\right)\right] \\
	&\geq&\E\left[F_m\left(\left(\bar{Y}_{\frac{kT}{m}}\right)_{0\leq k\leq m}\right)\right]\E\left[G_m\left(\left(\bar{Y}_{\frac{kT}{m}}\right)_{0\leq k\leq m}\right)\right]=\E\left[F\left(\bar{Y}^m\right)\right]\E\left[G\left(\bar{Y}^m\right)\right].
\end{eqnarray*}	

Note that if $F$ and $G$ are $C$-continuous with polynomial growth, so is $FG$. We derive from Theorem~\ref{weak}  that
$$\E\left[FG\left(\bar{Y}^m\right)\right]\underset{m\rightarrow\infty}{\longrightarrow}\E FG(Y), \quad \E\left[F\left(\bar{Y}^m\right)\right]\underset{m\rightarrow\infty}{\longrightarrow}\E F(Y), \quad \E\left[G\left(\bar{Y}^m\right)\right]\underset{m\rightarrow\infty}{\longrightarrow}\E G(Y),$$
therefore
$$\Cov\left(F(Y),G(Y)\right)\geq0.$$

To conclude the proof, we need to go back to the process $X$ by using the inverse Lamperti transform. Indeed, for every $t\in[0,T]$, $X_t=L^{-1}(t,Y_t)$, where $Y$ satisfies (\ref{edsY}). Let $F:\D([0,T],\R)\rightarrow\R$ $C$-continuous. Set
$$
\forall\alpha\in{\cal C}([0,T],\R), \quad \widetilde{F}(\alpha):=F\left(\left(L^{-1}(t,\alpha_t)\right)_{t\in[0,T]}\right).
$$
Assume first that $F$ and $G$ are bounded. The functional $\widetilde{F}$ is $C$-continuous owing to Proposition~\ref{prop: LamperCont}, non-decreasing (resp. non-increasing) since $L^{-1}(t,.)$ is for every $t\in [0,T]$ and is bounded. Consequently,
$$
\Cov\left(F(X),G(X)\right)=\Cov\left(\widetilde{F}(Y),\widetilde{G}(Y)\right)\geq0.
$$
To conclude we approximate $F$ and $G$ in a robust way with respect to the ``constraints'',  by a canonical truncation procedure, say
$$
F_M:=\max\Big((-M), \min\big(F,M\big)\Big), \quad M\in \N.
$$
If $F$ and $G$ have polynomial growth, it is clear that $\Cov\left(F_M(X),G_M(X)\right)\to \Cov\left(F(X),G(X)\right)$ as $M\to \infty$. \hfill$\cqfd$

\medskip \noindent {\bf Examples of admissible diffusions.} $\bullet$ {\em The Bachelier model}: This simply means that $X_t = \mu t +\sigma W_t$, $\sigma>0$, clearly fulfills the assumptions of Theorem~\ref{thm:IneqCovDiff}.

	\medskip	
	$\bullet$ {\em The Black-Scholes model}: The diffusion process $X$ is a geometric Brownian motion, solution to the SDE
	$$dX_t=rX_tdt+\vartheta X_tdW_t, \quad X_0=x_0>0,	$$
	where $r\in\R$ and $\vartheta>0$ are real numbers. The geometric Brownian motion lives in the open interval $I=(0,+\infty)$ and  $\beta(y)=\frac{r}{\vartheta}-\frac{\vartheta}{2}$ is  constant. One checks that $L(x)= \frac{1}{\sigma} \log\Big(\frac{x}{x_1}\Big)$ where $x_1\in (0,+\infty)$ is fixed.

	\medskip	
	$\bullet$ {\em The Hull-White model}: It is an elementary improvement of the Black-Scholes model where $\vartheta:[0,T]\to (0,+\infty)$  is a deterministic positive function $i.e.$ the diffusion process $X$ is a geometric Brownian motion solution of the SDE
	$$dX_t=rX_tdt+\vartheta(t) X_tdW_t, \quad X_0=x_0>0.$$
Then, elementary stochastic calculus shows that
\[
X_t =x_0 e^{rt-\frac 12\int_0^t \vartheta^2(s)ds +\int_0^t \vartheta(s)dW_s}= x_0e^{rt-\frac 12\int_0^t \vartheta^2(s)ds +B_{\int_0^t \vartheta^2(s)ds}}
\]
where $(B_u)_{u\ge 0}$ is a standard Brownian motion (the second equality follows form the Dambins-Dubins-Schwarz theorem).

Consequently $\displaystyle X_t= \varphi\Big(t,B_{\int_0^t \vartheta^2(s)ds}\Big)$ where the functional  $\displaystyle \xi\mapsto \Big(t\mapsto\varphi\Big(t,\xi\Big(\int_0^.\vartheta^2(s)ds\Big)\Big)\Big)$ defined on $\D([0,T_{\vartheta}],\R)$, $T_{\vartheta}= \int_0^T\vartheta^2(t)dt$, is $C$-continuous on ${\cal C}([0,T_{\vartheta}],\R)$. Hence for any $C$-continuous  $\R$-functional on $\D([0,T], \R)$, the $\R$-valued functional $\widetilde F$ defined by $\displaystyle \widetilde F(\xi) = F\Big(\varphi\Big(t,\xi\Big(\int_0^.\vartheta^2(s)ds\Big)\Big)\Big)$ is $C$-continuous on $\D([0,T_{\vartheta}],\R)$. Then, on can transfer the co-monotony property from $B$ to $X$.
	
	\medskip	
 	$\bullet$ {\em Local volatility model (elliptic case)}: More generally, it applies still with $I=(0,+\infty)$ to some usual extensions like the models with local volatility
	$$
	dX_t=rX_tdt+\vartheta(X_t) X_tdW_t, \quad X_0=x_0>0,
	$$	
	where $\vartheta:\R\rightarrow(\vartheta_0,+\infty)$, $\vartheta_0>0$, is a bounded, twice differentiable function satisfying $\left|\vartheta'(x)\right|\leq\frac{C}{1+ |x|}$ and $\left|\vartheta''(x)\right|\leq\frac{C}{1+|x|^2}$, $x\in (0,+\infty)$.

	\smallskip
	In this case $I= (0,+\infty)$ and, $x_1\in I$ being fixed, one has for every $x\in I$,   
	\[
	L(x) =\int_{x_1}^x \frac{d\xi}{\xi\vartheta(\xi)}
	\] 
	which clearly  defines an increasing homeomorphism from $I$ onto $\R$ since $\vartheta$ is bounded. 

Furthermore, one easily derives from the explicit form~(\ref{eq:betaprime}) and the condition~(\ref{eq:betaprime2}) that $\beta$ is Lipschitz as soon as the function 
\[
x\mapsto rx\frac{\vartheta'}{\vartheta}(x)+\frac{x^2\vartheta\vartheta''(x)}{2} +x\vartheta\vartheta'(x)\;\mbox{ is bounded on $(0,\infty)$}
\]
which easily follows from the assumptions made on $\vartheta$.

\medskip \noindent {\bf Extension to other classes of diffusions and models.} This general approach does not embody all situations: thus the true CEV model  does not fulfill the above assumptions. The $CEV$ model is a diffusion process $X$ following the SDE
$$dX_t=rX_tdt+\vartheta X_t^{\alpha}dW_t, \quad X_0=x_0,$$
where $\vartheta>0$ and $0<\alpha<1$ are real numbers. 

So this $CEV$ model, for which $I=(0,+\infty)$,  does not fulfill Definition~${\cal A}_{b,\sigma}$-$(iii)$. As a consequence $L(t, I)\neq \R$ is an open interval (depending on the choice of $x_1$. To be precise, if $x_1\in (0,+\infty)$ is fixed,
\[
L(x)=\frac{1}{\vartheta(1-\alpha)}\big(x^{1-\alpha}-x_1^{1-\alpha}\big),\quad x\in (0,+\infty)
\]
so that, if we set 
\[
J_{x_1}:= L(I)= \Big(-\frac{x_1^{1-\alpha}}{\vartheta(1-\alpha)},+\infty\Big),
\]
$L$ defines an homeomorphism from $I=(0,+\infty)$ onto $J_{x_1}$. Finally the function $\beta$ defined by 
\[
\beta(y) = \frac{r}{\vartheta}\big(\vartheta(1-\alpha)y+x_1^{1-\alpha}\big)-\frac{\alpha\vartheta}{2}\frac{1}{(\vartheta(1-\alpha)y+x_1^{1-\alpha})},\quad y\in J_{x_1}
\]
is non-decreasing with linear growth at $+\infty$. Now, tracing the lines of the above proof, in particular establishing weak existence and uniqueness of the solution  of the SDE~(\ref{edsweak}) in that setting,  leads to the same positive conclusion concerning the covariance inequalities for co-monotonic or anti-monotonic functionals.

\small
\bibliographystyle{plain}
\bibliography{biblio}

\end{document}